\documentclass[a4paper,11pt]{article}
\pdfoutput=1
\usepackage{jheppub}
\usepackage{graphicx}
\usepackage{epsfig}
\usepackage{subfig}
\usepackage{url}
\usepackage{comment}
\usepackage{dsfont}
\usepackage{amssymb}
\usepackage{slashed}
\usepackage{booktabs,multirow,tabularx}
\usepackage{braket}

\newcommand{\eg}{{\it e.g.}}
\newcommand{\ie}{{\it i.e.}}

\DeclareMathAlphabet{\pazocal}{OMS}{zplm}{m}{n}
\newcommand\sss{}
\newcommand\mydot{\!\cdot\!}

\newcommand\ep{\epsilon}
\def\abs#1{\left|#1\right|}

\def\p0{{\bigl.^3\hspace{-1mm}P^{[8]}_0}}

\def\to{\rightarrow}

\def\bqa{\begin{eqnarray}}
\def\eqa{\end{eqnarray}}
\def\bc{\begin{center}}
\def\bc{\end{center}}

\newcommand\allproco{\dot{{\cal R}}}

\newcommand\allprocnmoo{\allproco_{n-1}}
\newcommand\allprocno{\allproco_{n}}
\newcommand\nini{n_{\sss I}}
\newcommand\nlight{n_{\sss L}}
\newcommand\nlightB{\nlight^{\sss (B)}}
\newcommand\nlightR{\nlight^{\sss (R)}}

\newcommand\nheavy{n_{\sss H}}

\newcommand\avg{{\cal N}}
\newcommand\oavg{{\cal G}}
\newcommand\ident{{\cal I}}
\newcommand\amp{{\cal A}}
\newcommand\ampo{{\mathds A}}

\newcommand\ampnmoo{\ampo^{(n-1,0)}}
\newcommand\ampnto{\ampo^{(n,0)}}
\newcommand\ampnt{\amp^{(n,0)}}
\newcommand\ampnpot{\amp^{(n+1,0)}}
\newcommand\ampnl{\amp^{(n,1)}}
\newcommand\ampnll{\amp^{(n,\ell)}}
\newcommand\JetsB{J^{\nlightB}}

\newcommand\Quark{Q}
\newcommand\Antiquark{\bar{Q}^\prime}
\newcommand\QQ{Q\bar{Q}^\prime}
\newcommand\colorQ{c_{j_{\Quark}}}
\newcommand\colorQbar{c_{j_{\Antiquark}}}
\newcommand\colorQQ{c_{j_{\Quark}j_{\Antiquark}}}
\def\remove#1#2{#1\hspace{-#2truecm}\backslash}

\newcommand\isubrmv{\remove{i}{0.125}}

\newcommand\Qbarsubrmv{\remove{4}{0.125}}
\newcommand\Qbsubrmv{\remove{j_{\Antiquark}}{0.4}}

\newcommand\FKSpairs{{\cal P}_{\sss\rm FKS}}

\newcommand\xicut{\xi_{cut}}
\newcommand\ximax{\xi_{\rm max}}
\newcommand\deltaO{\delta_{\sss O}}
\newcommand\deltaI{\delta_{\sss I}}

\newcommand\Qop{\vec{Q}}
\newcommand\Fop{\vec{F}}
\newcommand\Jop{\vec{J}}
\newcommand\ampsq{{\cal M}}
\newcommand\ampsqo{{\mathds M}}
\newcommand\ampsqsoft{{\mathsf M}}

\newcommand\ampsqnmoo{\ampsqo^{(n-1,0)}}
\newcommand\ampsqnmootilde{\tilde{\ampsqo}^{(n-1,0)}}
\newcommand\ampsqsoftnmo{\ampsqsoft^{(n-1,0)}}
\newcommand\ampsqnt{\ampsq^{(n,0)}}
\newcommand\ampsqnto{\ampsqo^{(n,0)}}
\newcommand\ampsqnpot{\ampsq^{(n+1,0)}}
\newcommand\ampsqnl{\ampsq^{(n,1)}}

\newcommand\Ione{\ident_1}
\newcommand\Itwo{\ident_2}
\newcommand\xii{\xi_i}
\newcommand\yij{y_{ij}}
\newcommand\phii{\varphi_i}

\newcommand\xic{\left(\frac{1}{\xii}\right)_c}

\newcommand\omyijd{\left(\frac{1}{1-\yij}\right)_\delta}

\newcommand\Sfun{{\cal S}}
\newcommand\Sfunij{\Sfun_{ij}}

\newcommand\stepf{\Theta}
\newcommand\NC{N_{\sss c}}
\newcommand\CA{C_{\sss A}}
\newcommand\CF{C_{\sss F}}

\newcommand\BF{B_{\sss F}}
\newcommand\phsp{d\phi}
\newcommand\phispnmo{\phsp_{n-1}}
\newcommand\phspn{\phsp_{n}}

\newcommand\tphsp{d\widetilde{\phi}}

\newcommand\tphspnmoij{\tphsp_{n-1}^{ij}}

\newcommand{\madonia}{{\sc MadOnia}}
\newcommand{\helaconia}{{\sc HELAC-Onia}}
\newcommand{\mfks}{{\sc MadFKS}}
\newcommand{\mgamc}{{\sc MadGraph5\_aMC@NLO}}
\newcommand{\recola}{{\sc Recola}}
\newcommand{\sherpa}{{\sc Sherpa}}
\newcommand{\openloops}{{\sc OpenLoops}}
\newcommand{\gosam}{{\sc GoSam}}

\def\be{\begin{equation}}
\def\ee{\end{equation}}
\def\bea{\begin{eqnarray}}
\def\eea{\end{eqnarray}}

\def\dfrac{\displaystyle\frac}

\newcommand{\boldirrep}{\mathbf}
\newcommand{\irrepbase}[1]{\ensuremath{\boldirrep{#1}}}

\newlength{\irrepwidth}
\newlength{\irrepbarthickness}
\newlength{\irrepbarheight}
\newcommand{\irrepbarbase}[1]{%
    \settoheight{\irrepbarheight}{\irrepbase{#1}}%
    \settowidth{\irrepwidth}{\irrepbase{#1}}%
    \makebox[0pt][l]{\irrepbase{#1}}%
    \rule[1.2\irrepbarheight]{\irrepwidth}{\irrepbarthickness}%
}

\def\primes#1#2{\count0=#1 \loop \ifnum\count0>0 \advance\count0 by -1 #2\repeat}
\newcommand{\irrep}[2][0]{\ensuremath{\irrepbase{#2}^{\primes{#1}{\prime}}}}
\newcommand{\irrepbar}[2][0]{\ensuremath{\irrepbarbase{#2}^{\primes{#1}{\prime}}}}

\usepackage{tikz}
\usetikzlibrary{positioning,arrows}
\usetikzlibrary{decorations.pathmorphing}
\usetikzlibrary{decorations.markings}
\usetikzlibrary{shapes.geometric}
\usepackage{endnotes}
\tikzset{
    vector/.style={decorate, decoration={snake}, draw},
    provector/.style={decorate, decoration={snake,amplitude=2.5pt}, draw},
    antivector/.style={decorate, decoration={snake,amplitude=-2.5pt}, draw},
    fermion/.style={draw=black,
      postaction={decorate},decoration={markings,mark=at position .55
        with {\arrow[draw=black]{>}}}},
    fermionbar/.style={draw=black, postaction={decorate},
                       decoration={markings,mark=at position .55 with {\arrow[draw=black]{<}}}},
    fermionnoarrow/.style={draw=black},
    gluon/.style={decorate, draw=black,decoration={coil,amplitude=4pt, segment length=6pt}},
    scalar/.style={dashed,draw=black,
      postaction={decorate},decoration={markings,mark=at position .55
        with {\arrow[draw=black]{>}}}},
    scalarbar/.style={dashed,draw=black,
      postaction={decorate},decoration={markings,mark=at position .55
        with {\arrow[draw=black]{<}}}},
    scalarnoarrow/.style={dashed,draw=black},
    electron/.style={draw=black,
      postaction={decorate},decoration={markings,mark=at position .55
        with {\arrow[draw=black]{>}}}},
    bigvector/.style={decorate, decoration={snake,amplitude=4pt}, draw},
}

\normalbaselines

\title{FKS subtraction for quarkonium production at NLO}

\author{Ajjath A H, Hua-Sheng Shao, and Lukas Simon}

\affiliation{Laboratoire de Physique Th\'eorique et Hautes Energies (LPTHE), UMR 7589, Sorbonne Universit\'e et CNRS, 4 place Jussieu, 75252 Paris Cedex 05, France}

\emailAdd{aabdulhameed@lpthe.jussieu.fr}
\emailAdd{huasheng.shao@lpthe.jussieu.fr}
\emailAdd{lsimon@lpthe.jussieu.fr}

\abstract{We extend the local infrared-divergence subtraction formalism, originally proposed by Frixione, Kunszt and Signer (FKS), to calculate short-distance (differential) cross section for any inclusive process involving a quarkonium particle in non-relativistic QCD (NRQCD) factorisation at next-to-leading order (NLO) accuracy in the strong coupling constant $\alpha_s$. The new formulas are generally applicable to the production of an S- or P-wave quarkonium state in association with any number of elementary particles. The main new ingredients derived in this paper are the local and integrated soft counterterms for the colour-singlet and colour-octet P-wave bound states. It, therefore, paves the way to the automation of the NLO calculations for heavy quarkonium inclusive and associated production processes.}

\keywords{NLO Computations, IR divergences, Quarkonium, QCD, NRQCD}

\begin{document}
\maketitle
\flushbottom
\section{Introduction}

The theoretical interpretations of the analysed LHC data nowadays heavily rely on precision calculations of short-distance cross sections, as well as precise Monte Carlo event simulations in the context of both the Standard Model (SM) and its extensions. As of today, next-to-leading order (NLO) QCD calculations and their interface to general-purpose parton-shower Monte Carlo programmes have been automated, as seen, \eg, in the \mgamc~\cite{Alwall:2014hca} framework, for elementary-particle production processes~\footnote{Strictly speaking, this statement only applies to processes with not-too-high particle multiplicity, given the limitations of computing resources. It is also possible that special issues may arise for particular problems that have not yet been addressed in the automated codes. Additionally, this consideration does not cover processes involving loops at their lowest order. In the context of this paper, when we refer to a similar statement, we mean it in a loose sense.} in the SM and in a large class of new physics models. NLO electroweak corrections have also been automated in recent years by several collaborations, such as \mgamc~\cite{Frederix:2018nkq} and \sherpa~\cite{Schonherr:2017qcj} along with external one-loop matrix element providers like \recola~\cite{Biedermann:2017yoi}, \openloops~\cite{Kallweit:2017khh} or \gosam~\cite{Chiesa:2017gqx}. Some low-particle-multiplicity processes, solved using customised methods, have been extended to next-to-NLO (NNLO) and even next-to-NNLO (N$^3$LO) accuracies. However, these significant theoretical developments are currently restricted to point-like elementary particles, and they cannot be directly applied to non-relativistic bound states like heavy quarkonia. This limitation can be roughly understood as the latter case intrinsically involving multiple scales and thus requiring simultaneous consideration in relativistic quantum field theories (QFT), such as QCD, and their non-relativistic low-energy effective field theories (EFT), \eg, non-relativistic QCD (NRQCD)~\cite{Bodwin:1994jh}. This introduces additional conceptual and technical challenges on the theory side. 

Therefore, theoretical progress in perturbative calculations for heavy quarkonium is far less advanced. Fairly speaking, automation has been achieved only for tree-level quarkonium processes (single quarkonium in \madonia~\cite{Artoisenet:2007qm} and one or more quarkonia in \helaconia~\cite{Shao:2012iz,Shao:2015vga}). NLO and even higher-order calculations, in many cases, are indispensable not only for precision or accuracy but also for a qualitative understanding. Due to conservation laws at the quantum level, short-distance cross sections of quarkonium production often receive giant $K$ factors~\footnote{Thanks to recent advancements in parton showers~\cite{Cooke:2023ldt}, there is a chance that the problem of giant $K$ factors can be addressed through the matching and merging of matrix elements and parton showers.} from high-order radiative corrections (see, \eg, ref.~\cite{Shao:2018adj} and references therein). This places the theoretical interpretations of measured quarkonium data on shaky ground if higher-order radiative corrections are not well under control.

The physics that we can learn from quarkonium is, however, no less interesting. In fact, quarkonia provide powerful and sometimes even unique tools that allow us to conduct rich particle and nuclear physics studies~\cite{Chapon:2020heu}. For instance, they can be used to determine the structures of free nucleons~\cite{Boer:2011fh,denDunnen:2014kjo,Lansberg:2017dzg,Jones:2015nna,Jones:2016ldq,Lansberg:2014swa,Shao:2016wor,Lansberg:2016rcx,Lansberg:2016muq,Lansberg:2017chq,Shao:2019qob,Flett:2020duk} and nuclei~\cite{Kusina:2017gkz,Shao:2020acd,Shao:2020kgj,Guzey:2013xba}. Due to their sequential binding energies, quarkonia are widely used as a thermometer of quark-gluon plasma produced in heavy-ion collisions~\cite{Matsui:1986dk,Digal:2001ue} to probe the hot-and-dense QCD. They signify the presence of a QCD phase transition by either disappearing~\cite{Matsui:1986dk,Digal:2001ue} or being abundantly
produced, hinting at collective heavy-quark effects~\cite{Braun-Munzinger:2000csl,Thews:2000rj}. A golden channel of searching for QCD instantons, arising from the non-trivial topological structure of QCD vacuum which is believed to be crucial in understanding quark confinement, was suggested to study charmonium decays~\cite{Bjorken:2000ni,Zetocha:2002as}. Quarkonia were also proposed as a good system to investigate the non-linear dynamics of QCD, also known as parton saturation~\cite{Kharzeev:2008nw}, in addition to the well-known DGLAP and BFKL dynamics. They
have been readily used to extract the fundamental SM parameters, \eg, the strong coupling
constant $\alpha_s$~\cite{Brambilla:2007cz}, the Higgs-charm Yukawa coupling~\cite{Bodwin:2013gca,ATLAS:2015vss,ATLAS:2018xfc}, the CKM matrix elements~\cite{Belle:2001zzw,BaBar:2001pki}, as well as the masses of the charm and bottom quarks~\cite{Mateu:2017hlz}. Some exotic QCD hadrons, such as the fully-charmed tetraquark $X(6900)$~\cite{LHCb:2020bwg,ATLAS:2023bft,CMS:2023owd} and the first-observed pentaquark states $P_c^+(4380)$ and $P_c^+(4450)$~\cite{LHCb:2015yax}, were also discovered in final states with quarkonia.

It has been well understood that the perturbative calculations at NLO and beyond in QFT encounter ultraviolet (UV) and infrared (soft and collinear) divergences in the intermediate steps. While the UV divergences are removed through the renormalisation procedure, handling infrared (IR) divergences is more intricate. First, IR divergences can only be cancelled for so-called IR-safe observables, thanks to the Kinoshita-Lee-Nauenberg (KLN) theorem~\cite{Kinoshita:1962ur,Lee:1964is} and factorisation theorems/conjectures. Second, in a generic situation, phase-space integration must be carried out numerically using Monte Carlo importance sampling methods. However, this is hindered by the IR singularities present in real radiative corrections. Both IR
subtraction and phase-space slicing approaches are employed to overcome such complications in the real-emission contributions, with IR subtraction methods known to outperform slicing approaches. Therefore, the former serves as the backbone of contemporary NLO automation codes. The two widely adopted NLO subtraction methods were originally proposed by Frixione, Kunszt and Signer
(FKS)~\cite{Frixione:1995ms,Frixione:1997np}, and Catani and Seymour~\cite{Catani:1996jh,Catani:1996vz}. They are usually referred to as the FKS and dipole subtraction schemes, respectively. Both methods work well for both simple and complicated processes involving elementary particles. Initially devised for massless coloured particles, they have been generalised to include massive coloured particles~\cite{Phaf:2001gc,Catani:2002hc,Frederix:2009yq}. Both schemes have been implemented in various public computer programmes~\cite{Gleisberg:2007md,Frederix:2008hu,Frederix:2009yq,Czakon:2009ss,Hasegawa:2009tx,Alioli:2010xd}.

On the other hand, regarding the problem of heavy quarkonium production in NRQCD, NLO calculations in the literature are almost exclusively carried out using the slicing methods~\cite{Harris:2001sx}, with only a few exceptions. The earliest exception~\footnote{In simple cases, such as $2\to 1$ or $1\to 2$ underlying Born processes, fully analytical $d$-dimensional phase space integrations of real emissions are possible, similar to the approach taken in ref.~\cite{Petrelli:1997ge}.} pertains to inclusive colour-singlet S-wave quarkonium production at hadron colliders~\cite{Campbell:2007ws}, where the dipole counterterms for massless quarks and gluons suffice, as the colour-singlet S-wave quarkonium does not exhibit any IR singularities. The remaining exceptional NLO calculations~\cite{Qiu:2020xum,Butenschoen:2022wld} employ the dipole formalism~\footnote{The formalism was derived under the circumstance of a $2\to2$ underlying Born process.} developed for processes featuring an S- or P-wave quarkonium alongside massless quarks and gluons, as outlined in ref.~\cite{Butenschoen:2019lef,Butenschoen:2020mzi}. 
If one considers a process involving both a P-wave quarkonium and massive partons, such as the associated production processes of quarkonium and heavy quarks, new (yet unknown) dipole terms may be required. The aim of this paper is to incorporate heavy quarkonium into the FKS subtraction scheme. As we demonstrate later, the formalism is general enough to be applied to arbitrary processes involving a quarkonium and massless/massive partons. Therefore, the scope of the phenomenological applications using our formalism is anticipated to be broader than that of the dipole formalism derived in refs.~\cite{Butenschoen:2019lef,Butenschoen:2020mzi}.

Since our ultimate goal is to automate NLO computations for quarkonium production processes within the \mgamc\ framework, we will closely adhere to the notations and conventions of the original FKS formulation~\cite{Frixione:1995ms} and the \mfks\ paper~\cite{Frederix:2009yq}. For the sake of completeness and self-consistency, we will reproduce some known equations from the literature. We hope this will aid in improving the readability of the article, especially for readers who may not be familiar with the two aforementioned papers. 

The remaining context of this paper is organised as follows. In section \ref{sec:quarkonium}, we elucidate how to obtain short-distance cross sections for quarkonium production within NRQCD factorisation. This enables us to establish a few notations used throughout the paper. We derive the soft limit of the (squared) amplitudes in the real radiative corrections for a single quarkonium production in section \ref{sec:softlimit}. The local and integrated FKS subtraction counterterms are
given in section \ref{sec:FKS}. We perform a few cross-checks to ensure the validity of our formalism in section \ref{sec:checks}, and finally draw our conclusions in section \ref{sec:summary}. Appendix \ref{sec:eikonal} presents the analytic expressions for eikonal tensor integrals that appear in the integrated counterterms. The universal IR poles of one-loop matrix elements can be found in appendix \ref{sec:IRpoles4virt}.

\section{Quarkonium production in NRQCD factorisation\label{sec:quarkonium}}

In NRQCD factorisation~\cite{Bodwin:1994jh}, the inclusive production of a heavy quarkonium
factorises into the perturbative short-distance cross section and
the non-perturbative long-distance matrix elements (LDMEs):
\begin{equation}
d\sigma(AB\rightarrow H+X) = \sum_n \Big(
\sum_{a,b,X}\int dx_a dx_b f_{a/A}(x_a) f_{b/B}(x_b) d\hat{\sigma}(ab \rightarrow \QQ[n] + X) 
           \Big) \braket{{\mathcal O}^H_n},                          
\end{equation}
where $f_{a/A}$ and $f_{b/B}$ are parton distribution functions (PDFs) of partons $a$ and $b$ in the initial hadrons $A$ and $B$.  $d\hat{\sigma}(ab \rightarrow \QQ[n] + X)$ describes the short distance production of a $\QQ$ pair~\footnote{Note that the flavours of the two constituent heavy quarks do not have to be identical. For instance, the constituent quarks of $B_c^+$ are a charm quark and a bottom antiquark.} in a specific colour irreducible representation $C$, with spin $S$ and orbital angular momentum state $L$ denoted as $n=\bigl.^{2S+1}\hspace{-1mm}L^{[C]}_J$ following the usual spectroscopic notation, and the LDME, $\braket{{\mathcal O}^H_n}$, represents the hadronisation of the heavy quark pair into the physical quarkonium state $H$. An important consequence of NRQCD factorisation is the prediction that the LDMEs do not depend on the details of the
hard process, and their values can be extracted from experiments, lattice QCD calculations~\cite{Bodwin:1996tg,Brambilla:2021abf} or potential models~\cite{Eichten:1995ch}.

In principle, for a specific quarkonium, there is an infinite number of Fock states $n$ and an infinite number of LDMEs $\braket{{\mathcal O}^H_n}$ to be determined, which limits the prediction power. Thanks  to the power counting rules in NRQCD, only a limited number of Fock states should be involved in the calculations up to
a specific order of $v$, where $v$ ($v\ll 1$) is the relative velocity of the heavy quark pair $\QQ$. We express a given Fock state using the spectroscopic notation $\bigl.^{2S+1}\hspace{-1mm}L^{[C]}_J$.

Our focus is to evaluate the perturbative short-distance coefficients, which can be determined from the amplitudes of $\Quark$ and $\Antiquark$ production with necessary operations in order to constrain the heavy quark pair $\QQ$ into a specific quantum state $n$. A convenient way to do so is by performing projections. Let us consider a general $2\to n$ process involving only open $\Quark$ and $\Antiquark$ quarks~\footnote{We do not restrict the numbers of $\Quark$ and $\Antiquark$ appearing in the process.}, denoted as $\ident_1\ident_2\to \ident_3\ident_4\cdots \ident_{n+2}$, where the identity of the $k$-th particle is denoted by $\ident_k$, and $\ident_1=a, \ident_2=b, \ident_3=\Quark$, and $\ident_4=\Antiquark$. Following the same notation as ref.~\cite{Frederix:2009yq}, we can write the process as $r=\left(\ident_1,\ldots,\ident_{n+2}\right)$. We denote the corresponding (tree-level) amplitude as $\ampnt(r)$. Let us also define the amputated amplitude $\Gamma^{(n,0)}(r)$ by removing the external wavefunctions of $\Quark$ and $\Antiquark$ that will form a bound state $\QQ$, \ie,
\begin{eqnarray}
\ampnt(r)&=&\bar{u}_{\lambda_\Quark}(k_\Quark)\Gamma^{(n,0)}(r)v_{\lambda_{\Antiquark}}(k_{\Antiquark}),
\end{eqnarray}
where $u_\lambda$ and $v_\lambda$ are Dirac spinors and $\lambda_{\Quark/\Antiquark}$ are helicities of $\Quark$ and $\Antiquark$, respectively.

Since $\irrep{3}\otimes\irrepbar{3}=\irrep{1}\oplus \irrep{8}$, we only have colour singlet $C=1$ and colour octet $C=8$ in the decomposition of $\ampnt(r)$. The colour projectors are :
\begin{eqnarray}
\mathbb{P}_{C=1}&=&\frac{\delta_{c_4c_3}}{\sqrt{\NC}}, \nonumber \\ 
\mathbb{P}_{C=8} &=&\sqrt{2} t_{c_4c_3}^{c_{34}},
\end{eqnarray}
where $c_3,c_4$ are the colour indices of $\Quark$ and $\Antiquark$, and $t^{c_{34}}$ is the Gell-Mann matrix. In other words, we can define the following two amplitudes from $\ampnt(r)$:
\begin{eqnarray}
\ampnt_{\left\{[C=1]\right\}}(r)&=&\sum_{c_3,c_4}{\mathbb{P}_{C=1}\ampnt(r)},\nonumber\\
\ampnt_{\left\{[C=8]\right\}}(r)&=&\sum_{c_3,c_4}{\mathbb{P}_{C=8}\ampnt(r)},
\end{eqnarray}
where we have explicitly summed over the colour indices $c_3, c_4$.

Similarly, we only have spin singlet $S=0$ and spin triplet $S=1$ for the $\QQ$ pair. The spin projectors for the heavy quark momenta $k_{\Quark}^\mu=\frac{m_\Quark}{m_{\Quark}+m_{\Antiquark}}K^\mu + q^\mu$ and $k_{\Antiquark}^\mu = \frac{m_{\Antiquark}}{m_{\Quark}+m_{\Antiquark}}K^\mu - q^\mu$  are given by
\begin{eqnarray}
\mathbb{P}_{S=0} &= & \frac{1}{2\sqrt{2m_{\Quark}m_{\Antiquark}}} \bar{v}_{\lambda_{\Antiquark}}(k_{\Antiquark})\gamma_5  u_{\lambda_{\Quark}}(k_{\Quark}), \nonumber \\
\mathbb{P}_{S=1} &= & \frac{1}{2\sqrt{2 m_{\Quark}m_{\Antiquark}}} \bar{v}_{\lambda_{\Antiquark}}(k_{\Antiquark})\slashed{\varepsilon}_{\lambda_s}^*(K)u_{\lambda_{\Quark}}(k_{\Quark}),
\end{eqnarray} 
where $\varepsilon_{\lambda_s}^*(K)$ is the polarisation vector for the spin-$1$ $\QQ$ with its spin quantum number as $\lambda_s=\pm1,0$.
Here, $K$ is the four-momentum of the $\QQ$ pair, $q$ is the relative momentum between the two constituent heavy quarks, and $m_{\Quark}, m_{\Antiquark}$ are  the heavy quark masses of $\Quark$ and $\Antiquark$, respectively. For simplicitly, we just denote $\tilde{\gamma}_0=\gamma_5$ and $\tilde{\gamma}_1=\slashed{\varepsilon}_{\lambda_s}^*(K)$. Thus, the amplitudes can be further decomposed into two spin configurations
\begin{eqnarray}
\ampnt_{\left\{S\right\}}(r)&=&\sum_{\lambda_{\Quark},\lambda_{\Antiquark}}{\mathbb{P}_{S}\ampnt(r)}\nonumber\\
&=&\sum_{\lambda_{\Quark},\lambda_{\Antiquark}}{\frac{1}{2\sqrt{2m_{\Quark}m_{\Antiquark}}} \bar{v}_{\lambda_{\Antiquark}}(k_{\Antiquark})\tilde{\gamma}_Su_{\lambda_{\Quark}}(k_{\Quark})\bar{u}_{\lambda_\Quark}(k_\Quark)\Gamma^{(n,0)}(r)v_{\lambda_{\Antiquark}}(k_{\Antiquark})}\nonumber\\
&=&\frac{1}{2\sqrt{2m_{\Quark}m_{\Antiquark}}}{\rm Tr}_{\gamma}\left[\left(\slashed{k}_{\Antiquark}-m_{\Antiquark}\right)\tilde{\gamma}_S \left(\slashed{k}_{\Quark}+m_{\Quark}\right)\Gamma^{(n,0)}(r)\right].
\end{eqnarray}
Together with both spin and colour configurations, the amplitude takes the form:
\begin{eqnarray}
\ampnt_{\left\{[C],S\right\}}(r)&=&\sum_{\lambda_{\Quark},\lambda_{\Antiquark}}{\mathbb{P}_{S}\ampnt_{\left\{[C]\right\}}(r)}\nonumber\\
&=&\sum_{\lambda_{\Quark},\lambda_{\Antiquark}}{\sum_{c_3,c_4}{\mathbb{P}_{S}\mathbb{P}_{[C]}\ampnt(r)}}\nonumber\\
&=&\sum_{c_3,c_4}{\mathbb{P}_{[C]}\frac{1}{2\sqrt{2m_{\Quark}m_{\Antiquark}}}{\rm Tr}_{\gamma}\left[\left(\slashed{k}_{\Antiquark}-m_{\Antiquark}\right)\tilde{\gamma}_S \left(\slashed{k}_{\Quark}+m_{\Quark}\right)\Gamma^{(n,0)}(r)\right]}.
\end{eqnarray}
Here, the spin projection operator commutes with the colour projection operator. Note that the trace ${\rm Tr}_{\gamma}$ in the Dirac spinor space does not necessarily represent a real trace that we need to compute. Its evaluation depends on how the other fermion lines are organised in the amputated amplitude $\Gamma^{(n,0)}(r)$.

The non-relativistic nature, \ie, in the rest frame of $\QQ$, $q\ll \sqrt{m_{\Quark}m_{\Antiquark}}$, allows us to expand the amplitudes into the series of $v\sim q/\sqrt{m_{\Quark}m_{\Antiquark}}\ll 1$. This gives us the eigenfunctions of the orbital angular momentum operator. The projection on a state with orbital angular momentum $L$ is obtained by differentiating $L=0,1,\ldots$ (\textit{a la} $S,P,\ldots$ waves)
times the spin-colour projected amplitude with respect to the relative momentum $q$ of the heavy quarks
in the $\QQ$ rest frame, and then setting $q\rightarrow 0$. Considering only $L=0,1$ states, which we are only interested in at this stage, the amplitude takes the form:
\begin{eqnarray}
\ampnt_{\left\{[C], S, L\right\}}(r) & = & \left[\left(\varepsilon^{\mu,*}_{\lambda_l}(K)\frac{d}{dq^\mu}\right)^L \ampnt_{[C],S}(r)\right]_{q=0},
\label{AmpQ}
\end{eqnarray}
where $\varepsilon^{\mu,*}_{\lambda_l}(K)$ is the polarisation vector for $L=1$ orbital angular momentum with $\lambda_l=\pm1,0$.
Since the spin projectors depend on the relative momentum $q$, the orbital angular momentum expansion must be carried out after projecting onto the given spin configuration. 

Finally, the total angular momentum $J$ is uniquely determined by $L$ or $S$ unless $L\neq 0$ and $S\neq 0$. In our specific case of interest, this can only occur when $L=1$ and $S=1$. In the latter case, we know how to determine $J=0,1,2$ and $\lambda_j=-J,-J+1,\ldots, J-1,J$ from quantum mechanics, \ie,
\begin{eqnarray}
\varepsilon^{\mu \nu,*}_{J,\lambda_j}(K)&=&\sum_{\lambda_s,\lambda_l}{\langle J, \lambda_j| 1,\lambda_l; 1,\lambda_s\rangle \varepsilon^{\mu*}_{\lambda_l}(K)\varepsilon^{\nu*}_{\lambda_s}(K)},\label{eq:projLS2J}
\end{eqnarray}
where $\langle J, \lambda_j| 1,\lambda_l; 1,\lambda_s\rangle$ is the Clebsch-Gordan coefficient. For $J=0, 1$, the expressions are
\begin{eqnarray}
\varepsilon^{\mu \nu,*}_{0,0}(K)&=&\frac{1}{\sqrt{3}}\left(g^{\mu \nu}-\frac{K^\mu K^\nu}{K^2}\right),\nonumber\\
\varepsilon^{\mu \nu,*}_{1,\lambda_j}(K)&=&-\frac{i}{\sqrt{2}}\epsilon^{\mu \nu \alpha \beta }\frac{K_{\alpha}}{\sqrt{K^2}}\varepsilon^{*}_{\lambda_{j},\beta}(K),
\end{eqnarray}
where $\epsilon^{\mu \nu \alpha \beta }$ is the Levi-Civita tensor.
Thus, we obtain the amplitude $\ampnt_{\left\{[C], S, L,J\right\}}(r)$ for a given quantum number $\bigl.^{2S+1}\hspace{-1mm}L^{[C]}_J$. When the product $L S=0$, we find $\ampnt_{\left\{[C], S, L,J\right\}}(r)=\ampnt_{\left\{[C], S, L\right\}}(r)$. On the other hand, if $L=S=1$, we have
\begin{eqnarray}
\ampnt_{\left\{[C],1,1,J\right\}}(r)&=&\sum_{\lambda_s,\lambda_l}{\langle J, \lambda_j| 1,\lambda_l; 1,\lambda_s\rangle \ampnt_{\left\{[C],1,1\right\}}(r)}\nonumber\\
&=&\frac{1}{2\sqrt{2m_\Quark m_{\Antiquark}}}\sum_{c_3,c_4}\mathbb{P}_{[C]}\varepsilon_{J,\lambda_j}^{\mu \nu,*}(K)\nonumber\\
&&\times\left.\frac{d}{dq^\mu}{\rm Tr}_\gamma\left[\left(\slashed{k}_{\Antiquark}-m_{\Antiquark}\right)\gamma_\nu \left(\slashed{k}_{\Quark}+m_{\Quark}\right)\Gamma^{(n,0)}(r)\right]\right|_{q=0}\,.
\end{eqnarray}

After all of the above preparations, we can now glue $\ident_3=\Quark$ and $\ident_4=\Antiquark$ as a new single particle $\ident_{3\oplus 4}=\QQ[\bigl.^{2S+1}\hspace{-1mm}L^{[C]}_J]$ with four-momentum $K^\mu$ and invariant mass $\sqrt{K^2}=m_{\QQ}=m_{\Quark}+m_{\Antiquark}$ in the non-relativistic limit. The new process is denoted as
\begin{eqnarray}
\dot{r}&=&r^{3\oplus 4,\Qbarsubrmv}=
\left(\ident_1,\ident_2,\ident_{3\oplus 4},\remove{\ident}{0.2}_{4},\ldots\ident_{n+2}\right)\,.
\label{QQproc}
\end{eqnarray}
The amplitude for the process $\dot{r}=r^{3\oplus 4,\Qbarsubrmv}$ is
\begin{eqnarray}
\ampnmoo(\dot{r})&=&\ampnt_{\left\{[C], S, L,J\right\}}(r).
\end{eqnarray}
Note that the final state symmetry must be applied at the level of $\dot{r}$ not the initial $r$, while the phase space integration should also be carried out at the level of $\dot{r}$. The partonic cross section can be written as
\begin{eqnarray}
d\hat{\sigma}(\dot{r})&=&\frac{1}{\avg(\dot{r})}\underbrace{\frac{1}{(2J+1)N_{[C]}}\frac{m_{\Quark}+m_{\Antiquark}}{2m_{\Quark}m_{\Antiquark}}}_{=\oavg(\dot{r})}\left(\ampsqnmoo(\dot{r})\JetsB\right)\phispnmo(\dot{r}),
\end{eqnarray}
where the amplitude square is given by
\begin{eqnarray}
\ampsqnmoo(\dot{r})&=&\frac{1}{2s}\frac{1}{\omega(\Ione)\omega(\Itwo)}
\mathop{\sum_{\rm colour}}_{\rm spin}\abs{\ampnmoo(\dot{r})}^2\nonumber\\
=\ampsqnt_{\left\{[C],S, L,J\right\}}(r)&=&\frac{1}{2s}\frac{1}{\omega(\Ione)\omega(\Itwo)}
\mathop{\sum_{\rm colour}}_{\rm spin}\abs{\ampnt_{\left\{[C],S, L,J\right\}}(r)}^2,
\end{eqnarray}
$\avg(\dot{r})$ is the final state symmetry factor, $N_{[C=1]}=2\NC, N_{[C=8]}=\NC^2-1$ with $N_c=3$ in QCD, and $\phispnmo(\dot{r})$ is the $(n-1)$-body phase space measure. The Mandelstam variable $s=(k_1+k_2)^2=2k_1\cdot k_2$, 
and $\omega(\ident)$ is the product of spin and colour degrees of 
freedom for the particle $\ident$. The condition with the measurement $\nlightB$-jet function $\JetsB$, where $\nlightB$ is the number of the light partons in the underlying Born, is sufficient to prevent the appearance of phase-space singularities in the Born-like quantities. Without losing generality, we can always assume these cuts to be equivalent to the request of having either $\nlightB$ and $\nlightB+1$ jets in the final state for an NLO computation. The same procedure can be iterated if we have more-than-one quarkonia.

\section{Soft limit\label{sec:softlimit}}

At NLO, we have contributions coming from one-loop virtual corrections and real emissions besides Born. Due to the complexities introduced by bound states, it is necessary to derive new local and integrated FKS counterterms to handle the IR divergences in real contributions at NLO. We observe that, since the constituent quarks $\Quark$ and $\Antiquark$ are massive, we can recycle the counterterms for collinear and soft-collinear origins. What we need to deal with beyond elementary particle production is the soft but non-collinear part. This includes two new components. The first involves the usual soft counterterms that locally cancel singularities of real emissions and their one-body phase space integrated counterparts. The second consists of additional integrated counterterms resulting from the renormalisation of LDMEs, analogous to the usual initial/final collinear counterterms but originating from a soft origin. Similar to the latter, which are necessary to cancel the remaining IR divergences protected by the collinear factorisation in perturbative QCD, the former are a consequence of the NRQCD factorisation formalism. In this section, we begin by considering the soft limit~\footnote{The earlier analysis of the soft limit tailored for specific processes can be found in the literature, such as in sect. 4 of ref.~\cite{Petrelli:1997ge} for $2\to 1$ processes.} of the quarkonium real emission amplitudes and their squares, following the procedure outlined before.

\subsection{Soft limit at the amplitude level\label{sec:softamp}}

Let us consider a $2\to n+1$ real process denoted as $r=(\ident_1,\ident_2,\ident_3,\ident_4,\ldots,\ident_{n+3})$. Following the approach in ref.~\cite{Frederix:2009yq}, and without loss of generality, we can always reorder the final-state particles such that the final massless (anti-)quarks and gluons are $3\leq i\leq \nlightR+2$ and the strongly-interacting particles are $\nini\leq j \leq \nlightR+\nheavy+2$, where $\nini=1, 2, 3$ for hadron-hadron, lepton-hadron, and lepton-lepton collisions. Here, $\nheavy$ represents the number of massive coloured partons in the final state. Simultaneously, we have $\ident_{\nlightR+\nheavy+1}=\Quark$ and $\ident_{\nlightR+\nheavy+2}=\Antiquark$. Consequently, we always have $\nheavy\geq 2$. For simplicity, in the following discussion, we denote $j_\Quark=n_L^{(R)}+n_H+1$ and $j_{\Antiquark}=n_L^{(R)}+n_H+2$. In the soft limit of $\ident_i=g$ with $3\leq i\leq \nlightR+2$, the real amplitude can be expressed in terms of the reduced Born amplitude multiplied with an additional eikonal factor. If the soft gluon with four-momentum $k_i$ is emitted from an external leg $\ident_j$ ($j\neq i$) with four-momentum $k_j$, the real emission amplitude takes the form in the soft or eikonal approximation: 
\begin{align}
\lim_{k_i \rightarrow 0}{\ampnpot(r)} = g_s\frac{k_j \cdot \varepsilon_{\lambda_i}^*(k_i)}{k_j\cdot k_i} \Qop(\ident_j)  \ampnt(r^{\isubrmv}),
\end{align}
where the reduced process is $r^{\isubrmv}=\left(\ident_1,\ldots\remove{\ident}{0.2}_i,\ldots\ident_j,
\ldots\ident_{n+3}\right)$, and $\Qop(\ident) $ represents the colour generator associated with the particle $\ident$, and $g_s=\sqrt{4\pi\alpha_s}$ is the strong coupling. Depending on the particle species $\ident$, we have
\begin{align}\label{SUN}
\Qop(\ident) = \{t^a\}_{a=1}^{8} ,\quad  \{-t^{aT}\}_{a=1}^8, \quad \{T^a\}_{a=1}^8 \quad \ident\in {\irrep{3},\irrepbar{3},\irrep{8}}  
\end{align}
with $t^a$ and $T^a$ being the SU(3) generators in the fundamental and adjoint representations, respectively. We take the final quark or initial antiquark as $\irrep{3}$, while the initial quark and final antiquark are taken as $\irrepbar{3}$. The matrix element of the adjoint representation is $T^a_{bc}=-if_{abc}$ with $f_{abc}$ being the anti-symmetric structure constants. 

After performing the quantum number projection, as described in sect.~\ref{sec:quarkonium}, we obtain a similar eikonal decomposition as long as $j\neq j_{\Quark}, j_{\Antiquark}$, \ie,
\begin{align}
\lim_{k_i \rightarrow 0}{\ampnpot_{\left\{[C],S, L,J\right\}}(r)} = g_s\frac{k_j \cdot \varepsilon_{\lambda_i}^*(k_i)}{k_j\cdot k_i} \Qop(\ident_j)  \ampnt_{\left\{[C],S, L,J\right\}}(r^{\isubrmv}).
\end{align}
However, when $j=j_{\Quark}$ or $j=j_{\Antiquark}$, we should pay special attention to it. It will be always convenient to sum the contributions of $j=j_{\Quark}$ and $j=j_{\Antiquark}$ together, which we will adopt in the following. Now, let us consider the case of $j=j_{\Quark},j_{\Antiquark}$.

\subsubsection{Colour projection}

With the procedure outlined in sect.~\ref{sec:quarkonium}, the colour projected amplitudes are
\begin{eqnarray}
&&\lim_{k_i \rightarrow 0}{\ampnpot_{\left\{[C=1]\right\}}(r)} \nonumber\\
&=& g_s\sum_{\colorQ,\colorQbar}{\mathbb{P}_{[C=1]}\left[\frac{k_{j_{\Quark}} \cdot \varepsilon_{\lambda_i}^*(k_i)}{k_{j_{\Quark}}\cdot k_i} \Qop(\ident_{j_{\Quark}})+\frac{k_{j_{\Antiquark}} \cdot \varepsilon_{\lambda_i}^*(k_i)}{k_{j_{\Antiquark}}\cdot k_i} \Qop(\ident_{j_{\Antiquark}}) \right] \ampnt(r^{\isubrmv})}\nonumber\\
&=&g_s\sum_{\colorQ,\colorQbar}{\frac{\delta_{\colorQ \colorQbar}}{\sqrt{\NC}}\left[\frac{k_{j_{\Quark}} \cdot \varepsilon_{\lambda_i}^*(k_i)}{k_{j_{\Quark}}\cdot k_i}t^a_{\colorQ \colorQ^\prime}\ampnt_{\colorQ^\prime \colorQbar}(r^{\isubrmv})-\frac{k_{j_{\Antiquark}} \cdot \varepsilon_{\lambda_i}^*(k_i)}{k_{j_{\Antiquark}}\cdot k_i}t^a_{\colorQbar^\prime\colorQbar}\ampnt_{\colorQ \colorQbar^\prime}(r^{\isubrmv})\right]}\nonumber\\
&=&g_s\left[\frac{k_{j_{\Quark}} \cdot \varepsilon_{\lambda_i}^*(k_i)}{k_{j_{\Quark}}\cdot k_i}-\frac{k_{j_{\Antiquark}} \cdot \varepsilon_{\lambda_i}^*(k_i)}{k_{j_{\Antiquark}}\cdot k_i}\right]\frac{\delta_{a \colorQQ^\prime}}{\sqrt{2\NC}}\ampnt_{\left\{[C=8]\right\}}(r^{\isubrmv}),
\end{eqnarray}
and
\begin{eqnarray}
&&\lim_{k_i \rightarrow 0}{\ampnpot_{\left\{[C=8]\right\}}(r)}\nonumber\\
&=& g_s\sum_{\colorQ,\colorQbar}{\mathbb{P}_{[C=8]}\left[\frac{k_{j_{\Quark}} \cdot \varepsilon_{\lambda_i}^*(k_i)}{k_{j_{\Quark}}\cdot k_i} \Qop(\ident_{j_{\Quark}})+\frac{k_{j_{\Antiquark}} \cdot \varepsilon_{\lambda_i}^*(k_i)}{k_{j_{\Antiquark}}\cdot k_i} \Qop(\ident_{j_{\Antiquark}}) \right] \ampnt(r^{\isubrmv})}\nonumber\\
&=&g_s\!\!\sum_{\colorQ,\colorQbar}{\sqrt{2}t^{\colorQQ}_{\colorQbar\colorQ}\!\!\!\left[\frac{k_{j_{\Quark}} \cdot \varepsilon_{\lambda_i}^*(k_i)}{k_{j_{\Quark}}\cdot k_i}t^a_{\colorQ \colorQbar^\prime}\ampnt_{\colorQ^\prime \colorQbar}\!(r^{\isubrmv})-\frac{k_{j_{\Antiquark}} \cdot \varepsilon_{\lambda_i}^*(k_i)}{k_{j_{\Antiquark}}\cdot k_i}t^a_{\colorQbar^\prime\colorQbar}\ampnt_{\colorQ \colorQbar^\prime}\!(r^{\isubrmv})\right]}\nonumber\\
&=&g_s\left[\frac{k_{j_{\Quark}} \cdot \varepsilon_{\lambda_i}^*(k_i)}{k_{j_{\Quark}}\cdot k_i}-\frac{k_{j_{\Antiquark}} \cdot \varepsilon_{\lambda_i}^*(k_i)}{k_{j_{\Antiquark}}\cdot k_i}\right]\left[\frac{\delta_{a \colorQQ}}{\sqrt{2\NC}}\ampnt_{\left\{[C=1]\right\}}(r^{\isubrmv})+\frac{1}{2}d_{a\colorQQ \colorQQ^\prime}\ampnt_{\left\{[C=8]\right\}}(r^{\isubrmv})\right]\nonumber\\
&&+g_s\left[\frac{k_{j_{\Quark}} \cdot \varepsilon_{\lambda_i}^*(k_i)}{k_{j_{\Quark}}\cdot k_i}+\frac{k_{j_{\Antiquark}} \cdot \varepsilon_{\lambda_i}^*(k_i)}{k_{j_{\Antiquark}}\cdot k_i}\right]\left(-\frac{i}{2}f_{a \colorQQ \colorQQ^\prime}\right)\ampnt_{\left\{[C=8]\right\}}(r^{\isubrmv}),
\end{eqnarray} 
where $d_{abc}$'s are the symmetric structure constants. Moreover, we have used the following relation of Gell-Mann matrices
\begin{align}
(t^at^b)_{kl} = \dfrac{\delta_{ab}}{2\NC}\delta_{kl} + \dfrac{1}{2}(d_{abc} + if_{abc})t^c_{kl}
\end{align}
and have assumed the colour index for $\QQ$ in the real process $r$ (the reduced Born process $r^{\isubrmv}$) to be $\colorQQ$ ($\colorQQ^\prime$). We can put the two equations into the compact matrix form by using the colour nonet $\irrep{9}$ index $\textbf{b}=0,1,2,\ldots,8$, where $\textbf{b}=0$ corresponds to $\ampnpot_{\left\{[C=1]\right\}}(r)$ and $\ampnt_{\left\{[C=1]\right\}}(r^{\isubrmv})$, while $\textbf{b}=1,\ldots,8$ are $\ampnpot_{\left\{[C=8]\right\}}(r)$ and $\ampnt_{\left\{[C=8]\right\}}(r^{\isubrmv})$ with the colour index of the $\QQ$ as $\textbf{b}$. In other words, with the emitters being $\Quark$ and $\Antiquark$, we have
\begin{eqnarray}
\lim_{k_i \rightarrow 0}{\left(\begin{array}{c}\ampnpot_{\left\{[C=1]\right\}}(r) \\
\ampnpot_{\left\{[C=8]\right\},\textbf{b}=1}(r)\\
\vdots\\
\ampnpot_{\left\{[C=8]\right\},\textbf{b}=8}(r)
\end{array}\right)}&=&g_s\left\{\left[\frac{k_{j_{\Quark}} \cdot \varepsilon_{\lambda_i}^*(k_i)}{k_{j_{\Quark}}\cdot k_i}-\frac{k_{j_{\Antiquark}} \cdot \varepsilon_{\lambda_i}^*(k_i)}{k_{j_{\Antiquark}}\cdot k_i}\right]\Qop_1(\QQ)\right.\nonumber\\
&&\left.+\left[\frac{k_{j_{\Quark}} \cdot \varepsilon_{\lambda_i}^*(k_i)}{k_{j_{\Quark}}\cdot k_i}+\frac{k_{j_{\Antiquark}} \cdot \varepsilon_{\lambda_i}^*(k_i)}{k_{j_{\Antiquark}}\cdot k_i}\right]\Qop_2(\QQ)\right\}\nonumber\\
&&\times\left(\begin{array}{c}\ampnt_{\left\{[C=1]\right\}}(r^{\isubrmv})\\
\ampnt_{\left\{[C=8]\right\},\textbf{b}=1}(r^{\isubrmv})\\
\vdots\\
\ampnt_{\left\{[C=8]\right\},\textbf{b}=8}(r^{\isubrmv})
\end{array}\right),
\end{eqnarray}
where the colour generators are
\begin{eqnarray}
\Qop_1(\QQ)&=&\left(\begin{array}{cccc} 0 & \frac{\delta_{a1}}{\sqrt{2\NC}} & \cdots &  \frac{\delta_{a8}}{\sqrt{2\NC}} \\
\frac{\delta_{a1}}{\sqrt{2\NC}} & \ddots & & \reflectbox{$\ddots$} \\
\vdots & & \frac{D^a}{2} & \\
\frac{\delta_{a8}}{\sqrt{2\NC}} & \reflectbox{$\ddots$} & & \ddots\\
\end{array}\right),\\
\Qop_2(\QQ)&=&\left(\begin{array}{cccc} 0 & 0 & \cdots &  0 \\
0 & \ddots & & \reflectbox{$\ddots$} \\
\vdots & & \frac{T^a}{2} & \\
0 & \reflectbox{$\ddots$} & & \ddots \\
\end{array}\right),
\end{eqnarray}
with the elements of the matrix $D^a$ as $D^a_{bc}=d_{abc}$.

\subsubsection{Spin and orbital angular momentum projections}

The next step is to perform the projection for the spin and the orbital angular momentum via
\begin{eqnarray}
&&\lim_{k_i \rightarrow 0}{\left(\begin{array}{c}\ampnpot_{\left\{[C=1],S,L\right\}}(r) \\
\ampnpot_{\left\{[C=8],S,L\right\},\textbf{b}=1}(r)\\
\vdots\\
\ampnpot_{\left\{[C=8],S,L\right\},\textbf{b}=8}(r)
\end{array}\right)}\nonumber\\
&=&\frac{g_s}{2\sqrt{2m_{\Quark}m_{\Antiquark}}}\left\{\left(\varepsilon_{\lambda_l}^{\mu,*}(K)\frac{d}{dq^\mu}\right)^L\left[\left(\frac{k_{j_{\Quark}} \cdot \varepsilon_{\lambda_i}^*(k_i)}{k_{j_{\Quark}}\cdot k_i}-\frac{k_{j_{\Antiquark}} \cdot \varepsilon_{\lambda_i}^*(k_i)}{k_{j_{\Antiquark}}\cdot k_i}\right)\Qop_1(\QQ)\right.\right.\nonumber\\
&&\left.+\left(\frac{k_{j_{\Quark}} \cdot \varepsilon_{\lambda_i}^*(k_i)}{k_{j_{\Quark}}\cdot k_i}+\frac{k_{j_{\Antiquark}} \cdot \varepsilon_{\lambda_i}^*(k_i)}{k_{j_{\Antiquark}}\cdot k_i}\right)\Qop_2(\QQ)\right]\nonumber\\
&&\times \left.\sum_{\colorQ,\colorQbar}{\left(\begin{array}{c} \frac{\delta_{\colorQ \colorQbar}}{\sqrt{\NC}}\\
\sqrt{2}t^{1}_{\colorQbar \colorQ}\\ \vdots \\ \sqrt{2}t^{8}_{\colorQbar \colorQ}\\\end{array}\right){\rm Tr}_{\gamma}\left(\left(\slashed{k}_{\Antiquark}-m_{\Antiquark}\right)\tilde{\gamma}_S \left(\slashed{k}_{\Quark}+m_{\Quark}\right)\Gamma^{(n,0)}(r^{\isubrmv})\right)}\right\}_{q=0}.
\end{eqnarray}
When $L=0$, we just set the relative momentum $q$ to be zero. The coefficient of $\Qop_1(\QQ)$ vanishes. This implies the following two consequences:
\begin{itemize}
\item For the colour singlet $C=1$ with $L=0$ (S-wave), regardless of the value of the spin $S$, there are no soft divergences. Thus, we can treat a colour-singlet S-wave state as any other elementary colour-singlet particle, such as $Z$ and $H$ bosons, from the IR perspective.
\item For the colour octet $C=8$ S-wave states, we have the following soft limit relation
\begin{eqnarray}
\lim_{k_i \rightarrow 0}{\ampnpot_{\left\{[8],S,0\right\}}(r)}&=&g_s\frac{K \cdot \varepsilon_{\lambda_i}^*(k_i)}{K\cdot k_i}\Qop(\QQ[\bigl.^{2S+1}\hspace{-1mm}S^{[8]}_J])\ampnt_{\left\{[8],S,0\right\}}(r^{\isubrmv}),\label{eq:softOctetS}
\end{eqnarray}
where now $\QQ[\bigl.^{2S+1}\hspace{-1mm}S^{[8]}_J] \in \irrep{8}$ and $\Qop(\QQ[\bigl.^{2S+1}\hspace{-1mm}S^{[8]}_J])=\left\{T^a\right\}_{a=1}^{8}$. It means that, in the soft limit, a colour-octet S-wave state behaves as if $\QQ$ were an elementary colour-octet particle, akin to a sgluon or a Kaluza-Klein massive gluon.
\end{itemize}

For P-wave ($L=1$) Fock states, we have
\begin{eqnarray}
\lim_{k_i \rightarrow 0}{\left(\begin{array}{c}\ampnpot_{\left\{[C=1],S,1\right\}}(r) \\
\ampnpot_{\left\{[C=8],S,1\right\},\textbf{b}=1}(r)\\
\vdots\\
\ampnpot_{\left\{[C=8],S,1\right\},\textbf{b}=8}(r)
\end{array}\right)}&=&g_s\frac{K \cdot \varepsilon_{\lambda_i}^*(k_i)}{K\cdot k_i}2\Qop_2(\QQ)\left(\begin{array}{c}\ampnt_{\left\{[C=1],S,1\right\}}(r^{\isubrmv}) \\
\ampnt_{\left\{[C=8],S,1\right\},\textbf{b}=1}(r^{\isubrmv})\\
\vdots\\
\ampnt_{\left\{[C=8],S,1\right\},\textbf{b}=8}(r^{\isubrmv})
\end{array}\right)\nonumber\\
&&+g_s\left[\frac{\varepsilon_{\lambda_l}^*(K)\cdot \varepsilon_{\lambda_i}^*(k_i)}{K\cdot k_i}-\frac{K \cdot \varepsilon_{\lambda_i}^*(k_i) k_i\cdot \varepsilon_{\lambda_l}^*(K)}{\left(K\cdot k_i\right)^2}\right]\nonumber\\
&&\times\!\left[\left(2+\frac{m_{\Quark}}{m_{\Antiquark}}+\frac{m_{\Antiquark}}{m_{\Quark}}\right)\Qop_{1}(\QQ)\!+\!\left(\frac{m_{\Antiquark}}{m_{\Quark}}\!-\!\frac{m_{\Quark}}{m_{\Antiquark}}\right)\Qop_{2}(\QQ)\right]\nonumber\\
&&\times\left(\begin{array}{c}\ampnt_{\left\{[C=1],S,0\right\}}(r^{\isubrmv}) \\
\ampnt_{\left\{[C=8],S,0\right\},\textbf{b}=1}(r^{\isubrmv})\\
\vdots\\
\ampnt_{\left\{[C=8],S,0\right\},\textbf{b}=8}(r^{\isubrmv})
\end{array}\right),
\end{eqnarray}
because of
\begin{eqnarray}
\varepsilon_{\lambda_l}^{\mu,*}(K)\frac{d}{dq^\mu}\frac{k_{j_{\Quark}} \cdot \varepsilon_{\lambda_i}^*(k_i)}{k_{j_{\Quark}}\cdot k_i}&=&\frac{\varepsilon_{\lambda_l}^*(K)\cdot \varepsilon_{\lambda_i}^*(k_i)}{k_{j_{\Quark}}\cdot k_i}-\frac{k_{j_{\Quark}} \cdot \varepsilon_{\lambda_i}^*(k_i) k_i\cdot \varepsilon_{\lambda_l}^*(K)}{\left(k_{j_{\Quark}}\cdot k_i\right)^2}\nonumber\\
&\overset{q=0}{=}&\frac{m_{\Quark}+m_{\Antiquark}}{m_{\Quark}}\left[\frac{\varepsilon_{\lambda_l}^*(K)\cdot \varepsilon_{\lambda_i}^*(k_i)}{K\cdot k_i}-\frac{K \cdot \varepsilon_{\lambda_i}^*(k_i) k_i\cdot \varepsilon_{\lambda_l}^*(K)}{\left(K\cdot k_i\right)^2}\right],\nonumber\\
\varepsilon_{\lambda_l}^{\mu,*}(K)\frac{d}{dq^\mu}\frac{k_{j_{\Antiquark}} \cdot \varepsilon_{\lambda_i}^*(k_i)}{k_{j_{\Antiquark}}\cdot k_i}&=&-\frac{\varepsilon_{\lambda_l}^*(K)\cdot \varepsilon_{\lambda_i}^*(k_i)}{k_{j_{\Antiquark}}\cdot k_i}+\frac{k_{j_{\Antiquark}} \cdot \varepsilon_{\lambda_i}^*(k_i) k_i\cdot \varepsilon_{\lambda_l}^*(K)}{\left(k_{j_{\Antiquark}}\cdot k_i\right)^2}\nonumber\\
&\overset{q=0}{=}&-\frac{m_{\Quark}+m_{\Antiquark}}{m_{\Antiquark}}\left[\frac{\varepsilon_{\lambda_l}^*(K)\cdot \varepsilon_{\lambda_i}^*(k_i)}{K\cdot k_i}-\frac{K \cdot \varepsilon_{\lambda_i}^*(k_i) k_i\cdot \varepsilon_{\lambda_l}^*(K)}{\left(K\cdot k_i\right)^2}\right].\nonumber\\
\end{eqnarray}
This essentially means that
\begin{itemize}
\item For the colour-singlet ($C=1$) P-wave states, we get
\begin{eqnarray}
\lim_{k_i \rightarrow 0}{\ampnpot_{\left\{[C=1],S,1\right\}}(r)}&=&g_s\left[\frac{\varepsilon_{\lambda_l}^*(K)\cdot \varepsilon_{\lambda_i}^*(k_i)}{K\cdot k_i}-\frac{K \cdot \varepsilon_{\lambda_i}^*(k_i) k_i\cdot \varepsilon_{\lambda_l}^*(K)}{\left(K\cdot k_i\right)^2}\right]\nonumber\\
&&\times\left(2+\frac{m_{\Quark}}{m_{\Antiquark}}+\frac{m_{\Antiquark}}{m_{\Quark}}\right)\frac{\delta_{a \colorQQ^\prime}}{\sqrt{2\NC}}\ampnt_{\left\{[C=8],S,0\right\}}(r^{\isubrmv}).\label{eq:softSingletP1}
\end{eqnarray}
The object $\ident_{j_{\Quark}\oplus j_{\Antiquark}\oplus i}$ may be interpreted as a new colour-octet S-wave particle (with colour index $\colorQQ^\prime=a$) with a non-standard eikonal factor.
\item For the colour-octet ($C=8$) P-wave states, the soft limit yields a more complicated expression
\begin{eqnarray}
\lim_{k_i \rightarrow 0}{\ampnpot_{\left\{[C=8],S,1\right\}}(r)}&=&g_s\frac{K \cdot \varepsilon_{\lambda_i}^*(k_i)}{K\cdot k_i}\Qop(\QQ[\bigl.^{2S+1}\hspace{-1mm}P^{[8]}_J])\ampnt_{\left\{[C=8],S,1\right\}}(r^{\isubrmv})\nonumber\\
&&+g_s\left[\frac{\varepsilon_{\lambda_l}^*(K)\cdot \varepsilon_{\lambda_i}^*(k_i)}{K\cdot k_i}-\frac{K \cdot \varepsilon_{\lambda_i}^*(k_i) k_i\cdot \varepsilon_{\lambda_l}^*(K)}{\left(K\cdot k_i\right)^2}\right]\nonumber\\
&&\times\left[\left(2+\frac{m_{\Quark}}{m_{\Antiquark}}+\frac{m_{\Antiquark}}{m_{\Quark}}\right)\left(\frac{\delta_{a \colorQQ}}{\sqrt{2\NC}}\ampnt_{\left\{[C=1],S,0\right\}}(r^{\isubrmv})\right.\right.\nonumber\\
&&\left.+\frac{d_{a\colorQQ\colorQQ^\prime}}{2}\ampnt_{\left\{[C=8],S,0\right\}}(r^{\isubrmv})\right)\nonumber\\
&&\left.+\frac{1}{2}\left(\frac{m_{\Antiquark}}{m_{\Quark}}-\frac{m_{\Quark}}{m_{\Antiquark}}\right)\Qop(\QQ[\bigl.^{2S+1}\hspace{-1mm}S^{[8]}_J])\ampnt_{\left\{[C=8],S,0\right\}}(r^{\isubrmv})\right].\nonumber\\
\label{eq:softOctetP1}
\end{eqnarray}
\end{itemize}

If we define the following effective colour generators
\begin{eqnarray}
\Qop_{{\rm eff}}(\QQ_{[18]})&=&\left\{\left(2+\frac{m_{\Quark}}{m_{\Antiquark}}+\frac{m_{\Antiquark}}{m_{\Quark}}\right)\frac{\Delta^{a}}{\sqrt{2\NC}}\right\}_{a=1}^{8},\nonumber\\
\Qop_{{\rm eff}}(\QQ_{[81]})&=&\left\{\left(2+\frac{m_{\Quark}}{m_{\Antiquark}}+\frac{m_{\Antiquark}}{m_{\Quark}}\right)\frac{\Delta^{aT}}{\sqrt{2\NC}}\right\}_{a=1}^{8},\nonumber\\
\Qop_{{\rm eff}}(\QQ_{[88]})&=&\left\{\left(2+\frac{m_{\Quark}}{m_{\Antiquark}}+\frac{m_{\Antiquark}}{m_{\Quark}}\right)\frac{D^a}{2}+\left(\frac{m_{\Antiquark}}{m_{\Quark}}-\frac{m_{\Quark}}{m_{\Antiquark}}\right)\frac{T^a}{2}\right\}_{a=1}^{8},
\end{eqnarray}
where $\Delta^a_{\circ b}=\delta_{ab}$ and $\circ$ represents no colour index for a colour-singlet state, the soft-limit expressions can be written in a more compact form
\begin{eqnarray}
\lim_{k_i \rightarrow 0}{\ampnpot_{\left\{[C=1],S,1\right\}}(r)}&=&g_s\left[\frac{\varepsilon_{\lambda_l}^*(K)\cdot \varepsilon_{\lambda_i}^*(k_i)}{K\cdot k_i}-\frac{K \cdot \varepsilon_{\lambda_i}^*(k_i) k_i\cdot \varepsilon_{\lambda_l}^*(K)}{\left(K\cdot k_i\right)^2}\right]\nonumber\\
&&\times\Qop_{{\rm eff}}(\QQ_{[18]})\ampnt_{\left\{[C=8],S,0\right\}}(r^{\isubrmv}),\label{eq:softSingletP2}\\
\lim_{k_i \rightarrow 0}{\ampnpot_{\left\{[C=8],S,1\right\}}(r)}&=&g_s\frac{K \cdot \varepsilon_{\lambda_i}^*(k_i)}{K\cdot k_i}\Qop(\QQ[\bigl.^{2S+1}\hspace{-1mm}P^{[8]}_J])\ampnt_{\left\{[C=8],S,1\right\}}(r^{\isubrmv})\nonumber\\
&&+g_s\left[\frac{\varepsilon_{\lambda_l}^*(K)\cdot \varepsilon_{\lambda_i}^*(k_i)}{K\cdot k_i}-\frac{K \cdot \varepsilon_{\lambda_i}^*(k_i) k_i\cdot \varepsilon_{\lambda_l}^*(K)}{\left(K\cdot k_i\right)^2}\right]\nonumber\\
&&\times\left[\Qop_{{\rm eff}}(\QQ_{[81]})\ampnt_{\left\{[C=1],S,0\right\}}(r^{\isubrmv})+\Qop_{{\rm eff}}(\QQ_{[88]})\ampnt_{\left\{[C=8],S,0\right\}}(r^{\isubrmv})\right].\nonumber\\
\label{eq:softOctetP2}
\end{eqnarray}
Here, we have defined the effective Casimir constants through the following relations
\begin{eqnarray}
C_{{\rm eff}}
(\QQ_{[18]})&=&\Qop_{{\rm eff}}(\QQ_{[18]})\mydot \Qop_{{\rm eff}}(\QQ_{[18]})=\left(2+\frac{m_{\Quark}}{m_{\Antiquark}}+\frac{m_{\Antiquark}}{m_{\Quark}}\right)^2\frac{1}{2\NC}\,,\nonumber\\
C_{{\rm eff}}
(\QQ_{[81]})&=&\Qop_{{\rm eff}}(\QQ_{[81]})\mydot \Qop_{{\rm eff}}(\QQ_{[81]})=\left(2+\frac{m_{\Quark}}{m_{\Antiquark}}+\frac{m_{\Antiquark}}{m_{\Quark}}\right)^2\CF\,, \nonumber\\
C_{{\rm eff}}
(\QQ_{[88]})&=&\Qop_{{\rm eff}}(\QQ_{[88]})\mydot \Qop_{{\rm eff}}(\QQ_{[88]})=\left(2+\frac{m_{\Quark}}{m_{\Antiquark}}+\frac{m_{\Antiquark}}{m_{\Quark}}\right)\nonumber\\
&&~~~~~~~~~~~~~~~~~~~~~~~~~~~~~~~~~~~~\times\left(-\frac{2}{\NC}+\frac{\NC^2-2}{2\NC}\left(\frac{m_{\Antiquark}}{m_{\Quark}}+\frac{m_{\Quark}}{m_{\Antiquark}}\right)\right).\nonumber\\
\end{eqnarray}
In addition, we also have
\begin{eqnarray}
\Qop_{{\rm eff}}(\QQ_{[81]})\mydot\Qop_{{\rm eff}}(\QQ_{[88]})&=&\Qop_{{\rm eff}}(\QQ_{[88]})\mydot \Qop_{{\rm eff}}(\QQ_{[81]})=0,\nonumber\\
\Qop(\QQ[\bigl.^{2S+1}\hspace{-1mm}P^{[8]}_J])\mydot \Qop_{{\rm eff}}(\QQ_{[81]})&=&\Qop_{{\rm eff}}(\QQ_{[81]})\mydot \Qop(\QQ[\bigl.^{2S+1}\hspace{-1mm}P^{[8]}_J])=0\,,\nonumber\\
\Qop(\QQ[\bigl.^{2S+1}\hspace{-1mm}P^{[8]}_J])\mydot \Qop_{{\rm eff}}(\QQ_{[88]})&=&\Qop_{{\rm eff}}(\QQ_{[88]})\mydot \Qop(\QQ[\bigl.^{2S+1}\hspace{-1mm}P^{[8]}_J])=\frac{1}{2}\left(\frac{m_{\Antiquark}}{m_{\Quark}}-\frac{m_{\Quark}}{m_{\Antiquark}}\right)\CA\,.\nonumber\\
\end{eqnarray}

Finally, in order to get a given total angular momentum $J$ when $L=S=1$, we need to use eq.\eqref{eq:projLS2J} at the amplitude level: 
\begin{eqnarray}
\lim_{k_i \rightarrow 0}{\ampnpot_{\left\{[C=1],1,1,J\right\}}(r)}&=&\sum_{\lambda_l,\lambda_s}{\langle J, \lambda_j|1,\lambda_l; 1, \lambda_s\rangle g_s\left[\frac{\varepsilon_{\lambda_l}^*(K)\cdot \varepsilon_{\lambda_i}^*(k_i)}{K\cdot k_i}-\frac{K \cdot \varepsilon_{\lambda_i}^*(k_i) k_i\cdot \varepsilon_{\lambda_l}^*(K)}{\left(K\cdot k_i\right)^2}\right]}\nonumber\\
&&\times\Qop_{{\rm eff}}(\QQ_{[18]})\ampnt_{\left\{[C=8],1,0\right\}}(r^{\isubrmv}),\label{eq:softSingletP3}
\end{eqnarray}
and
\begin{eqnarray}
\lim_{k_i \rightarrow 0}{\ampnpot_{\left\{[C=8],1,1,J\right\}}(r)}&=&g_s\frac{K \cdot \varepsilon_{\lambda_i}^*(k_i)}{K\cdot k_i}\Qop(\QQ[\bigl.^{3}\hspace{-1mm}P^{[8]}_J])\ampnt_{\left\{[C=8],1,1,J\right\}}(r^{\isubrmv})\nonumber\\
&&+\sum_{\lambda_l,\lambda_s}{\langle J,\lambda_j|1,\lambda_l; 1,\lambda_s\rangle g_s\left[\frac{\varepsilon_{\lambda_l}^*(K)\cdot \varepsilon_{\lambda_i}^*(k_i)}{K\cdot k_i}-\frac{K \cdot \varepsilon_{\lambda_i}^*(k_i) k_i\cdot \varepsilon_{\lambda_l}^*(K)}{\left(K\cdot k_i\right)^2}\right]}\nonumber\\
&&\times\left[\Qop_{{\rm eff}}(\QQ_{[81]})\ampnt_{\left\{[C=1],S,0\right\}}(r^{\isubrmv})+\Qop_{{\rm eff}}(\QQ_{[88]})\ampnt_{\left\{[C=8],S,0\right\}}(r^{\isubrmv})\right].\nonumber\\
&&\label{eq:softOctetP3}
\end{eqnarray}
To rewrite this in a unified compact form, we can further define the following effective identity (called ``flavour") operators
\begin{eqnarray}
\Fop_{{\rm eff}}(\QQ_{[18]})&:&\QQ[\bigl.^{2S+1}\hspace{-1mm}P^{[1]}_J]\to \QQ[\bigl.^{2S+1}\hspace{-1mm}S^{[8]}_{S}],\nonumber\\
\Fop_{{\rm eff}}(\QQ_{[81]})&:&\QQ[\bigl.^{2S+1}\hspace{-1mm}P^{[8]}_J]\to \QQ[\bigl.^{2S+1}\hspace{-1mm}S^{[1]}_{S}],\nonumber\\
\Fop_{{\rm eff}}(\QQ_{[88]})&:&\QQ[\bigl.^{2S+1}\hspace{-1mm}P^{[8]}_J]\to \QQ[\bigl.^{2S+1}\hspace{-1mm}S^{[8]}_{S}].
\end{eqnarray}
The eikonal current operator for an elementary coloured particle $\ident_j$ is
\begin{eqnarray}
\Jop(\ident_j)&=&\frac{k_j\cdot \varepsilon^*_{\lambda_i}(k_i)}{k_j\cdot k_i}\Qop(\ident_j),\label{eq:eikonalcurrent1}
\end{eqnarray}
while for bound states it is
\begin{eqnarray}
\Jop(\QQ[\bigl.^{2S+1}\hspace{-1mm}S^{[1]}_{S}])&=&0,\nonumber\\
\Jop(\QQ[\bigl.^{2S+1}\hspace{-1mm}S^{[8]}_{S}])&=&\frac{K\cdot \varepsilon^*_{\lambda_i}(k_i)}{K\cdot k_i}\Qop(\QQ[\bigl.^{2S+1}\hspace{-1mm}S^{[8]}_{S}]),\nonumber\\
\Jop(\QQ[\bigl.^{1}\hspace{-1mm}P^{[1]}_{1}])&=&\left[\frac{\varepsilon_{\lambda_l}^*(K)\cdot \varepsilon_{\lambda_i}^*(k_i)}{K\cdot k_i}-\frac{K \cdot \varepsilon_{\lambda_i}^*(k_i) k_i\cdot \varepsilon_{\lambda_l}^*(K)}{\left(K\cdot k_i\right)^2}\right]\Qop_{{\rm eff}}(\QQ_{[18]})\Fop_{{\rm eff}}(\QQ_{[18]}),\nonumber\\
\Jop(\QQ[\bigl.^{1}\hspace{-1mm}P^{[8]}_{1}])&=&\frac{K \cdot \varepsilon_{\lambda_i}^*(k_i)}{K\cdot k_i}\Qop(\QQ[\bigl.^{2S+1}\hspace{-1mm}P^{[8]}_J])\nonumber\\
&&+\left[\frac{\varepsilon_{\lambda_l}^*(K)\cdot \varepsilon_{\lambda_i}^*(k_i)}{K\cdot k_i}-\frac{K \cdot \varepsilon_{\lambda_i}^*(k_i) k_i\cdot \varepsilon_{\lambda_l}^*(K)}{\left(K\cdot k_i\right)^2}\right]\nonumber\\
&&\times\left[\Qop_{{\rm eff}}(\QQ_{[81]})\Fop_{{\rm eff}}(\QQ_{[81]})+\Qop_{{\rm eff}}(\QQ_{[88]})\Fop_{{\rm eff}}(\QQ_{[88]})\right],\nonumber\\
\Jop(\QQ[\bigl.^{3}\hspace{-1mm}P^{[1]}_{J}])&=&\sum_{\lambda_l,\lambda_s}{\langle J,\lambda_j|1,\lambda_1;1,\lambda_s\rangle\left[\frac{\varepsilon_{\lambda_l}^*(K)\cdot \varepsilon_{\lambda_i}^*(k_i)}{K\cdot k_i}-\frac{K \cdot \varepsilon_{\lambda_i}^*(k_i) k_i\cdot \varepsilon_{\lambda_l}^*(K)}{\left(K\cdot k_i\right)^2}\right]}\nonumber\\
&&\times\Qop_{{\rm eff}}(\QQ_{[18]})\Fop_{{\rm eff}}(\QQ_{[18]}),\nonumber\\
\Jop(\QQ[\bigl.^{3}\hspace{-1mm}P^{[8]}_{J}])&=&\frac{K \cdot \varepsilon_{\lambda_i}^*(k_i)}{K\cdot k_i}\Qop(\QQ[\bigl.^{2S+1}\hspace{-1mm}P^{[8]}_J])\nonumber\\
&&+\sum_{\lambda_l,\lambda_s}{\left[\frac{\varepsilon_{\lambda_l}^*(K)\cdot \varepsilon_{\lambda_i}^*(k_i)}{K\cdot k_i}-\frac{K \cdot \varepsilon_{\lambda_i}^*(k_i) k_i\cdot \varepsilon_{\lambda_l}^*(K)}{\left(K\cdot k_i\right)^2}\right]}\nonumber\\
&&\times\langle J,\lambda_j|1,\lambda_l;1,\lambda_s\rangle\!\!\left[\Qop_{{\rm eff}}(\QQ_{[81]})\Fop_{{\rm eff}}(\QQ_{[81]})\!+\!\Qop_{{\rm eff}}(\QQ_{[88]})\Fop_{{\rm eff}}(\QQ_{[88]})\right].\nonumber\\ \label{eq:eikonalcurrent2}
\end{eqnarray}
In such a case, the amplitude in the soft limit of $k_i\to 0$ can be rewritten as
\begin{eqnarray}
\lim_{k_i\to 0}{\ampnto(\dot{r})}&=&g_s\mathop{\sum_{j=\nini}}_{j\neq i}^{\nlightR+\nheavy+1}{\Jop(\ident_j)\ampnmoo(\dot{r}^{\isubrmv})},\label{eq:softamp}
\end{eqnarray}
where we have glued the $\QQ$ pair into a single particle in processes with a dot, \ie,
\begin{eqnarray}
\dot{r}&=&r^{j_{\Quark}\oplus j_{\Antiquark}, \Qbsubrmv}=\left(\ident_1,\ldots,\ident_i,\ldots,\ident_j,\ldots,\ident_{\nlightR+\nheavy},\QQ[\bigl.^{2S+1}\hspace{-1mm}L^{[C]}_J],\remove{\Antiquark}{0.3},
\ldots,\ident_{n+3}\right),\nonumber\\
\dot{r}^{\isubrmv}&=&\left(\ident_1,\ldots,\remove{\ident}{0.2}_i,\ldots,\ident_j,\ldots,\ident_{\nlightR+\nheavy},\QQ[\bigl.^{2S+1}\hspace{-1mm}L^{[C]}_J],\remove{\Antiquark}{0.3},
\ldots,\ident_{n+3}\right).
\end{eqnarray}
Note that $\dot{r}$ is essentially equivalent to eq.\eqref{QQproc} except that we have reordered the final particles, and the total number of external parton legs is $n+3$ instead of $n+2$.

\subsection{Soft limit of real matrix elements\label{sec:softampsquare}}

We can now examine the soft limit of the real amplitude square by considering the $\ell$-loop amplitude $\ampnll(r)$ for a generic $2\to n$ process $r$. To facilitate our later discussion, we introduce the following amplitude squares:
\begin{eqnarray}
\ampsqnto(\dot{r})&=&\ampsqnpot_{\left\{[C],S,L,J\right\}}(r)=\frac{1}{2s}\frac{1}{\omega(\Ione)\omega(\Itwo)}
\mathop{\sum_{\rm colour}}_{\rm spin}\abs{\ampnpot_{\left\{[C],S,L,J\right\}}(r)}^2,
\label{MtreenpoOnium}
\\
\ampsqnto_J(\dot{r})^{\mu\nu}&=&\frac{1}{2s}\frac{1}{\omega(\Ione)\omega(\Itwo)}\nonumber\\
&&\times\mathop{\sum_{\rm colour}}_{\rm spin}\left[\left(\sum_{\lambda_l,\lambda_s}{\langle J,\lambda_j| 1,\lambda_l;1,\lambda_s\rangle\varepsilon^{\mu,*}_{\lambda_l}(K)\ampnmoo(\dot{r})}\right)\right.\nonumber\\
&&\left.\times\left(\sum_{\lambda_l^\prime,\lambda_s^\prime}{\langle J,\lambda_j| 1,\lambda_l^\prime;1,\lambda_s^\prime\rangle\varepsilon^{\nu,*}_{\lambda_l^\prime}(K)\ampnmoo(\dot{r})}\right)^\star\right],
\label{MtreenpoOniumJ}
\\
\ampsqnmoo(\dot{r})&=&\ampsqnt_{\left\{[C],S,L,J\right\}}(r)=\frac{1}{2s}\frac{1}{\omega(\Ione)\omega(\Itwo)}
\mathop{\sum_{\rm colour}}_{\rm spin}\abs{\ampnt_{\left\{[C],S,L,J\right\}}(r)}^2,
\label{MtreenOnium}
\\
\ampsqnt_{\left\{[C],S,L,J\right\},kl}(r)&=&-\frac{1}{2s}
\frac{2-\delta_{kl}}{\omega(\Ione)\omega(\Itwo)}\nonumber\\
&&\times\mathop{\sum_{\rm colour}}_{\rm spin}
\ampnt_{\left\{[C],S,L,J\right\}}(r) \Qop(\ident_k)\mydot\Qop(\ident_l)
{\ampnt_{\left\{[C],S,L,J\right\}}(r)}^{\star},\phantom{aa}
\label{MlinkedOnium1}
\\
\ampsqnmoo_{kl}(\dot{r})&=&-\frac{1}{2s}
\frac{2-\delta_{kl}}{\omega(\Ione)\omega(\Itwo)}\nonumber\\
&&\times\mathop{\sum_{\rm colour}}_{\rm spin}
\ampnmoo(\dot{r}) \Qop(\ident_k)\mydot\Qop(\ident_l)
{\ampnmoo(\dot{r})}^{\star},\phantom{aa}
\label{MlinkedOnium2}
\\
\ampsqnmoo_{k[C_1C_2]}(\dot{r}_1,\dot{r}_2)^{\mu}&=&-\frac{1}{2s}
\frac{1}{\omega(\Ione)\omega(\Itwo)}\nonumber\\
&&\mathop{\sum_{\rm colour}}_{\rm spin}
2\Re{\left\{\varepsilon^{*,\mu}_{\lambda_l}(K)\ampnmoo(\dot{r}_1)^{\star}\Qop(\ident_k)\mydot\Qop_{{\rm eff}}(\QQ_{[C_1C_2]})
{\ampnmoo(\dot{r}_2)}\right\}},\nonumber\\[-10pt]
\label{MlinkedOnium3}
\\
\ampsqnmoo_{J,k[C_1C_2]}(\dot{r}_1,\dot{r}_2)^{\mu}&=&-\frac{1}{2s}
\frac{1}{\omega(\Ione)\omega(\Itwo)}\nonumber\\
&&\mathop{\sum_{\rm colour}}_{\rm spin}
2\Re{\left\{\ampnmoo(\dot{r}_1)^{\star}\Qop(\ident_k)\mydot\Qop_{{\rm eff}}(\QQ_{[C_1C_2]})\right.}\nonumber\\
&&\left.\times\sum_{\lambda_l,\lambda_s}{\langle J,\lambda_j|1,\lambda_l;1,\lambda_s\rangle\varepsilon^{*,\mu}_{\lambda_l}(K)\ampnmoo(\dot{r}_2)}\right\},\phantom{aa}
\label{MlinkedOnium4J}
\\
\ampsqnl(r)&=&\frac{1}{2s}\frac{1}{\omega(\Ione)\omega(\Itwo)}
\mathop{\sum_{\rm colour}}_{\rm spin}
2\Re\left\{\ampnt(r){\ampnl(r)}^{\star}\right\}.
\label{Moneloop}
\end{eqnarray}
Once again, in these equations, $s=(k_1+k_2)^2=2k_1\cdot k_2$, 
and $\omega(\ident)$ represents the product of spin and colour degrees of 
freedom for particle $\ident$. In $d=4-2\epsilon$ dimensions, the average factors are given by
$\omega(q)=\omega(\bar{q})=2\NC=6$ and $\omega(g)=2(1-\ep)(\NC^2-1)=16(1-\ep)$.
These average factors are utilised to fully specify the divergent part of the one-loop
contribution within the conventional dimensional regularisation (CDR) scheme. It is important to note that while these factors include $\epsilon$ for completeness, in numerical calculations, $d=4$ tree-level amplitudes are typically evaluated, and the $\epsilon$ dependence is dropped.

In general, we can write down the soft limit of the amplitude square in the form
\begin{eqnarray}
\lim_{k_i\to 0}{\ampsqnto(\dot{r})}&=&g_s^2\mathop{\sum_{k,l=\nini}}_{k,l\neq i, k\leq l}^{\nlightR+\nheavy+1}{\ampsqsoftnmo_{{\rm soft}, kl}(\dot{r}^{\isubrmv})}.\label{eq:softampsq0}
\end{eqnarray}
It is well known that if both $\ident_k$ and $\ident_l$ are elementary particles, we have
\begin{eqnarray}
\ampsqsoftnmo_{{\rm soft}, kl}(\dot{r}^{\isubrmv})&=&\frac{k_k\mydot k_l}{k_k\mydot k_ik_l\mydot k_i}\ampsqnmoo_{kl}(\dot{r}^{\isubrmv}).\label{eq:softampsq}
\end{eqnarray}
In the following, we will try to derive $\ampsqsoftnmo_{{\rm soft}, kl}(\dot{r}^{\isubrmv})$ when at least $\ident_k$ or $\ident_l$ is a quarkonium state.

\subsubsection{Colour-singlet S-wave state}

We have derived the soft limit of colour-singlet S-wave states at the amplitude level in sect.~\ref{sec:softamp} as
\begin{eqnarray}
\lim_{k_i\to 0}{\ampnpot_{\left\{[1],S,0,J=S\right\}}(r)}&=&\mathop{\sum_{j=\nini}}_{j\neq i}^{\nlightR+\nheavy}{g_s\frac{k_j\cdot \varepsilon_{\lambda_i}^*(k_i)}{k_j\cdot k_i}\Qop(\ident_j)\ampnt_{\left\{[1],S,0,J=S\right\}}(r^{\isubrmv})}.
\end{eqnarray}
It follows that the amplitude square in the soft limit is given by
\begin{eqnarray}
\lim_{k_i\to 0}{\ampsqnto(\dot{r})}=\lim_{k_i\to 0}{\ampsqnpot_{\left\{[1],S,0,J=S\right\}}(r)}&=&g_s^2\mathop{\sum_{k,l=\nini}}_{k,l\neq i, k\leq l}^{\nlightR+\nheavy}{\frac{k_k\cdot k_l}{k_k\cdot k_i k_l\cdot k_i}\ampsqnt_{\left\{[1],S,0,J=S\right\},kl}(r^{\isubrmv})}\nonumber\\
&=&g_s^2\mathop{\sum_{k,l=\nini}}_{k,l\neq i, k\leq l}^{\nlightR+\nheavy}{\frac{k_k\cdot k_l}{k_k\cdot k_i k_l\cdot k_i}\ampsqnmoo_{kl}(\dot{r}^{\isubrmv})},\label{eq:softCSSwave}
\end{eqnarray}
where
\begin{eqnarray}
\dot{r}^{\isubrmv}&=&\left(\ident_1,\ldots,\remove{\ident}{0.2}_i,\ldots,\ident_j,\ldots,\ident_{\nlightR+\nheavy},\QQ[\bigl.^{2S+1}\hspace{-1mm}S^{[1]}_J],\remove{\Antiquark}{0.3},
\ldots,\ident_{n+3}\right).
\end{eqnarray}
In eq.\eqref{eq:softCSSwave}, we have used the light-cone axial gauge
\begin{eqnarray}
\sum_{\lambda_i}{\varepsilon^{\mu *}_{\lambda_i}(k_i)\varepsilon^{\nu}_{\lambda_i}(k_i)}&=&-g^{\mu\nu}+\frac{k_i^\mu n^\nu+k_i^\nu n^\mu}{k_i\cdot n},\label{eq:axialgaugesum}
\end{eqnarray}
where $n^\mu$ is a light-like auxiliary vector ($n^2=0$). The terms involving $n^{\mu}$ cancel out in eq.\eqref{eq:softCSSwave} as a result of colour conservation. Therefore, we have, for $\nini\leq k \leq \nlightR+\nheavy+1$,
\begin{eqnarray}
\ampsqsoftnmo_{{\rm soft}, kj_{\Quark}}(\dot{r}^{\isubrmv})&=&0.
\end{eqnarray}

\subsubsection{Colour-octet S-wave state}

For colour-octet S-wave states, we have derived the soft limit at the amplitude level in sect.~\ref{sec:softamp} ({\it cf.} eq.\eqref{eq:softOctetS}) as
\begin{eqnarray}
\lim_{k_i\to 0}{\ampnpot_{\left\{[8],S,0,J=S\right\}}(r)}&=&\mathop{\sum_{j=\nini}}_{j\neq i}^{\nlightR+\nheavy}{g_s\frac{k_j\cdot \varepsilon_{\lambda_i}^*(k_i)}{k_j\cdot k_i}\Qop(\ident_j)\ampnt_{\left\{[8],S,0,J=S\right\}}(r^{\isubrmv})}\nonumber\\
&&+g_s\frac{K\cdot \varepsilon_{\lambda_i}^*(k_i)}{K\cdot k_i}\Qop(\QQ[\bigl.^{2S+1}\hspace{-1mm}S^{[8]}_J])\ampnt_{\left\{[8],S,0,J=S\right\}}(r^{\isubrmv})\,.
\end{eqnarray}
The amplitude square in the soft limit is
\begin{eqnarray}
\lim_{k_i\to 0}{\ampsqnto(\dot{r})}=\lim_{k_i\to 0}{\ampsqnpot_{\left\{[8],S,0,J=S\right\}}(r)}&=&g_s^2\!\mathop{\sum_{k,l=\nini}}_{k,l\neq i,k\leq l}^{\nlightR+\nheavy+1}\!\!{\frac{\tilde{k}_k\cdot \tilde{k}_l}{\tilde{k}_k\cdot k_i \tilde{k}_l\cdot k_i}\ampsqnmoo_{kl}(\dot{r}^{\isubrmv})}\,,~~~\label{eq:softCOSwave}
\end{eqnarray}
where $\tilde{k}_{\nlightR+\nheavy+1}=K$ and $\tilde{k}_j=k_j$ when $j\neq j_{\Quark}, j_{\Antiquark}$ and 
\begin{eqnarray}
\dot{r}^{\isubrmv}&=&\left(\ident_1,\ldots,\remove{\ident}{0.2}_i,\ldots,\ident_j,\ldots,\ident_{\nlightR+\nheavy},\QQ[\bigl.^{2S+1}\hspace{-1mm}S^{[8]}_J],\remove{\Antiquark}{0.3},
\ldots,\ident_{n+3}\right)\,.
\end{eqnarray}
Thanks to colour conservation, the $n^\mu$ dependent terms stemming from eq.\eqref{eq:axialgaugesum} vanish on the right-hand side (r.h.s) of eq.\eqref{eq:softCOSwave}. Thus, we have, for $\nini\leq k \leq \nlightR+\nheavy$, 
\begin{eqnarray}
\ampsqsoftnmo_{{\rm soft}, kj_{\Quark}}(\dot{r}^{\isubrmv})&=&\frac{k_k\cdot K}{k_k\cdot k_i K\cdot k_i}\ampsqnmoo_{kj_{\Quark}}(\dot{r}^{\isubrmv})\,,\nonumber\\
\ampsqsoftnmo_{{\rm soft}, j_{\Quark}j_{\Quark}}(\dot{r}^{\isubrmv})&=&-\frac{K^2}{\left(K\cdot k_i\right)^2}\CA \ampsqnmoo(\dot{r}^{\isubrmv})\,.
\end{eqnarray}

\subsubsection{Colour-singlet spin-singlet P-wave state}

With eq.\eqref{eq:softSingletP2} in sect.~\ref{sec:softamp}, the soft limit of the amplitude for a colour-singlet spin-singlet P-wave quarkonium is
\begin{eqnarray}
\lim_{k_i\to 0}{\ampnpot_{\left\{[1],0,1,1\right\}}(r)}&=&\mathop{\sum_{j=\nini}}_{j\neq i}^{\nlightR+\nheavy}{g_s\frac{k_j\cdot \varepsilon_{\lambda_i}^*(k_i)}{k_j\cdot k_i}\Qop(\ident_j)\ampnt_{\left\{[1],0,1,1\right\}}(r^{\isubrmv})}\nonumber\\
&&+g_s\left[\frac{\varepsilon_{\lambda_l}^*(K)\cdot \varepsilon_{\lambda_i}^*(k_i)}{K\cdot k_i}-\frac{K \cdot \varepsilon_{\lambda_i}^*(k_i) k_i\cdot \varepsilon_{\lambda_l}^*(K)}{\left(K\cdot k_i\right)^2}\right]\nonumber\\
&&\times\Qop_{{\rm eff}}(\QQ_{[18]})\ampnt_{\left\{[8],0,0,0\right\}}(r^{\isubrmv}).
\end{eqnarray}
Then, the amplitude square in the soft limit becomes
\begin{eqnarray}
&&\lim_{k_i\to 0}{\ampsqnto(\dot{r})}=\lim_{k_i\to 0}{\ampsqnpot_{\left\{[1],0,1,1\right\}}(r)}\nonumber\\
&=&g_s^2\mathop{\sum_{k,l=\nini}}_{k,l\neq i, k\leq l}^{\nlightR+\nheavy}{\frac{k_k\cdot k_l}{k_k\cdot k_i k_l\cdot k_i}\ampsqnmoo_{kl}(\dot{r}^{\isubrmv})}\nonumber\\
&&+g_s^2\mathop{\sum_{k=\nini}}_{k\neq i}^{\nlightR+\nheavy}{\left[\frac{k_{k,\mu}}{k_k\cdot k_i K\cdot k_i}-\frac{K\cdot k_k k_{i,\mu}}{k_k\cdot k_i \left(K\cdot k_i\right)^2}\right]\ampsqnmoo_{k[18]}(\dot{r}^{\isubrmv},\dot{r}_1^{\isubrmv})^\mu}\nonumber\\
&&-g_s^2\frac{2\epsilon-2}{\left(K\cdot k_i\right)^2}C_{{\rm eff}}(\QQ_{[18]})\ampsqnmoo(\dot{r}_1^{\isubrmv}),\label{eq:softCSPwavesinglet}
\end{eqnarray}
where 
\begin{eqnarray}
\dot{r}^{\isubrmv}&=&\left(\ident_1,\ldots,\remove{\ident}{0.2}_i,\ldots,\ident_j,\ldots,\ident_{\nlightR+\nheavy},\QQ[\bigl.^{1}\hspace{-1mm}P^{[1]}_1],\remove{\Antiquark}{0.3},
\ldots,\ident_{n+3}\right),\nonumber\\
\dot{r}_1^{\isubrmv}&=&\left(\ident_1,\ldots,\remove{\ident}{0.2}_i,\ldots,\ident_j,\ldots,\ident_{\nlightR+\nheavy},\QQ[\bigl.^{1}\hspace{-1mm}S^{[8]}_0],\remove{\Antiquark}{0.3},
\ldots,\ident_{n+3}\right),
\end{eqnarray}
and we have used the relations
\begin{eqnarray}
K\cdot \varepsilon^{(*)}_{\lambda_l}(K)&=&0,\label{eq:Kdotepsis0}\\
\sum_{\lambda_l}{\varepsilon^{\mu,*}_{\lambda_l}(K)\varepsilon^{\nu}_{\lambda_l}(K)}&=&-g^{\mu\nu}+\frac{K^\mu K^\nu}{K^2}.\label{eq:polrelation1}
\end{eqnarray}
In numerical calculations, setting the dimensional regulator $\epsilon$ to zero in the prefactor of the soft matrix element ensures consistency with the use of $4$-dimensional external wave functions. The terms dependent on $n^\mu$ from eq.\eqref{eq:axialgaugesum} are absent on the r.h.s of eq.\eqref{eq:softCSPwavesinglet} due to colour conservation and 
\begin{eqnarray}
\left[\frac{\varepsilon_{\lambda_l}^*(K)\cdot \varepsilon_{\lambda_i}^*(k_i)}{K\cdot k_i}-\frac{K \cdot \varepsilon_{\lambda_i}^*(k_i) k_i\cdot \varepsilon_{\lambda_l}^*(K)}{\left(K\cdot k_i\right)^2}\right]_{\varepsilon_{\lambda_i}^*(k_i)\to k_i}&=&0.
\end{eqnarray}
The same observation holds for the other P-wave states discussed later, and there is no need to reiterate this point.

Thus, we have, for $\nini\leq k \leq \nlightR+\nheavy$, 
\begin{eqnarray}
\ampsqsoftnmo_{{\rm soft}, kj_{\Quark}}(\dot{r}^{\isubrmv})&=&\left[\frac{k_{k,\mu}}{k_k\cdot k_i K\cdot k_i}-\frac{K\cdot k_k k_{i,\mu}}{k_k\cdot k_i \left(K\cdot k_i\right)^2}\right]\ampsqnmoo_{k[18]}(\dot{r}^{\isubrmv},\dot{r}_1^{\isubrmv})^\mu,\nonumber\\
\ampsqsoftnmo_{{\rm soft}, j_{\Quark}j_{\Quark}}(\dot{r}^{\isubrmv})&=&-\frac{2\epsilon-2}{\left(K\cdot k_i\right)^2}C_{{\rm eff}}(\QQ_{[18]})\ampsqnmoo(\dot{r}_1^{\isubrmv}).
\end{eqnarray}

\subsubsection{Colour-octet spin-singlet P-wave state}

For the production of a colour-octet spin-singlet P-wave quarkonium, we have established the soft limit at the amplitude level ({\it cf.} eq.\eqref{eq:softOctetP2} in sect.~\ref{sec:softamp}) as follows:
\begin{eqnarray}
\lim_{k_i\to 0}{\ampnpot_{\left\{[8],0,1,1\right\}}(r)}&=&\mathop{\sum_{j=\nini}}_{j\neq i}^{\nlightR+\nheavy}{g_s\frac{k_j\cdot \varepsilon_{\lambda_i}^*(k_i)}{k_j\cdot k_i}\Qop(\ident_j)\ampnt_{\left\{[8],0,1,1\right\}}(r^{\isubrmv})}\nonumber\\
&&+g_s\frac{K \cdot \varepsilon_{\lambda_i}^*(k_i)}{K\cdot k_i}\Qop(\QQ[\bigl.^{1}\hspace{-1mm}P^{[8]}_1])\ampnt_{\left\{[8],0,1,1\right\}}(r^{\isubrmv})\nonumber\\
&&+g_s\left[\frac{\varepsilon_{\lambda_l}^*(K)\cdot \varepsilon_{\lambda_i}^*(k_i)}{K\cdot k_i}-\frac{K \cdot \varepsilon_{\lambda_i}^*(k_i) k_i\cdot \varepsilon_{\lambda_l}^*(K)}{\left(K\cdot k_i\right)^2}\right]\nonumber\\
&&\times\left[\Qop_{{\rm eff}}(\QQ_{[81]})\ampnt_{\left\{[1],0,0,0\right\}}(r^{\isubrmv})+\Qop_{{\rm eff}}(\QQ_{[88]})\ampnt_{\left\{[8],0,0,0\right\}}(r^{\isubrmv})\right].\nonumber\\
\end{eqnarray}
The amplitude square in the soft limit is
\begin{eqnarray}
&&\lim_{k_i\to 0}{\ampsqnto(\dot{r})}=\lim_{k_i\to 0}{\ampsqnpot_{\left\{[8],0,1,1\right\}}(r)}\nonumber\\
&=&g_s^2\mathop{\sum_{k,l=\nini}}_{k,l\neq i, k\leq l}^{\nlightR+\nheavy+1}{\frac{\tilde{k}_k\cdot \tilde{k}_l}{\tilde{k}_k\cdot k_i \tilde{k}_l\cdot k_i}\ampsqnmoo_{kl}(\dot{r}^{\isubrmv})}\nonumber\\
&&+g_s^2\mathop{\sum_{k=\nini}}_{k\neq i}^{\nlightR+\nheavy+1}{\left[\frac{\tilde{k}_{k,\mu}}{\tilde{k}_k\cdot k_i K\cdot k_i}-\frac{K\cdot \tilde{k}_k k_{i,\mu}}{\tilde{k}_k\cdot k_i \left(K\cdot k_i\right)^2}\right]\left[\ampsqnmoo_{k[88]}(\dot{r}^{\isubrmv},\dot{r}_1^{\isubrmv})^\mu+
\ampsqnmoo_{k[81]}(\dot{r}^{\isubrmv},\dot{r}_2^{\isubrmv})^\mu\right]}\nonumber\\
&&-g_s^2\frac{2\epsilon-2}{\left(K\cdot k_i\right)^2}\left[C_{{\rm eff}}
(\QQ_{[88]})\ampsqnmoo(\dot{r}_1^{\isubrmv})+C_{{\rm eff}}
(\QQ_{[81]})\ampsqnmoo(\dot{r}_2^{\isubrmv})\right],
\end{eqnarray}
where 
\begin{eqnarray}
\dot{r}^{\isubrmv}&=&\left(\ident_1,\ldots,\remove{\ident}{0.2}_i,\ldots,\ident_j,\ldots,\ident_{\nlightR+\nheavy},\QQ[\bigl.^{1}\hspace{-1mm}P^{[8]}_1],\remove{\Antiquark}{0.3},
\ldots,\ident_{n+3}\right),\nonumber\\
\dot{r}_1^{\isubrmv}&=&\left(\ident_1,\ldots,\remove{\ident}{0.2}_i,\ldots,\ident_j,\ldots,\ident_{\nlightR+\nheavy},\QQ[\bigl.^{1}\hspace{-1mm}S^{[8]}_0],\remove{\Antiquark}{0.3},
\ldots,\ident_{n+3}\right),\nonumber\\
\dot{r}_2^{\isubrmv}&=&\left(\ident_1,\ldots,\remove{\ident}{0.2}_i,\ldots,\ident_j,\ldots,\ident_{\nlightR+\nheavy},\QQ[\bigl.^{1}\hspace{-1mm}S^{[1]}_0],\remove{\Antiquark}{0.3},
\ldots,\ident_{n+3}\right),
\end{eqnarray}
and $\tilde{k}_{\nlightR+\nheavy+1}=K$ and $\tilde{k}_j=k_j$ when $j\neq j_{\Quark}, j_{\Antiquark}$. Additionally, we have used the relations eq.\eqref{eq:polrelation1}. 


Thus, we have, for $\nini\leq k \leq \nlightR+\nheavy$, 
\begin{eqnarray}
\ampsqsoftnmo_{{\rm soft}, kj_{\Quark}}(\dot{r}^{\isubrmv})&=&\frac{k_k\cdot K}{k_k\cdot k_i K\cdot k_i}\ampsqnmoo_{kj_{\Quark}}(\dot{r}^{\isubrmv})\nonumber\\
&&+\left[\frac{k_{k,\mu}}{k_k\cdot k_i K\cdot k_i}-\frac{K\cdot k_k k_{i,\mu}}{k_k\cdot k_i \left(K\cdot k_i\right)^2}\right]\left[\ampsqnmoo_{k[88]}(\dot{r}^{\isubrmv},\dot{r}_1^{\isubrmv})^\mu+
\ampsqnmoo_{k[81]}(\dot{r}^{\isubrmv},\dot{r}_2^{\isubrmv})^\mu\right],\nonumber\\
\ampsqsoftnmo_{{\rm soft}, j_{\Quark}j_{\Quark}}(\dot{r}^{\isubrmv})&=&-\frac{K^2}{\left(K\cdot k_i\right)^2}\CA\ampsqnmoo(\dot{r}^{\isubrmv})\nonumber\\
&&+\left[\frac{K_{\mu}}{\left(K\cdot k_i\right)^2}-\frac{K^2 k_{i,\mu}}{\left(K\cdot k_i\right)^3}\right]\left[\ampsqnmoo_{j_{\Quark}[88]}(\dot{r}^{\isubrmv},\dot{r}_1^{\isubrmv})^\mu+\ampsqnmoo_{j_{\Quark}[81]}(\dot{r}^{\isubrmv},\dot{r}_2^{\isubrmv})^\mu\right]\nonumber\\
&&-\frac{2\epsilon-2}{\left(K\cdot k_i\right)^2}\left[C_{{\rm eff}}
(\QQ_{[88]})\ampsqnmoo(\dot{r}_1^{\isubrmv})+C_{{\rm eff}}
(\QQ_{[81]})\ampsqnmoo(\dot{r}_2^{\isubrmv})\right].
\end{eqnarray}

\subsubsection{Colour-singlet spin-triplet P-wave state}

We have derived the soft limit at the amplitude level for a colour-singlet spin-triplet P-wave state in sect.~\ref{sec:softamp} ({\it cf.} eq.\eqref{eq:softSingletP3}) as
\begin{eqnarray}
\lim_{k_i\to 0}{\ampnpot_{\left\{[1],1,1,J\right\}}(r)}&=&\mathop{\sum_{j=\nini}}_{j\neq i}^{\nlightR+\nheavy}{g_s\frac{k_j\cdot \varepsilon_{\lambda_i}^*(k_i)}{k_j\cdot k_i}\Qop(\ident_j)\ampnt_{\left\{[1],1,1,J\right\}}(r^{\isubrmv})}\nonumber\\
&&+g_s\sum_{\lambda_l,\lambda_s}{\langle J,\lambda_j|1,\lambda_l;1,\lambda_s\rangle \left[\frac{\varepsilon_{\lambda_l}^*(K)\cdot \varepsilon_{\lambda_i}^*(k_i)}{K\cdot k_i}-\frac{K \cdot \varepsilon_{\lambda_i}^*(k_i) k_i\cdot \varepsilon_{\lambda_l}^*(K)}{\left(K\cdot k_i\right)^2}\right]}\nonumber\\
&&\times\Qop_{{\rm eff}}(\QQ_{[18]})\ampnt_{\left\{[8],1,0,1\right\}}(r^{\isubrmv})\,.
\end{eqnarray}
The amplitude square in the soft limit is
\begin{eqnarray}
&&\lim_{k_i\to 0}{\ampsqnto(\dot{r})}=\lim_{k_i\to 0}{\ampsqnpot_{\left\{[1],1,1,J\right\}}(r)}\nonumber\\
&=&g_s^2\mathop{\sum_{k,l=\nini}}_{k,l\neq i, k\leq l}^{\nlightR+\nheavy}{\frac{k_k\cdot k_l}{k_k\cdot k_i k_l\cdot k_i}\ampsqnmoo_{kl}(\dot{r}^{\isubrmv})}\nonumber\\
&&+g_s^2\mathop{\sum_{k=\nini}}_{k\neq i}^{\nlightR+\nheavy}{\left[\frac{k_{k,\mu}}{k_k\cdot k_i K\cdot k_i}-\frac{K\cdot k_k k_{i,\mu}}{k_k\cdot k_i \left(K\cdot k_i\right)^2}\right]\ampsqnmoo_{J,k[18]}(\dot{r},\dot{r}_1)^{\mu}}\nonumber\\
&&-g_s^2\left[\frac{g_{\mu\nu}}{\left(K\cdot k_i\right)^2}+\frac{K^2 k_{i,\mu}k_{i,\nu}}{\left(K\cdot k_i\right)^4}\right]C_{{\rm eff}}(\QQ_{[18]})\ampsqnmoo_J(\dot{r}_1^{\isubrmv})^{\mu\nu}\,,
\end{eqnarray}
where 
\begin{eqnarray}
\dot{r}^{\isubrmv}&=&\left(\ident_1,\ldots,\remove{\ident}{0.2}_i,\ldots,\ident_j,\ldots,\ident_{\nlightR+\nheavy},\QQ[\bigl.^{3}\hspace{-1mm}P^{[1]}_J],\remove{\Antiquark}{0.3}\,,
\ldots,\ident_{n+3}\right),\nonumber\\
\dot{r}_1^{\isubrmv}&=&\left(\ident_1,\ldots,\remove{\ident}{0.2}_i,\ldots,\ident_j,\ldots,\ident_{\nlightR+\nheavy},\QQ[\bigl.^{3}\hspace{-1mm}S^{[8]}_1],\remove{\Antiquark}{0.3}\,,
\ldots,\ident_{n+3}\right),
\end{eqnarray}
and we have used the relations
\begin{eqnarray}
K\cdot \varepsilon^{(*)}_{\lambda_l}(K)&=&K\cdot \varepsilon^{(*)}_{\lambda_s}(K)=0\,,\nonumber\\
\sum_{\lambda_l}{\varepsilon^{\mu,*}_{\lambda_l}(K)\varepsilon^{\nu}_{\lambda_l}(K)}&=&-g^{\mu\nu}+\frac{K^\mu K^\nu}{K^2}=\Pi^{\mu\nu}(K)\,,\nonumber\\
\varepsilon_{0,0}^{\mu\nu,*}(K)\varepsilon_{0,0}^{\alpha\beta}(K)&=&\frac{1}{3}\Pi^{\mu\nu}(K)\Pi^{\alpha\beta}(K)\,,\nonumber\\
\sum_{\lambda_j=-1}^{1}{\varepsilon_{1,\lambda_j}^{\mu\nu,*}(K)\varepsilon_{1,\lambda_j}^{\alpha\beta}(K)}&=&\frac{1}{2}\left[\Pi^{\mu\alpha}(K)\Pi^{\nu\beta}(K)-\Pi^{\mu\beta}(K)\Pi^{\nu\alpha}(K)\right]\,,\nonumber\\
\sum_{\lambda_j=-2}^{2}{\varepsilon_{2,\lambda_j}^{\mu\nu,*}(K)\varepsilon_{2,\lambda_j}^{\alpha\beta}(K)}&=&\frac{1}{2}\left[\Pi^{\mu\alpha}(K)\Pi^{\nu\beta}(K)+\Pi^{\mu\beta}(K)\Pi^{\nu\alpha}(K)\right]-\frac{1}{3}\Pi^{\mu\nu}(K)\Pi^{\alpha\beta}(K)\,.\nonumber\\
\end{eqnarray}

Then, we can easily derive, 
\begin{eqnarray}
\ampsqsoftnmo_{{\rm soft}, kj_{\Quark}}(\dot{r}^{\isubrmv})&=&\left[\frac{k_{k,\mu}}{k_k\cdot k_i K\cdot k_i}-\frac{K\cdot k_k k_{i,\mu}}{k_k\cdot k_i \left(K\cdot k_i\right)^2}\right]\ampsqnmoo_{J,k[18]}(\dot{r}^{\isubrmv},\dot{r}_1^{\isubrmv})^\mu\,,\nonumber\\
\ampsqsoftnmo_{{\rm soft}, j_{\Quark}j_{\Quark}}(\dot{r}^{\isubrmv})&=&-\left[\frac{g_{\mu\nu}}{\left(K\cdot k_i\right)^2}+\frac{K^2 k_{i,\mu}k_{i,\nu}}{\left(K\cdot k_i\right)^4}\right]C_{{\rm eff}}(\QQ_{[18]})\ampsqnmoo_J(\dot{r}_1^{\isubrmv})^{\mu\nu}\,,
\end{eqnarray}
with $\nini\leq k \leq \nlightR+\nheavy$.

\subsubsection{Colour-octet spin-triplet P-wave state}

Finally, for the colour-octet spin-triplet P-wave state production, the soft limit of the amplitude from sect.~\ref{sec:softamp} ({\it cf.} eq.\eqref{eq:softOctetP3}) is
\begin{eqnarray}
\lim_{k_i\to 0}{\ampnpot_{\left\{[8],1,1,J\right\}}(r)}&=&\mathop{\sum_{j=\nini}}_{j\neq i}^{\nlightR+\nheavy}{g_s\frac{k_j\cdot \varepsilon_{\lambda_i}^*(k_i)}{k_j\cdot k_i}\Qop(\ident_j)\ampnt_{\left\{[8],1,1,J\right\}}(r^{\isubrmv})}\nonumber\\
&&+g_s\frac{K \cdot \varepsilon_{\lambda_i}^*(k_i)}{K\cdot k_i}\Qop(\QQ[\bigl.^{3}\hspace{-1mm}P^{[8]}_J])\ampnt_{\left\{[8],1,1,J\right\}}(r^{\isubrmv})\nonumber\\
&&+g_s\sum_{\lambda_l,\lambda_s}{\langle J,\lambda_j|1,\lambda_l; 1, \lambda_s\rangle\left[\frac{\varepsilon_{\lambda_l}^*(K)\cdot \varepsilon_{\lambda_i}^*(k_i)}{K\cdot k_i}-\frac{K \cdot \varepsilon_{\lambda_i}^*(k_i) k_i\cdot \varepsilon_{\lambda_l}^*(K)}{\left(K\cdot k_i\right)^2}\right]}\nonumber\\
&&\times\left[\Qop_{{\rm eff}}(\QQ_{[81]})\ampnt_{\left\{[1],1,0,1\right\}}(r^{\isubrmv})+\Qop_{{\rm eff}}(\QQ_{[88]})\ampnt_{\left\{[8],1,0,1\right\}}(r^{\isubrmv})\right].
\end{eqnarray}
The amplitude square in the soft limit is
\begin{eqnarray}
&&\lim_{k_i\to 0}{\ampsqnto(\dot{r})}=\lim_{k_i\to 0}{\ampsqnpot_{\left\{[8],1,1,J\right\}}(r)}\nonumber\\
&=&g_s^2\mathop{\sum_{k,l=\nini}}_{k,l\neq i, k\leq l}^{\nlightR+\nheavy+1}{\frac{\tilde{k}_k\cdot \tilde{k}_l}{\tilde{k}_k\cdot k_i \tilde{k}_l\cdot k_i}\ampsqnmoo_{kl}(\dot{r}^{\isubrmv})}\nonumber\\
&&+g_s^2\mathop{\sum_{k=\nini}}_{k\neq i}^{\nlightR+\nheavy+1}{\left[\frac{\tilde{k}_{k,\mu}}{\tilde{k}_k\cdot k_i K\cdot k_i}-\frac{K\cdot \tilde{k}_k k_{i,\mu}}{\tilde{k}_k\cdot k_i \left(K\cdot k_i\right)^2}\right]\left[\ampsqnmoo_{J,k[88]}(\dot{r}^{\isubrmv},\dot{r}_1^{\isubrmv})^\mu+
\ampsqnmoo_{J,k[81]}(\dot{r}^{\isubrmv},\dot{r}_2^{\isubrmv})^\mu\right]}\nonumber\\
&&-g_s^2\left[\frac{g_{\mu\nu}}{\left(K\cdot k_i\right)^2}+\frac{K^2 k_{i,\mu}k_{i,\nu}}{\left(K\cdot k_i\right)^4}\right]\left[C_{{\rm eff}}
(\QQ_{[88]})\ampsqnmoo_{J}(\dot{r}_1^{\isubrmv})^{\mu\nu}+C_{{\rm eff}}
(\QQ_{[81]})\ampsqnmoo_{J}(\dot{r}_2^{\isubrmv})^{\mu\nu}\right],\nonumber\\
\end{eqnarray}
where 
\begin{eqnarray}
\dot{r}^{\isubrmv}&=&\left(\ident_1,\ldots,\remove{\ident}{0.2}_i,\ldots,\ident_j,\ldots,\ident_{\nlightR+\nheavy},\QQ[\bigl.^{3}\hspace{-1mm}P^{[8]}_J],\remove{\Antiquark}{0.3},
\ldots,\ident_{n+3}\right),\nonumber\\
\dot{r}_1^{\isubrmv}&=&\left(\ident_1,\ldots,\remove{\ident}{0.2}_i,\ldots,\ident_j,\ldots,\ident_{\nlightR+\nheavy},\QQ[\bigl.^{3}\hspace{-1mm}S^{[8]}_1],\remove{\Antiquark}{0.3},
\ldots,\ident_{n+3}\right),\nonumber\\
\dot{r}_2^{\isubrmv}&=&\left(\ident_1,\ldots,\remove{\ident}{0.2}_i,\ldots,\ident_j,\ldots,\ident_{\nlightR+\nheavy},\QQ[\bigl.^{3}\hspace{-1mm}S^{[1]}_1],\remove{\Antiquark}{0.3},
\ldots,\ident_{n+3}\right),
\end{eqnarray}
and $\tilde{k}_{\nlightR+\nheavy+1}=K$ and $\tilde{k}_j=k_j$ when $j\neq j_{\Quark}, j_{\Antiquark}$. Once again, we have used the relations eq.\eqref{eq:polrelation1}.

Thus, we can easily get, for $\nini\leq k \leq \nlightR+\nheavy$, 
\begin{eqnarray}
\ampsqsoftnmo_{{\rm soft}, kj_{\Quark}}(\dot{r}^{\isubrmv})&=&\frac{k_k\cdot K}{k_k\cdot k_i K\cdot k_i}\ampsqnmoo_{kj_{\Quark}}(\dot{r}^{\isubrmv})\nonumber\\
&&+\left[\frac{k_{k,\mu}}{k_k\cdot k_i K\cdot k_i}-\frac{K\cdot k_k k_{i,\mu}}{k_k\cdot k_i \left(K\cdot k_i\right)^2}\right]\left[\ampsqnmoo_{J,k[88]}(\dot{r}^{\isubrmv},\dot{r}_1^{\isubrmv})^\mu+
\ampsqnmoo_{J,k[81]}(\dot{r}^{\isubrmv},\dot{r}_2^{\isubrmv})^\mu\right],\nonumber\\
\ampsqsoftnmo_{{\rm soft}, j_{\Quark}j_{\Quark}}(\dot{r}^{\isubrmv})&=&-\frac{K^2}{\left(K\cdot k_i\right)^2}\CA\ampsqnmoo(\dot{r}^{\isubrmv})\nonumber\\
&&+\left[\frac{K_{\mu}}{\left(K\cdot k_i\right)^2}-\frac{K^2 k_{i,\mu}}{\left(K\cdot k_i\right)^3}\right]\left[\ampsqnmoo_{J,j_{\Quark}[88]}(\dot{r}^{\isubrmv},\dot{r}_1^{\isubrmv})^\mu+\ampsqnmoo_{J,j_{\Quark}[81]}(\dot{r}^{\isubrmv},\dot{r}_2^{\isubrmv})^\mu\right]\nonumber\\
&&-\left[\frac{g_{\mu\nu}}{\left(K\cdot k_i\right)^2}+\frac{K^2 k_{i,\mu}k_{i,\nu}}{\left(K\cdot k_i\right)^4}\right]\left[C_{{\rm eff}}
(\QQ_{[88]})\ampsqnmoo_{J}(\dot{r}_1^{\isubrmv})^{\mu\nu}\right.\nonumber\\
&&\left.+C_{{\rm eff}}
(\QQ_{[81]})\ampsqnmoo_{J}(\dot{r}_2^{\isubrmv})^{\mu\nu}\right]\,.
\end{eqnarray}

\section{FKS subtraction for single quarkonium production\label{sec:FKS}}

\subsection{FKS pairs and partition functions}

With the squared real amplitudes in the soft limit at hand, we will now structure our approach using the FKS formalism. To categorise the IR divergences and facilitate their subtraction, we introduce a set of ordered pairs for any given process $\dot{r}\in \allprocno$. This set is denoted as the set of FKS pairs
\begin{eqnarray}
\FKSpairs(\dot{r})&=&\Big\{(i,j)\;\Big|\;3\le i\le\nlightR+2\,, 
\nini\le j\le\nlightR+\nheavy+1\,, 
i\ne j\,,
\nonumber\\*&&\phantom{aaa}
\ampsqnto(\dot{r})\JetsB\to\infty~~{\rm if}~~\tilde{k}_i^0\to 0~~
{\rm or}~~\tilde{k}_j^0\to 0~~{\rm or}~~\vec{\tilde{k}}_i\parallel \vec{\tilde{k}}_j\Big\}.
\phantom{aaaaa}
\label{PFKSdef}
\end{eqnarray}
This implies that a pair of particles belongs to the set of FKS pairs if they induce soft or collinear singularities (or both) in the $n$-body real matrix elements. It is important to note that $\tilde{k}_j^0\to 0$ is irrelevant when $j=1,2$. In the calculation of an NLO cross section within the FKS formalism, each pair belonging to $\FKSpairs$ corresponds to a set of subtractions of soft and collinear singularities.

In the FKS formalism, we multiply the real emission matrix elements with the so-called measurement/partition function $\Sfunij$, so that the phase space is partitioned into different kinematic regions where each region contains at most one soft and one collinear singularity. The partitioning is accomplished through the introduction of a set of positive-definite functions
\begin{eqnarray}
\Sfunij(\dot{r})\,,\;\;\;\;\;\;(i,j)\in\FKSpairs(\dot{r})\,,
\end{eqnarray}
where the argument $\dot{r}\in \allprocno$ means that we can choose different $\Sfun$ for different processes. As described in ref.~\cite{Frederix:2009yq}, $\Sfunij$ is defined in different regions as following:
\begin{eqnarray}
\!\!\!\sum_{(i,j)\in \FKSpairs(\dot{r})}{\Sfunij(\dot{r})} &=&1, \nonumber \\
\lim_{\vec{\tilde{k}}_i\parallel\vec{\tilde{k}}_j}{\Sfunij(\dot{r})} &=& h_{ij}\left(\frac{\tilde{k}_i^0}{\tilde{k}_i^0+\tilde{k}_j^0}\right),\,~~~~{\rm if}~~\tilde{m}_i=\sqrt{\tilde{k}_i^2}=\tilde{m}_j=\sqrt{\tilde{k}_j^2}=0\,,\nonumber \\
\lim_{\tilde{k}_i^0 \to 0}{\Sfunij(\dot{r})} &=& c_{ij}\,,\phantom{}\;
{\rm if}~~~\ident_i=g\,,\,~~~~{\rm with}~~0<c_{ij}\le 1\qquad{\rm and}\,\,\,\,\,
\mathop{\sum_{j}}_{(i,j)\in\FKSpairs(\dot{r})}\!\!\!\!\!c_{ij}=1\,,\nonumber\\
\lim_{\vec{\tilde{k}}_k\parallel\vec{\tilde{k}}_l}{\Sfunij(\dot{r})}&=&0~~\forall\,
\{k,l\}\ne\{i,j\}~~{\rm with}~~(k,l)\in\FKSpairs(\dot{r})~~{\land}~~
\tilde{m}_k=\tilde{m}_l=0\,,\nonumber\\
\lim_{\tilde{k}_k^0\to 0}{\Sfunij(\dot{r})}&=&0~~\forall k~~{\rm with}~~\ident_k=g
~~{\rm and}~~\exists l~~{\rm with}~~(k,l)\in\FKSpairs(\dot{r})~{\lor}~
(l,k)\in\FKSpairs(\dot{r}).
\nonumber\\&&
\label{eq:SfunC}
\end{eqnarray}
In other words, $\Sfunij$ goes to zero in all regions of the phase space where the real emission matrix elements diverge, except if this involves particle $i$ being soft, or particles $i$ and $j$ being collinear. The functions $h_{ij}(z)$ introduced in eq.\eqref{eq:SfunC} 
are defined in \mbox{$0\le z\le 1$}, and have the following properties:
\begin{eqnarray}
h_{ij}(z)&=&1\,,~~~~~~~~~{\rm if}~~\nini\le j\le 2\,,
\label{eq:hdef1}
\\
h_{ij}(z)&=&h(z)\,,~~~~~{\rm if}~~3\le j\le \nlightR+\nheavy+1\,,
\label{eq:hdef2}
\end{eqnarray}
with $h(z)$ a positive-definite function such that
\begin{equation}
\lim_{z\to 0}{h(z)}=1\,,\;\;\;\;\;\;
\lim_{z\to 1}{h(z)}=0\,,\;\;\;\;\;\;
h(z)+h(1-z)=1\,.
\label{eq:hdef3}
\end{equation}
Note that in $\dot{r}$, we have glued $\Quark$ and $\Antiquark$ to a single particle $\QQ[n]$. This means that the number of strongly interacting massive particles (except $n$ is a colour-singlet S-wave state) is $\nheavy-1$ instead of $\nheavy$.

The real matrix elements can be rewritten 
\begin{eqnarray}
\ampsqnto(\dot{r})&=&\sum_{(i,j)\in \FKSpairs(\dot{r})}{\Sfunij(\dot{r})\ampsqnto(\dot{r})},
\end{eqnarray}
which will only be singular for a given term on the r.h.s if particle $i$ is soft and/or particles $i$ and $j$ are collinear after applying $\JetsB$.

\subsection{Local subtraction counterterms}

The IR divergence subtracted real cross sections can be formulated in the center of mass frame of the incoming partons:
\begin{eqnarray}
k_1=\tilde{k}_1=\frac{\sqrt{s}}{2}(1,0,0,1)\,,\;\;\;\;\;
k_2=\tilde{k}_2=\frac{\sqrt{s}}{2}(1,0,0,-1)\,.
\end{eqnarray}
In this frame, for each pair $(i,j)\in \FKSpairs(\dot{r})$, we can introduce the variables $\xii$ and $\yij$, where
\begin{eqnarray}
\tilde{k}_i^0&=&\frac{\sqrt{s}}{2}\xii\,,
\label{eq:xiidef}
\\
\vec{\tilde{k}}_i\mydot\vec{\tilde{k}}_j&=&\abs{\vec{\tilde{k}}_i}\abs{\vec{\tilde{k}}_j}\yij\,.
\label{eq:yijdef}
\end{eqnarray}
Thus, $\xii$ is the rescaled energy of the FKS parton $i$, and $\yij$ is the cosine of the angle between the FKS parton $i$ and its sister $j$. The soft and collinear singularities of $\Sfunij(\dot{r})\ampsqnto(\dot{r})\JetsB$ correspond to $\xii=0$ and to $\yij=1$, respectively. The IR-divergence locally subtracted partonic cross section is
\begin{eqnarray}
d\hat{\sigma}_{{\rm FKS}}(\dot{r})&=&\sum_{(i,j)\in\FKSpairs(\dot{r})}{d\hat{\sigma}_{ij}(\dot{r})},
\end{eqnarray}
where the IR-divergence locally subtracted real partonic cross section is
\begin{equation}
d\hat{\sigma}_{ij}(\dot{r})=\xic\omyijd
\Big((1-\yij)\xii^2\ampsqnto(\dot{r})\Big)
\Sfunij(\dot{r})\frac{\JetsB}{\avg(\dot{r})}\oavg(\dot{r})\, 
d\xii d\yij d\phii\tphspnmoij(\dot{r})\,.
\label{eq:dsigijnpo}
\end{equation}
The variable $\phii$ is the azimuthal direction of the FKS parton. The quantity $\tphspnmoij$ is the reduced $(n-1)$-body phase space via the following relation:
\begin{eqnarray}
\phspn(\dot{r})&=&\xii^{1-2\epsilon} d\xii (1-\yij^2)^{-\epsilon}d\yij d\Omega_{i}^{(2-2\epsilon)} \tphspnmoij(\dot{r})\,.
\label{eq:tphspdef}
\end{eqnarray}
The reduced phase space measure has the following limits:
\begin{eqnarray}
\lim_{\xii\to 0}{\tphspnmoij(\dot{r})}&=&\frac{s^{1-\epsilon}}{(4\pi)^{3-2\epsilon}}\phispnmo(\dot{r}^{\isubrmv})\,,\,~~~~~~~~~{\rm if}~~\tilde{m}_i=0\,,
\label{eq:tphisoft}
\\
\lim_{\yij\to 1}{\tphspnmoij(\dot{r})}&=&\frac{s^{1-\epsilon}}{(4\pi)^{3-2\epsilon}}\phispnmo(\dot{r}^{j\oplus i,\isubrmv})\,,~~~~~{\rm if}~~\tilde{m}_i=\tilde{m}_j=0\,.
\label{eq:tphicoll}
\end{eqnarray}
In $d=4$ dimensions, we can simply set $\epsilon=0$ and $d\Omega_i^{(2-2\epsilon)}=d\phii$. 
The distributions entering eq.\eqref{eq:dsigijnpo} are defined as 
follows, for any test functions $f()$ and $g()$:
\begin{eqnarray}
\int_0^{\ximax}{d\xii f(\xii)\xic}&=&
\int_0^{\ximax}{d\xii \frac{f(\xii)-f(0)\stepf(\xicut-\xii)}{\xii}}\,,
\label{eq:distrxii}
\\
\int_{-1}^1{d\yij g(\yij)\omyijd}&=&
\int_{-1}^1{d\yij \frac{g(\yij)-g(1)\stepf(\yij-1+\delta)}{1-\yij}}\,,
\label{eq:distryij}
\end{eqnarray}
where
\begin{equation}
\ximax=1-\frac{1}{s}\left(\sum_{k=3}^{n+2}\tilde{m}_k\right)^2\,,
\end{equation}
and $\stepf()$ is the Heaviside theta function.
Note that $\tilde{k}_j=k_j, \tilde{m}_{j}=\sqrt{\tilde{k}_j^2}=\sqrt{k_j^2}=m_j$ when $j\leq \nlightR+\nheavy$, $\tilde{k}_{\nlightR+\nheavy+1}=k_{\nlightR+\nheavy+1}+k_{\nlightR+\nheavy+2}=K,\tilde{m}_{\nlightR+\nheavy+1}=m_{\Quark}+m_{\Antiquark}$, and $\tilde{k}_j=k_{j+1},\tilde{m}_j=m_{j+1}$ when $j\geq \nlightR+\nheavy+3$.
In eqs.\eqref{eq:distrxii} and \eqref{eq:distryij}, $\xicut$ and $\delta$ are 
free parameters, that can be chosen in the ranges
\begin{equation}
0<\xicut\le\ximax\,,\;\;\;\;\;
0<\delta\le 2\,.
\end{equation}
In \mfks~\cite{Frederix:2009yq}, $\delta=\deltaI$ for the initial state collinear singularities $(i,j)\in \FKSpairs(\dot{r}), j\leq 2$, and $\delta=\deltaO$ for the final state collinear singularities $(i,j)\in \FKSpairs(\dot{r}) , j\geq 3$. Now, we can introduce the quantity in $d=4$ dimensions
\begin{eqnarray}
\Sigma_{ij}(\dot{r}; \xii, \yij)&=&\Big((1-\yij)\xii^2\ampsqnto(\dot{r})\Big)
\Sfunij(\dot{r})\frac{\JetsB}{\avg(\dot{r})}\oavg(\dot{r})\, 
d\phii\tphspnmoij(\dot{r}),
\end{eqnarray}
so that
\begin{eqnarray}
d\hat{\sigma}_{ij}(\dot{r})=\xic\omyijd \Sigma_{ij}(\dot{r}; \xii, \yij) d\xii d\yij.
\end{eqnarray}
If we expand the plus distributions, we have
\begin{eqnarray}
d\hat{\sigma}_{ij}(\dot{r})&=&
\frac{1}{\xii(1-\yij)}
\Big[\Sigma_{ij}(\dot{r}; \xii,\yij)-\Sigma_{ij}(\dot{r};\xii,1)\stepf(\yij-1+\delta)
\nonumber\\*&&-\Sigma_{ij}(\dot{r}; 0,\yij)\stepf(\xicut-\xii)
+\Sigma_{ij}(\dot{r}; 0,1)\stepf(\xicut-\xii)\stepf(\yij-1+\delta)
\Big]d\xii d\yij.
\nonumber\\*&&
\label{eq:dsigijnpoE}
\end{eqnarray}
The first term in the integrand, called ``events", contributes to the initial partonic real cross section, while the remaining three terms are the local subtraction counterterms, which are called ``collinear counterevent", ``soft counterevent", and ``soft-collinear counterevent", respectively. The collinear and soft-collinear local counterterms can be inferred from the collinear limit of the real-emission matrix elements
\begin{eqnarray}
\lim_{\vec{\tilde{k}}_i\parallel \vec{\tilde{k}}_j}{(1-\yij)\xii^2\ampsqnto(\dot{r})}&=&\frac{4}{s}g_s^2\mu^{2\epsilon}\xii P_{\ident_{j\oplus \bar{i}}\ident_j}^{<}(1-\xii,\epsilon)\ampsqnmoo(\dot{r}^{j\oplus \bar{i},\isubrmv})\nonumber\\
&&+\frac{4}{s}g_s^2\mu^{2\epsilon}\xi_i\underbrace{Q_{\ident_{j\oplus \bar{i}}^\star\ident_j}(1-\xii)\ampsqnmootilde_{ij}(\dot{r}^{j\oplus \bar{i},\isubrmv})}_{\equiv \Delta_{ij}},\label{eq:localinitialcoll0}\\
&&~~~~~~~~~~~~~~~~~~~~~~~~~~~~~~~~j=\nini,\ldots,2,\nonumber\\
\lim_{\vec{\tilde{k}}_i\parallel \vec{\tilde{k}}_j}{(1-\yij)\xii^2\ampsqnto(\dot{r})\Sfunij(\dot{r})}&=&\frac{4}{s}g_s^2\mu^{2\epsilon}\frac{1-z}{z}h(z)P_{\ident_j\ident_{j\oplus i}}^{<}(z,\epsilon)\ampsqnmoo(\dot{r}^{j\oplus i,\isubrmv})\nonumber\\
&&+\frac{4}{s}g_s^2\mu^{2\epsilon}\frac{1-z}{z}h(z)\underbrace{Q_{\ident_j\ident_{j\oplus i}^\star}(z)\ampsqnmootilde_{ij}(\dot{r}^{j\oplus i,\isubrmv})}_{\equiv \Delta_{ij}},\label{eq:localfinalcoll0}\\
&&~~~~~~~~~~~~~~~~~~~~~~~~~~~~~~~~j=3,\ldots, \nlightR+2, j\neq i,\nonumber
\end{eqnarray}
where $\bar{i}$ denotes the antiparticle of the particle $i$, $P_{\ident_k\ident_l}^{<}(z,\epsilon)$ is the unregularised Altarelli-Parisi kernel for $z<1$ in $d=4-2\epsilon$ dimensions that can be found in the literature (see, \eg, eqs.(D.15-D.18) in ref.~\cite{Frederix:2009yq}), and $Q_{\ident_{j\oplus \bar{i}}^\star\ident_j}$ and $Q_{\ident_j\ident_{j\oplus i}^\star}$ are given in eqs.(D.3-D.10) of ref.~\cite{Frederix:2009yq}. The reduced matrix element is defined as~\cite{Frixione:1995ms,Frederix:2009yq}
\begin{eqnarray}
\ampsqnmootilde_{ij}(\dot{r}^{j\oplus i,\isubrmv})&=&\frac{1}{2s}
\frac{1}{\omega(\Ione)\omega(\Itwo)}
\Re{\left\{\frac{\langle ij\rangle}{[ij]}\tilde{\mathop{\sum_{\rm colour}}_{\rm spin}}\!
\ampnmoo_+(\dot{r}^{j\oplus i,\isubrmv})
{\ampnmoo_-(\dot{r}^{j\oplus i,\isubrmv})}^{\star}\right\}}\,,~~~~~
\end{eqnarray}
where the spinor-helicity formalism takes the conventions of ref.~\cite{Mangano:1990by}, $\ampnmoo_{\pm}$ represents the helicity amplitude with the helicity of the parton $j\oplus i$ being $\pm$, and the sum with a tilde has summed over the colour and spin of the external states except the spin of the particle $j\oplus i$.
The $\Delta_{ij}$ term vanishes upon the integration of the azimuthal variable $d\phii$ of the parton $i$. In eq.\eqref{eq:localfinalcoll0}, we have assumed $\tilde{k}_i=(1-z)\left(\tilde{k}_i+\tilde{k}_j\right)$ and $\tilde{k}_j=z\left(\tilde{k}_i+\tilde{k}_j\right)$. This essentially gives us for the initial-state collinear singularities ($\nini\leq j\leq 2$)
\begin{eqnarray}
\Sigma_{ij}(\dot{r}; \xi_i, 1)&=&\frac{4}{\left(4\pi\right)^3}g_s^2\xii \left[P_{\ident_{j\oplus \bar{i}}\ident_j}^{<}(1-\xii,0)\ampsqnmoo(\dot{r}^{j\oplus \bar{i},\isubrmv})+\Delta_{ij}\right]\nonumber\\
&&\times\frac{\JetsB}{\avg(\dot{r})}\oavg(\dot{r})\, 
d\phii\phispnmo(\dot{r}^{j\oplus \bar{i},\isubrmv}),
\end{eqnarray}
and, for the final state collinear singularities ($3\leq j \leq \nlightR+2, j\neq i$), we have
\begin{eqnarray}
\Sigma_{ij}(\dot{r}; \xi_i, 1)&=&\frac{4}{\left(4\pi\right)^3}g_s^2\frac{1-z}{z}h(z)\left[P_{\ident_j\ident_{j\oplus i}}^{<}(z,0)\ampsqnmoo(\dot{r}^{j\oplus i,\isubrmv})+\Delta_{ij}\right]\nonumber\\
&&\times\frac{\JetsB}{\avg(\dot{r})}\oavg(\dot{r})\, 
d\phii\phispnmo(\dot{r}^{j\oplus i,\isubrmv})\,.
\end{eqnarray}
For the soft-collinear counterparts, we need to take $\xi_i\to 0$ and $z\to 1$ on the r.h.s of the above two equations. The soft local counterterm is
\begin{eqnarray}
\Sigma_{ij}(\dot{r}; 0, \yij)&=&\frac{s}{\left(4\pi\right)^3}(1-\yij)\Big(\lim_{\xii\to 0}{\xii^2\ampsqnto(\dot{r})}\Big)
c_{ij}\frac{\JetsB}{\avg(\dot{r})}\oavg(\dot{r})\, 
d\phii\phispnmo(\dot{r}^{\isubrmv}),
\end{eqnarray}
where the soft limit of the real matrix element has been derived in sect.~\ref{sec:softampsquare}.

\subsection{Integrated soft counterterm\label{sec:integsoftCTs}}

We can split the integrated soft counterterms into two parts:
\begin{eqnarray}
d\hat{\sigma}^{(S)}(\dot{r})&=&d\hat{\sigma}^{(S,1)}(\dot{r})+d\hat{\sigma}^{(S,2)}(\dot{r}).
\end{eqnarray}

The first one is simply stemming from the $(4-2\epsilon)$-dimensional counterpart of the function $\Sigma_{ij}(\dot{r}; 0, \yij)$ in the local subtraction counterterms appearing in the former section. It is
\begin{eqnarray}
&&d\hat{\sigma}^{(S,1)}(\dot{r})\nonumber\\
&=&\phispnmo(\dot{r}^{\isubrmv})\left(\frac{\mu^{2}}{s}\right)^\epsilon\frac{s}{(4\pi)^{3-2\epsilon}}\sum_{(i,j)\in\FKSpairs(\dot{r})}{c_{ij}\frac{\JetsB}{\avg(\dot{r})}\oavg(\dot{r})}\nonumber\\
&&\times\int{\xii^{-1-2\epsilon}(1-\yij^2)^{-\epsilon}\left(\lim_{\xii\to 0}{\xii^2\ampsqnto(\dot{r})}\right)\stepf(\xicut-\xii)d\xii d\yij d\Omega_i^{(2-2\epsilon)}}\nonumber\\
&=&\frac{\phispnmo(\dot{r}^{\isubrmv})}{8\pi^2}\frac{\JetsB}{\avg(\dot{r})}\oavg(\dot{r})\mathop{\sum_{i=3}}_{\ident_i=g}^{\nlightR+2}{\left[-\frac{\xicut^{-2\epsilon}}{2\epsilon}\frac{2^{2\epsilon}}{(2\pi)^{1-2\epsilon}}\left(\frac{s}{\mu^2}\right)^{-\epsilon}\int{\left(\lim_{\xii\to 0}{\left(\tilde{k}_i^0\right)^2\ampsqnto(\dot{r})}\right)d\Omega_i}\right]}\nonumber\\
&=&\frac{\alpha_s}{2\pi}\phispnmo(\dot{r}^{\isubrmv})\frac{\JetsB}{\avg(\dot{r})}\oavg(\dot{r})\nonumber\\
&&\times\mathop{\sum_{i=3}}_{\ident_i=g}^{\nlightR+2}{\mathop{\sum_{k,l=\nini}}_{k,l\neq i, k\leq l}^{\nlightR+\nheavy+1}{\left[-\frac{\xicut^{-2\epsilon}}{2\epsilon}\frac{2^{2\epsilon}}{(2\pi)^{1-2\epsilon}}\left(\frac{s}{\mu^2}\right)^{-\epsilon}\int{\left(\lim_{\xii\to 0}{\left(\tilde{k}_i^0\right)^2\ampsqsoftnmo_{{\rm soft},kl}(\dot{r}^{\isubrmv})}\right)d\Omega_i}\right]}}\nonumber\\
&=&\frac{\alpha_s}{2\pi}\phispnmo(\dot{r}^{\isubrmv})\frac{\JetsB}{\avg(\dot{r}^{\isubrmv})}\oavg(\dot{r}^{\isubrmv})\nonumber\\
&&\times\sum_{k=\nini}^{\nlightB+\nheavy+1}{\sum_{l=k}^{\nlightB+\nheavy+1}{\left[-\frac{\xicut^{-2\epsilon}}{2\epsilon}\frac{2^{2\epsilon}}{(2\pi)^{1-2\epsilon}}\left(\frac{s}{\mu^2}\right)^{-\epsilon}\int{\left(\lim_{\xii\to 0}{\left(\tilde{k}_i^0\right)^2\ampsqsoftnmo_{{\rm soft},kl}(\dot{r}^{\isubrmv})}\right)d\Omega_i}\right]}}\,,\nonumber\\
\end{eqnarray}
where the $d-1=3-2\epsilon$ dimensional solid angle measure is 
\begin{eqnarray}
d\Omega_i&=&(1-\yij^2)^{-\epsilon}d\yij d\Omega_i^{(2-2\epsilon)},\label{eq:solidanglemeasure}
\end{eqnarray}
and we have used eq.\eqref{eq:softampsq0} as well as the fact that $\lim_{\xii\to 0}{\xii^2\ampsqnto(\dot{r})}$ is independent of $\xii$. In the last equation, we have removed any final state gluon $\ident_i=g$ and have changed the final state symmetry factor accordingly.

The second one is from the soft singularities of the LDMEs in NRQCD, which we will give explicitly in the following.

\subsubsection{Soft counterterms of LDMEs}

The second term originates from the renormalisation of LDMEs and is analogous to the initial state collinear counterterm at hadron colliders or the final state collinear counterterm due to the presence of fragmentation functions~\cite{Frederix:2018nkq}. The soft counterterms of the LDMEs can be incorporated into either the real or the virtual matrix elements, depending on the preference. In this paper, we include them as part of the integrated soft subtraction terms for the real matrix element. We have the following perturbative corrections for the LDMEs~\cite{Petrelli:1997ge} in the $\overline{{\rm MS}}$ scheme, by defining $1/\bar{\epsilon}=1/\epsilon+\log{\left(4\pi\right)}-\gamma_E$,
\begin{eqnarray}
 \braket{{\mathcal O}^H_{\bigl.^{3}\hspace{-1mm}S^{[8]}_1}}(\mu)&=&\braket{{\mathcal O}^H_{\bigl.^{3}\hspace{-1mm}S^{[8]}_1}}+\frac{4\alpha_s}{3\pi m_{\Quark}m_{\Antiquark}}\left(\frac{1}{\bar{\epsilon}}+\log{\frac{\mu^2}{\mu_{{\rm NRQCD}}^2}}\right)\!\left[\frac{\CF}{2\NC}\!\sum_{J=0}^{2}{\braket{{\mathcal O}^H_{\bigl.^{3}\hspace{-1mm}P^{[1]}_J}}}\!+\!\BF\!\sum_{J=0}^{2}{\braket{{\mathcal O}^H_{\bigl.^{3}\hspace{-1mm}P^{[8]}_J}}}\right],\nonumber\\
\braket{{\mathcal O}^H_{\bigl.^{3}\hspace{-1mm}S^{[1]}_1}}(\mu)&=&\braket{{\mathcal O}^H_{\bigl.^{3}\hspace{-1mm}S^{[1]}_1}}+\frac{4\alpha_s}{3\pi m_{\Quark}m_{\Antiquark}}\left(\frac{1}{\bar{\epsilon}}+\log{\frac{\mu^2}{\mu_{{\rm NRQCD}}^2}}\right)\!\left[\sum_{J=0}^{2}{\braket{{\mathcal O}^H_{\bigl.^{3}\hspace{-1mm}P^{[8]}_J}}}\right],\nonumber\\
\braket{{\mathcal O}^H_{\bigl.^{1}\hspace{-1mm}S^{[8]}_0}}(\mu)&=&\braket{{\mathcal O}^H_{\bigl.^{1}\hspace{-1mm}S^{[8]}_0}}+\frac{4\alpha_s}{3\pi m_{\Quark}m_{\Antiquark}}\left(\frac{1}{\bar{\epsilon}}+\log{\frac{\mu^2}{\mu_{{\rm NRQCD}}^2}}\right)\!\left[\frac{\CF}{2\NC}\braket{{\mathcal O}^H_{\bigl.^{1}\hspace{-1mm}P^{[1]}_1}}+\BF\braket{{\mathcal O}^H_{\bigl.^{1}\hspace{-1mm}P^{[8]}_1}}\right],\nonumber\\
\braket{{\mathcal O}^H_{\bigl.^{1}\hspace{-1mm}S^{[1]}_0}}(\mu)&=&\braket{{\mathcal O}^H_{\bigl.^{1}\hspace{-1mm}S^{[1]}_0}}+\frac{4\alpha_s}{3\pi m_{\Quark}m_{\Antiquark}}\left(\frac{1}{\bar{\epsilon}}+\log{\frac{\mu^2}{\mu_{{\rm NRQCD}}^2}}\right)\!\left[\braket{{\mathcal O}^H_{\bigl.^{1}\hspace{-1mm}P^{[8]}_1}}\right],\label{eq:NRQCDCTs}
\end{eqnarray}
where $\gamma_E$ is the Euler-Mascheroni constant, $\mu_{{\rm NRQCD}}$ is the NRQCD cutoff scale and $\BF=(\NC^2-4)/(4\NC)=5/12$. If we use the Binoth Les Houches Accord convention by taking out the global prefactor $(4\pi)^\epsilon/\Gamma(1-\epsilon)$, $1/\bar{\epsilon}$ simply becomes $1/\epsilon$ in eq.\eqref{eq:NRQCDCTs}. Therefore, we have the additional integrated soft counterterms for P-waves:
\begin{itemize}
\item $\QQ[\bigl.^{1}\hspace{-1mm}P^{[1]}_1]$:
\begin{eqnarray}
d\hat{\sigma}^{(S,2)}(\dot{r})&=&\frac{\alpha_s}{2\pi}\phispnmo(\dot{r}^{\isubrmv})\frac{\JetsB}{\avg(\dot{r}^{\isubrmv})}\oavg(\dot{r}^{\isubrmv})\left[\frac{1}{2\NC}\ampsqnmoo(\dot{r}^{\isubrmv}_1)\right]\nonumber\\
&&\times\frac{8}{m_{\Quark}m_{\Antiquark}}\left(\frac{1}{\bar{\epsilon}}+\log{\frac{\mu^2}{\mu_{{\rm NRQCD}}^2}}\right)\,,
\end{eqnarray}
where 
\begin{eqnarray}
\dot{r}^{\isubrmv}&=&\left(\ident_1,\ldots,\remove{\ident}{0.2}_i,\ldots,\ident_j,\ldots,\ident_{\nlightR+\nheavy},\QQ[\bigl.^{1}\hspace{-1mm}P^{[1]}_1],\remove{\Antiquark}{0.3},
\ldots,\ident_{n+3}\right),\nonumber\\
\dot{r}_1^{\isubrmv}&=&\left(\ident_1,\ldots,\remove{\ident}{0.2}_i,\ldots,\ident_j,\ldots,\ident_{\nlightR+\nheavy},\QQ[\bigl.^{1}\hspace{-1mm}S^{[8]}_0],\remove{\Antiquark}{0.3},
\ldots,\ident_{n+3}\right),
\end{eqnarray}
and $\ident_i=g$.
\item $\QQ[\bigl.^{1}\hspace{-1mm}P^{[8]}_1]$:
\begin{eqnarray}
d\hat{\sigma}^{(S,2)}(\dot{r})&=&\frac{\alpha_s}{2\pi}\phispnmo(\dot{r}^{\isubrmv})\frac{\JetsB}{\avg(\dot{r}^{\isubrmv})}\oavg(\dot{r}^{\isubrmv})\left[\BF\ampsqnmoo(\dot{r}^{\isubrmv}_1)+\CF\ampsqnmoo(\dot{r}^{\isubrmv}_2)\right]\nonumber\\
&&\times\frac{8}{m_{\Quark}m_{\Antiquark}}\left(\frac{1}{\bar{\epsilon}}+\log{\frac{\mu^2}{\mu_{{\rm NRQCD}}^2}}\right),
\end{eqnarray}
where 
\begin{eqnarray}
\dot{r}^{\isubrmv}&=&\left(\ident_1,\ldots,\remove{\ident}{0.2}_i,\ldots,\ident_j,\ldots,\ident_{\nlightR+\nheavy},\QQ[\bigl.^{1}\hspace{-1mm}P^{[8]}_1],\remove{\Antiquark}{0.3},
\ldots,\ident_{n+3}\right),\nonumber\\
\dot{r}_1^{\isubrmv}&=&\left(\ident_1,\ldots,\remove{\ident}{0.2}_i,\ldots,\ident_j,\ldots,\ident_{\nlightR+\nheavy},\QQ[\bigl.^{1}\hspace{-1mm}S^{[8]}_0],\remove{\Antiquark}{0.3},
\ldots,\ident_{n+3}\right),\nonumber\\
\dot{r}_2^{\isubrmv}&=&\left(\ident_1,\ldots,\remove{\ident}{0.2}_i,\ldots,\ident_j,\ldots,\ident_{\nlightR+\nheavy},\QQ[\bigl.^{1}\hspace{-1mm}S^{[1]}_0],\remove{\Antiquark}{0.3},
\ldots,\ident_{n+3}\right).
\end{eqnarray}
\item $\QQ[\bigl.^{3}\hspace{-1mm}P^{[1]}_J]$:
\begin{eqnarray}
d\hat{\sigma}^{(S,2)}(\dot{r})&=&\frac{\alpha_s}{2\pi}\phispnmo(\dot{r}^{\isubrmv})\frac{\JetsB}{\avg(\dot{r}^{\isubrmv})}\oavg(\dot{r}^{\isubrmv})\left[\frac{2J+1}{2\NC}\ampsqnmoo(\dot{r}^{\isubrmv}_1)\right]\nonumber\\
&&\times\frac{8}{9 m_{\Quark}m_{\Antiquark}}\left(\frac{1}{\bar{\epsilon}}+\log{\frac{\mu^2}{\mu_{{\rm NRQCD}}^2}}\right)\,,
\end{eqnarray}
where 
\begin{eqnarray}
\dot{r}^{\isubrmv}&=&\left(\ident_1,\ldots,\remove{\ident}{0.2}_i,\ldots,\ident_j,\ldots,\ident_{\nlightR+\nheavy},\QQ[\bigl.^{3}\hspace{-1mm}P^{[1]}_J],\remove{\Antiquark}{0.3},
\ldots,\ident_{n+3}\right),\nonumber\\
\dot{r}_1^{\isubrmv}&=&\left(\ident_1,\ldots,\remove{\ident}{0.2}_i,\ldots,\ident_j,\ldots,\ident_{\nlightR+\nheavy},\QQ[\bigl.^{3}\hspace{-1mm}S^{[8]}_1],\remove{\Antiquark}{0.3},
\ldots,\ident_{n+3}\right).
\end{eqnarray}
\item $\QQ[\bigl.^{3}\hspace{-1mm}P^{[8]}_J]$:
\begin{eqnarray}
d\hat{\sigma}^{(S,2)}(\dot{r})&=&\frac{\alpha_s}{2\pi}\phispnmo(\dot{r}^{\isubrmv})\frac{\JetsB}{\avg(\dot{r}^{\isubrmv})}\oavg(\dot{r}^{\isubrmv})\left[\BF\ampsqnmoo(\dot{r}^{\isubrmv}_1)+\CF\ampsqnmoo(\dot{r}^{\isubrmv}_2)\right]\nonumber\\
&&\times\frac{8(2J+1)}{9 m_{\Quark}m_{\Antiquark}}\left(\frac{1}{\bar{\epsilon}}+\log{\frac{\mu^2}{\mu_{{\rm NRQCD}}^2}}\right),
\end{eqnarray}
where 
\begin{eqnarray}
\dot{r}^{\isubrmv}&=&\left(\ident_1,\ldots,\remove{\ident}{0.2}_i,\ldots,\ident_j,\ldots,\ident_{\nlightR+\nheavy},\QQ[\bigl.^{3}\hspace{-1mm}P^{[8]}_J],\remove{\Antiquark}{0.3},
\ldots,\ident_{n+3}\right),\nonumber\\
\dot{r}_1^{\isubrmv}&=&\left(\ident_1,\ldots,\remove{\ident}{0.2}_i,\ldots,\ident_j,\ldots,\ident_{\nlightR+\nheavy},\QQ[\bigl.^{3}\hspace{-1mm}S^{[8]}_1],\remove{\Antiquark}{0.3},
\ldots,\ident_{n+3}\right),\nonumber\\
\dot{r}_2^{\isubrmv}&=&\left(\ident_1,\ldots,\remove{\ident}{0.2}_i,\ldots,\ident_j,\ldots,\ident_{\nlightR+\nheavy},\QQ[\bigl.^{3}\hspace{-1mm}S^{[1]}_1],\remove{\Antiquark}{0.3},
\ldots,\ident_{n+3}\right).
\end{eqnarray}
\end{itemize}
It is clear that for S-waves $d\hat{\sigma}^{(S,2)}(\dot{r})=0$.

\subsubsection{Colour-singlet S-wave state}

The soft integrated partonic cross section for a colour-singlet S-wave quarkonium production is
\begin{eqnarray}
d\hat{\sigma}^{(S)}(\dot{r})&=&\frac{\alpha_s}{2\pi}\phispnmo(\dot{r}^{\isubrmv})\frac{\JetsB}{\avg(\dot{r}^{\isubrmv})}\oavg(\dot{r}^{\isubrmv})\sum_{k=\nini}^{\nlightB+\nheavy}{\sum_{l= k}^{\nlightB+\nheavy}{\bar{\mathcal{E}}(\{1,1\},\{k_k,k_l\})\ampsqnmoo_{kl}(\dot{r}^{\isubrmv})}}\,,\nonumber\\
\end{eqnarray}
where the eikonal integrals $\bar{\mathcal{E}}()$ can be found in app.~\ref{sec:eikonal}, and 
\begin{eqnarray}
\dot{r}^{\isubrmv}&=&\left(\ident_1,\ldots,\remove{\ident}{0.2}_i,\ldots,\ident_j,\ldots,\QQ[\bigl.^{2S+1}\hspace{-1mm}S^{[1]}_J],\remove{\Antiquark}{0.3},
\ldots,\ident_{n+3}\right)\label{eq:BornCSSwave}
\end{eqnarray}
with $\ident_i$ being a final state gluon. Now, the momenta $k_k, k_l$ are defined in the reduced Born process $\dot{r}^{\isubrmv}\in \allprocnmoo$ with $k_{\nlightB+\nheavy+1}=K$ (the quarkonium momentum).

\subsubsection{Colour-octet S-wave state}

The integrated soft counterterm for a colour-octet S-wave quarkonium is
\begin{eqnarray}
d\hat{\sigma}^{(S)}(\dot{r})&=&\frac{\alpha_s}{2\pi}\phispnmo(\dot{r}^{\isubrmv})\frac{\JetsB}{\avg(\dot{r}^{\isubrmv})}\oavg(\dot{r}^{\isubrmv})\sum_{k=\nini}^{\nlightB+\nheavy+1}{\sum_{l= k}^{\nlightB+\nheavy+1}{\bar{\mathcal{E}}(\{1,1\},\{k_k, k_l\})\ampsqnmoo_{kl}(\dot{r}^{\isubrmv})}}\,,\nonumber\\
\end{eqnarray}
where the eikonal integrals can be found in app.~\ref{sec:eikonal}, and 
\begin{eqnarray}
\dot{r}^{\isubrmv}&=&\left(\ident_1,\ldots,\remove{\ident}{0.2}_i,\ldots,\ident_j,\ldots,\QQ[\bigl.^{2S+1}\hspace{-1mm}S^{[8]}_J],\remove{\Antiquark}{0.3},
\ldots,\ident_{n+3}\right)\label{eq:BornCOSwave}
\end{eqnarray}
with $\ident_i$ being a gluon. The momenta $k_k, k_l$ are defined in the Born process $\dot{r}^{\isubrmv}\in \allprocnmoo$ with $k_{\nlightB+\nheavy+1}=K$ (the quarkonium momentum).

\subsubsection{Colour-singlet spin-singlet P-wave state}

For a colour-singlet spin-singlet P-wave state, the integrated soft counterterm is
\begin{eqnarray}
d\hat{\sigma}^{(S)}(\dot{r})&=&\frac{\alpha_s}{2\pi}\phispnmo(\dot{r}^{\isubrmv})\frac{\JetsB}{\avg(\dot{r}^{\isubrmv})}\oavg(\dot{r}^{\isubrmv})\Bigg[\sum_{k=\nini}^{\nlightB+\nheavy}{\sum_{l= k}^{\nlightB+\nheavy}{\bar{\mathcal{E}}(\{1,1\},\{k_k, k_l\})\ampsqnmoo_{kl}(\dot{r}^{\isubrmv})}}\nonumber\\
&&+\sum_{k=\nini}^{\nlightB+\nheavy}{\left(\bar{\mathcal{E}}(\{1,1\},\{k_k,K\})\frac{k_{k,\mu}}{K\cdot k_k}-\bar{\mathcal{E}}_{\mu}(\{1,2\},\{k_k,K\})\right)\ampsqnmoo_{k[18]}(\dot{r}^{\isubrmv},\dot{r}_1^{\isubrmv})^\mu}\nonumber\\
&&+\left(\frac{1}{2\NC}\frac{8}{m_{\Quark}m_{\Antiquark}}\left(\frac{1}{\bar{\epsilon}}+\log{\frac{\mu^2}{\mu^2_{\rm NRQCD}}}\right)-\frac{2\epsilon-2}{K^2}\bar{\mathcal{E}}(\{1,1\},\{K,K\})C_{{\rm eff}}(\QQ_{[18]})\right)\nonumber\\
&&\times\ampsqnmoo(\dot{r}_1^{\isubrmv})\Bigg]\,,
\end{eqnarray}
where the eikonal integrals can be found in app.~\ref{sec:eikonal}, and 
\begin{eqnarray}
\dot{r}^{\isubrmv}&=&\left(\ident_1,\ldots,\remove{\ident}{0.2}_i,\ldots,\ident_j,\ldots,\QQ[\bigl.^{1}\hspace{-1mm}P^{[1]}_1],\remove{\Antiquark}{0.3},
\ldots,\ident_{n+3}\right),\nonumber\\
\dot{r}_1^{\isubrmv}&=&\left(\ident_1,\ldots,\remove{\ident}{0.2}_i,\ldots,\ident_j,\ldots,\QQ[\bigl.^{1}\hspace{-1mm}S^{[8]}_0],\remove{\Antiquark}{0.3},
\ldots,\ident_{n+3}\right),\label{eq:BornCS1Pwave}
\end{eqnarray}
with $\ident_i=g$. Once again, we remind the readers that the momenta $k_k, k_l$ are defined in the reduced Born process $\dot{r}^{\isubrmv}\in \allprocnmoo$ with the quarkonium momentum $k_{\nlightB+\nheavy+1}=K$. Now, besides the usual scalar eikonal integrals, we also need to introduce the rank-$1$ tensor integrals $\bar{\mathcal{E}}_{\mu}(\{1,2\},\{k_k,K\})$.

\subsubsection{Colour-octet spin-singlet P-wave state}

The soft integrated counterterm for a colour-octet spin-singlet P-wave quarkonium is
\begin{eqnarray}
d\hat{\sigma}^{(S)}(\dot{r})&=&\frac{\alpha_s}{2\pi}\phispnmo(\dot{r}^{\isubrmv})\frac{\JetsB}{\avg(\dot{r}^{\isubrmv})}\oavg(\dot{r}^{\isubrmv})\left[\sum_{k=\nini}^{\nlightB+\nheavy+1}{\sum_{l= k}^{\nlightB+\nheavy+1}{\bar{\mathcal{E}}(\{1,1\},\{k_k, k_l\})\ampsqnmoo_{kl}(\dot{r}^{\isubrmv})}}\right.\nonumber\\
&&+\sum_{k=\nini}^{\nlightB+\nheavy+1}{\left(\bar{\mathcal{E}}(\{1,1\},\{k_k,K\})\frac{k_{k,\mu}}{K\cdot k_k}-\bar{\mathcal{E}}_{\mu}(\{1,2\},\{k_k,K\})\right)}\nonumber\\
&&\times\left(\ampsqnmoo_{k[88]}(\dot{r}^{\isubrmv},\dot{r}_1^{\isubrmv})^\mu+\ampsqnmoo_{k[81]}(\dot{r}^{\isubrmv},\dot{r}_2^{\isubrmv})^\mu\right)\nonumber\\
&&+\left(\BF\frac{8}{m_{\Quark}m_{\Antiquark}}\left(\frac{1}{\bar{\epsilon}}+\log{\frac{\mu^2}{\mu^2_{\rm NRQCD}}}\right)-\frac{2\epsilon-2}{K^2}\bar{\mathcal{E}}(\{1,1\},\{K,K\})C_{{\rm eff}}(\QQ_{[88]})\right)\nonumber\\
&&\times\ampsqnmoo(\dot{r}_1^{\isubrmv})\nonumber\\
&&+\left(\CF\frac{8}{m_{\Quark}m_{\Antiquark}}\left(\frac{1}{\bar{\epsilon}}+\log{\frac{\mu^2}{\mu^2_{\rm NRQCD}}}\right)-\frac{2\epsilon-2}{K^2}\bar{\mathcal{E}}(\{1,1\},\{K,K\})C_{{\rm eff}}(\QQ_{[81]})\right)\nonumber\\
&&\times\ampsqnmoo(\dot{r}_2^{\isubrmv})\Bigg]\,,
\end{eqnarray}
where the expressions of the eikonal integrals can be found in app.~\ref{sec:eikonal}, and 
\begin{eqnarray}
\dot{r}^{\isubrmv}&=&\left(\ident_1,\ldots,\remove{\ident}{0.2}_i,\ldots,\ident_j,\ldots,\QQ[\bigl.^{1}\hspace{-1mm}P^{[8]}_1],\remove{\Antiquark}{0.3},
\ldots,\ident_{n+3}\right),\nonumber\\
\dot{r}_1^{\isubrmv}&=&\left(\ident_1,\ldots,\remove{\ident}{0.2}_i,\ldots,\ident_j,\ldots,\QQ[\bigl.^{1}\hspace{-1mm}S^{[8]}_0],\remove{\Antiquark}{0.3},
\ldots,\ident_{n+3}\right),\nonumber\\
\dot{r}_2^{\isubrmv}&=&\left(\ident_1,\ldots,\remove{\ident}{0.2}_i,\ldots,\ident_j,\ldots,\QQ[\bigl.^{1}\hspace{-1mm}S^{[1]}_0],\remove{\Antiquark}{0.3},
\ldots,\ident_{n+3}\right),\label{eq:BornCO1Pwave}
\end{eqnarray}
and $\ident_i$ is a final state gluon.

\subsubsection{Colour-singlet spin-triplet P-wave state}

The integrated soft partonic cross section of a process with a colour-singlet spin-triplet P-wave quarkonium is
\begin{eqnarray}
d\hat{\sigma}^{(S)}(\dot{r})&=&\frac{\alpha_s}{2\pi}\phispnmo(\dot{r}^{\isubrmv})\frac{\JetsB}{\avg(\dot{r}^{\isubrmv})}\oavg(\dot{r}^{\isubrmv})\left[\sum_{k=\nini}^{\nlightB+\nheavy}{\sum_{l= k}^{\nlightB+\nheavy}{\bar{\mathcal{E}}(\{1,1\},\{k_k, k_l\})\ampsqnmoo_{kl}(\dot{r}^{\isubrmv})}}\right.\nonumber\\
&&+\sum_{k=\nini}^{\nlightB+\nheavy}{\left(\bar{\mathcal{E}}(\{1,1\},\{k_k,K\})\frac{k_{k,\mu}}{K\cdot k_k}-\bar{\mathcal{E}}_{\mu}(\{1,2\},\{k_k,K\})\right)\ampsqnmoo_{J,k[18]}(\dot{r}^{\isubrmv},\dot{r}_1^{\isubrmv})^\mu}\nonumber\\
&&-\left(\frac{g_{\mu\nu}}{K^2}\bar{\mathcal{E}}(\{1,1\},\{K,K\})+\bar{\mathcal{E}}_{\mu\nu}(\{2,2\},\{K,K\})\right)C_{{\rm eff}}(\QQ_{[18]})\ampsqnmoo_J(\dot{r}_1^{\isubrmv})^{\mu\nu}\nonumber\\
&&\left.+\frac{2J+1}{2\NC}\frac{8}{9m_{\Quark}m_{\Antiquark}}\left(\frac{1}{\bar{\epsilon}}+\log{\frac{\mu^2}{\mu^2_{\rm NRQCD}}}\right)\ampsqnmoo(\dot{r}_1^{\isubrmv})\right],
\end{eqnarray}
where the eikonal integrals can be found in app.~\ref{sec:eikonal}, and 
\begin{eqnarray}
\dot{r}^{\isubrmv}&=&\left(\ident_1,\ldots,\remove{\ident}{0.2}_i,\ldots,\ident_j,\ldots,\QQ[\bigl.^{3}\hspace{-1mm}P^{[1]}_J],\remove{\Antiquark}{0.3},
\ldots,\ident_{n+3}\right),\nonumber\\
\dot{r}_1^{\isubrmv}&=&\left(\ident_1,\ldots,\remove{\ident}{0.2}_i,\ldots,\ident_j,\ldots,\QQ[\bigl.^{3}\hspace{-1mm}S^{[8]}_1],\remove{\Antiquark}{0.3},
\ldots,\ident_{n+3}\right),\label{eq:BornCS3Pwave}
\end{eqnarray}
and $\ident_i$ is a final state gluon. In this case, we also need the rank-$2$ eikonal tensor integrals $\bar{\mathcal{E}}_{\mu\nu}(\{2,2\},\{K,K\})$.

\subsubsection{Colour-octet spin-triplet P-wave state}

Finally, the colour-octet spin-triplet P-wave quarkonium production has the following soft integrated counterterm
\begin{eqnarray}
d\hat{\sigma}^{(S)}(\dot{r})&=&\frac{\alpha_s}{2\pi}\phispnmo(\dot{r}^{\isubrmv})\frac{\JetsB}{\avg(\dot{r}^{\isubrmv})}\oavg(\dot{r}^{\isubrmv})\left[\sum_{k=\nini}^{\nlightB+\nheavy+1}{\sum_{l= k}^{\nlightB+\nheavy+1}{\bar{\mathcal{E}}(\{1,1\},\{k_k, k_l\})\ampsqnmoo_{kl}(\dot{r}^{\isubrmv})}}\right.\nonumber\\
&&+\sum_{k=\nini}^{\nlightB+\nheavy+1}{\left(\bar{\mathcal{E}}(\{1,1\},\{k_k,K\})\frac{k_{k,\mu}}{K\cdot k_k}-\bar{\mathcal{E}}_{\mu}(\{1,2\},\{k_k,K\})\right)}\nonumber\\
&&\times\left(\ampsqnmoo_{J,k[88]}(\dot{r}^{\isubrmv},\dot{r}_1^{\isubrmv})^\mu+\ampsqnmoo_{J,k[81]}(\dot{r}^{\isubrmv},\dot{r}_2^{\isubrmv})^\mu\right)\nonumber\\
&&-\left(\frac{g_{\mu\nu}}{K^2}\bar{\mathcal{E}}(\{1,1\},\{K,K\})+\bar{\mathcal{E}}_{\mu\nu}(\{2,2\},\{K,K\})\right)\nonumber\\
&&\times\left(C_{{\rm eff}}(\QQ_{[88]})\ampsqnmoo_J(\dot{r}_1^{\isubrmv})^{\mu\nu}+C_{{\rm eff}}(\QQ_{[81]})\ampsqnmoo_J(\dot{r}_2^{\isubrmv})^{\mu\nu}\right)\nonumber\\
&&\left.+\frac{8(2J+1)}{9m_{\Quark}m_{\Antiquark}}\!\left(\frac{1}{\bar{\epsilon}}+\log{\frac{\mu^2}{\mu^2_{\rm NRQCD}}}\right)\!\!\left(\BF\ampsqnmoo(\dot{r}_1^{\isubrmv})+\CF\ampsqnmoo(\dot{r}_2^{\isubrmv})\right)\right],
\end{eqnarray}
where the eikonal integrals can be found in app.~\ref{sec:eikonal}, and 
\begin{eqnarray}
\dot{r}^{\isubrmv}&=&\left(\ident_1,\ldots,\remove{\ident}{0.2}_i,\ldots,\ident_j,\ldots,\QQ[\bigl.^{3}\hspace{-1mm}P^{[8]}_J],\remove{\Antiquark}{0.3},
\ldots,\ident_{n+3}\right),\nonumber\\
\dot{r}_1^{\isubrmv}&=&\left(\ident_1,\ldots,\remove{\ident}{0.2}_i,\ldots,\ident_j,\ldots,\QQ[\bigl.^{3}\hspace{-1mm}S^{[8]}_1],\remove{\Antiquark}{0.3},
\ldots,\ident_{n+3}\right),\nonumber\\
\dot{r}_2^{\isubrmv}&=&\left(\ident_1,\ldots,\remove{\ident}{0.2}_i,\ldots,\ident_j,\ldots,\QQ[\bigl.^{3}\hspace{-1mm}S^{[1]}_1],\remove{\Antiquark}{0.3},
\ldots,\ident_{n+3}\right),\label{eq:BornCO3Pwave}
\end{eqnarray}
with $\ident_i$ being a final state gluon.

\subsection{Integrated collinear and soft-collinear counterterms}

As explained earlier, the integrated collinear and soft-collinear counterterms should be identical to the original case since quarkonia are massive and do not exhibit any collinear divergences. However, for the completeness, we still discuss them here and adjust to our notations with the risk that they are actually well understood.

Following the explanation of ref.~\cite{Frixione:1995ms}, it is straightforward to show that the sum of the integrated counterterms for collinear and soft-collinear emissions from an incoming initial state parton takes the form
\begin{eqnarray}
d\hat{\sigma}^{(in)}(\dot{r})&=&-\frac{\alpha_s}{2\pi}\frac{\JetsB}{\avg(\dot{r}^{\isubrmv})}\oavg(\dot{r}^{\isubrmv})\sum_{k=\nini}^{2}\int_0^{\ximax} d\xii\left(\frac{1}{\bar{\ep}}-\log\!\left(\frac{s \deltaI}{2\mu^2}\right)\right)\left[\left(\frac{1}{\xii}\right)_c-2\ep\left(\frac{\log(\xii)}{\xii}\right)_c\right]\nonumber\\
&&\times\xii P_{\ident_{k\oplus \bar{i}}\ident_k}^{<}(1-\xii,\ep)\ampsqo^{(n-1,0)}(\dot{r}^{k\oplus \bar{i},\isubrmv})\phispnmo(\dot{r}^{k\oplus \bar{i},\isubrmv})\,,
\label{eq:dsiginitial}
\end{eqnarray}
and that for emissions from an outgoing final state parton for a given underlying Born process $\dot{r}^{\isubrmv}$ the counterterms can be written as
\begin{eqnarray}
d\hat{\sigma}^{(out)}(\dot{r})&=&\frac{\alpha_s}{2\pi}\frac{\JetsB}{\avg(\dot{r}^{\isubrmv})}\oavg(\dot{r}^{\isubrmv})\sum_{k=3}^{\nlightB+2}\bigg\lbrace\frac{(4\pi)^\ep}{\Gamma(1-\ep)}
\left(\frac{\mu^2}{Q_{\rm ES}^2}\right)^\ep\frac{1}{\ep}\left[\gamma(\ident_k)+C(\ident_k)\log\!\left(\frac{\xicut^2s}{4E_k^2}\right)\right]\nonumber\\
&&+\left[\gamma^\prime(\ident_k)-\log\!\left(\dfrac{s\deltaO}{2Q_{\rm ES}^2}\right)\left(\gamma(\ident_k)-2C(\ident_k)\log\!\left(\dfrac{2E_k}{\xicut\sqrt{s}}\right)\right)\right.\nonumber\\
&&\left.+2C(\ident_k)\left(\log^2\!\left(\dfrac{2E_k}{\sqrt{s}}\right)-\log^2(\xicut)\right)-2\gamma(\ident_k)\log\!\left(\dfrac{2E_k}{\sqrt{s}}\right)\right]\bigg\rbrace\nonumber\\
&&\times\ampsqo^{(n-1,0)}(\dot{r}^{\isubrmv})\phispnmo(\dot{r}^{\isubrmv})\,,
\label{eq:dsigfinal}
\end{eqnarray}
where $Q_{\rm ES}$ is the Ellis-Sexton scale~\cite{Ellis:1985er}, the Casimir factors are
\begin{equation}
\begin{aligned}
C(\ident)&=\left\{\begin{array}{ll}
C_F, &~~{\rm if}~~\ident\in \irrep{3},\irrepbar{3}\\
C_A, &~~{\rm if}~~\ident\in \irrep{8}\\
\end{array}\right..
\end{aligned}
\end{equation}
The collinear anomalous dimensions are
\begin{equation}
\begin{aligned}
\gamma(\ident)&=\left\{\begin{array}{ll}
\frac{3}{2}C_F\,,\hfill&~~{\rm if}~~\ident=q,\bar{q}\\
\frac{11}{6}C_A-\frac{2}{3}T_Fn_f\,,&~~{\rm if}~~\ident=g\\
\end{array}\right.,
\end{aligned}
\end{equation}
with $T_F=\frac{1}{2}$ and $n_f$ being the number of massless quark flavours, and
\begin{equation}
\begin{aligned}
\gamma^\prime(\ident)&=\left\{\begin{array}{ll}
\left(\frac{13}{2}-\frac{2\pi^2}{3}\right)C_F\,,\hfill&~~{\rm if}~~\ident=q,\bar{q}\\
\left(\frac{67}{9}-\frac{2\pi^2}{3}\right)C_A-\frac{23}{9}T_Fn_f\,,&~~{\rm if}~~\ident=g\\
\end{array}\right..
\end{aligned}
\end{equation}
These integrated counterterms have to be combined with the counterterms stemming from the collinear renormalisation of PDFs which read
\begin{eqnarray}
d\hat{\sigma}^{(cnt)}(\dot{r})&=&\frac{\alpha_s}{2\pi}\frac{\JetsB}{\avg(\dot{r}^{\isubrmv})}\oavg(\dot{r}^{\isubrmv})\sum_{k=\nini}^{2}\int_{1-\ximax}^1 dz \left(\frac{1}{\bar{\ep}}P_{\ident_{k\oplus \bar{i}}\ident_k}(z,0)-K_{\ident_{k\oplus \bar{i}}\ident_k}(z)\right)\nonumber\\
&&\times\ampsqo^{(n-1,0)}(\dot{r}^{k\oplus \bar{i},\isubrmv})\phispnmo(\dot{r}^{k\oplus \bar{i},\isubrmv})
\end{eqnarray}
with the four dimensional regularised Altarelli-Parisi splitting kernels
\begin{equation}
    P_{ab}(z,0) = \frac{(1-z)P_{ab}^{<}(z,0)}{(1-z)_{+}} + \gamma(a)\delta_{ab}\,\delta(1-z)
\end{equation}
and the subtraction scheme dependent function $K_{ab}(z)$~\footnote{$K_{ab}(z)$ is trivially zero for $\overline{\rm MS}$ PDFs.}. A change of the integration variable, $z\to 1-\xii$, results in~\footnote{Note that $E_1=E_2=\sqrt{s}/2$ here.}
\begin{eqnarray}
d\hat{\sigma}^{(cnt)}(\dot{r})&=&\frac{\alpha_s}{2\pi}\frac{\JetsB}{\avg(\dot{r}^{\isubrmv})}\oavg(\dot{r}^{\isubrmv})\sum_{k=\nini}^{2}\left\lbrace\frac{1}{\bar{\ep}}\left[\gamma(\ident_k)+C(\ident_k)\log\!\left(\frac{\xicut^2s}{4 E_k^2}\right)\right]\ampsqo^{(n-1,0)}(\dot{r}^{\isubrmv})\phispnmo(\dot{r}^{\isubrmv})\right.\nonumber\\
&&+\int_0^{\ximax} d\xii \left[\frac{1}{\bar{\ep}}\left(\frac{1}{\xii}\right)_c\xii P_{\ident_{k\oplus \bar{i}}\ident_k}^{<}(1-\xii,0)-K_{\ident_{k\oplus \bar{i}}\ident_k}(1-\xii)\right]\ampsqo^{(n-1,0)}(\dot{r}^{k\oplus \bar{i},\isubrmv})\nonumber\\
&&\left.\times\phispnmo(\dot{r}^{k\oplus \bar{i},\isubrmv})\right\rbrace\,,
\label{eq:dsigpdf}
\end{eqnarray}
where we made use of the identity
\begin{equation}
    \delta(\xii)\xii P_{ab}^{<}(1-\xii,0) = 2C(a)\delta_{ab}\,\delta(\xii)\,.
\end{equation}
If we sum up eqs.(\ref{eq:dsiginitial}), (\ref{eq:dsigfinal}), and (\ref{eq:dsigpdf}), we obtain the sum of the integrated collinear and soft-collinear counterterms
\begin{eqnarray}
d\hat{\sigma}^{(C)}(\dot{r})&=&d\hat{\sigma}^{(in)}(\dot{r})+d\hat{\sigma}^{(out)}(\dot{r})+d\hat{\sigma}^{(cnt)}(\dot{r})\nonumber\\
&=&d\hat{\sigma}^{(C)}_{\rm sing.}(\dot{r})+d\hat{\sigma}^{(C,n-1)}_{\rm FIN}(\dot{r})+d\hat{\sigma}^{(C,n)}_{\rm FIN}(\dot{r})\,,
\end{eqnarray}
which we have decomposed into a singular term
\begin{eqnarray}
d\hat{\sigma}^{(C)}_{\rm sing.}(\dot{r})&=&\frac{\alpha_s}{2\pi}\phispnmo(\dot{r}^{\isubrmv})\frac{\JetsB}{\avg(\dot{r}^{\isubrmv})}\oavg(\dot{r}^{\isubrmv})\sum_{k=\nini}^{\nlightB+2}\frac{(4\pi)^\ep}{\Gamma(1-\ep)}
\left(\frac{\mu^2}{Q_{\rm ES}^2}\right)^\ep\frac{1}{\ep}\nonumber\\
&&\times\left[\gamma(\ident_k)+C(\ident_k)\log\!\left(\frac{\xicut^2s}{4E_k^2}\right)\right]\ampsqo^{(n-1,0)}(\dot{r}^{\isubrmv})\,,
\end{eqnarray}
a finite $(n-1)$-body contribution (equivalent to eqs.(4.5-4.6) in ref.~\cite{Frederix:2009yq})
\begin{eqnarray}
d\hat{\sigma}^{(C,n-1)}_{\rm FIN}(\dot{r})&=&\frac{\alpha_s}{2\pi}\phispnmo(\dot{r}^{\isubrmv})\frac{\JetsB}{\avg(\dot{r}^{\isubrmv})}\oavg(\dot{r}^{\isubrmv})\Bigg\lbrace-\log\!\left(\dfrac{\mu^2}{Q_{\rm ES}^2}\right)\sum_{k=\nini}^{2}\Big(\gamma(\ident_k)+2C(\ident_k)\log(\xicut)\Big)\nonumber\\
&&+\sum_{k= 3}^{\nlightB+2}\left[\gamma^\prime(\ident_k)-\log\!\left(\dfrac{s\deltaO}{2Q_{\rm ES}^2}\right)\left(\gamma(\ident_k)-2C(\ident_k)\log\!\left(\dfrac{2E_k}{\xicut\sqrt{s}}\right)\right)\right.\nonumber\\
&&\left.+2C(\ident_k)\left(\log^2\!\left(\dfrac{2E_k}{\sqrt{s}}\right)-\log^2(\xicut)\right)-2\gamma(\ident_k)\log\!\left(\dfrac{2E_k}{\sqrt{s}}\right)\right]\Bigg\rbrace\ampsqnmoo(\dot{r}^{\isubrmv})\,,\nonumber\\
\end{eqnarray}
and a finite contribution with a degenerated $n$-body phase space (equivalent to eqs.(4.40-4.42) in ref.~\cite{Frederix:2009yq}), where the angular dependence of the collinear emitted parton is integrated out, but its energy integration remains left,
\begin{eqnarray}
d\hat{\sigma}^{(C,n)}_{\rm FIN}(\dot{r})&=&\frac{\alpha_s}{2\pi}\frac{\JetsB}{\avg(\dot{r}^{\isubrmv})}\oavg(\dot{r}^{\isubrmv})\sum_{k=\nini}^{2}\int_0^{\ximax} d\xii\left\lbrace\left[\left(\frac{1}{\xii}\right)_c\log\!\left(\frac{s\deltaI}{2\mu^2}\right)+2\left(\frac{\log(\xii)}{\xii}\right)_c\right]\right.\nonumber\\
&&\left.\times\xii P_{\ident_{k\oplus \bar{i}}\ident_k}^{<}(1-\xii,0)-\left(\frac{1}{\xii}\right)_c\xii P_{\ident_{k\oplus \bar{i}}\ident_k}^{\,\prime<}(1-\xii,0)-K_{\ident_{k\oplus \bar{i}}\ident_k}(1-\xii)\right\rbrace\nonumber\\
&&\times\ampsqo^{(n-1,0)}(\dot{r}^{k\oplus \bar{i},\isubrmv})\phispnmo(\dot{r}^{k\oplus \bar{i},\isubrmv})\,.
\end{eqnarray}
$P_{ab}^{\,\prime<}(z,0)$ is the $\epsilon$ part of the $d=4-2\epsilon$ dimensional (unregularised) Altarelli-Parisi splitting kernels
\begin{eqnarray}
P_{ab}^{<}(z,\epsilon)&=&P_{ab}^{<}(z,0)+\epsilon P_{ab}^{\,\prime<}(z,0).
\end{eqnarray}

\section{Validations\label{sec:checks}}

\begin{figure}[htbp!]
\centering
\includegraphics[width=0.49\textwidth]{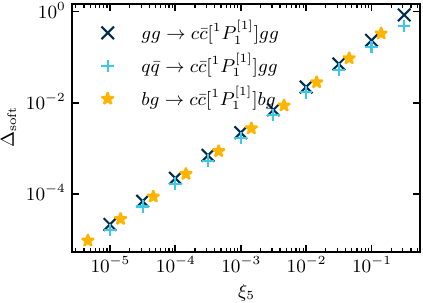}
\includegraphics[width=0.49\textwidth]{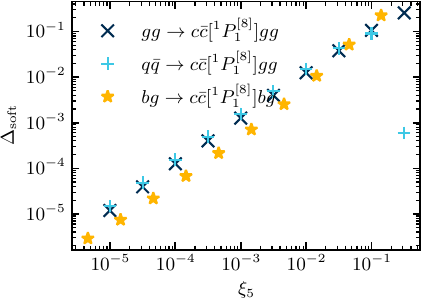}
\includegraphics[width=0.49\textwidth]{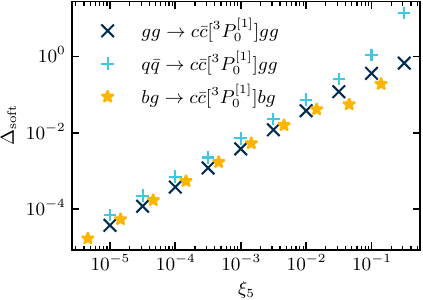}
\includegraphics[width=0.49\textwidth]{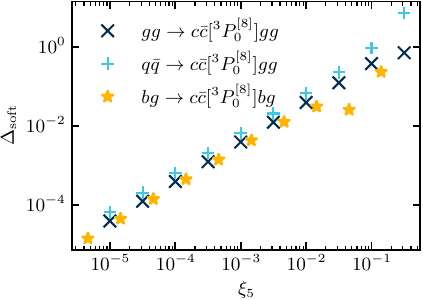}
\includegraphics[width=0.49\textwidth]{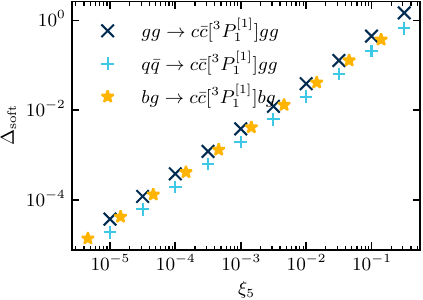}
\includegraphics[width=0.49\textwidth]{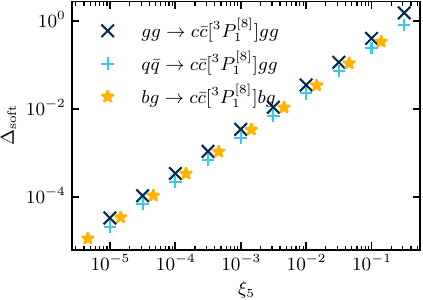}
\includegraphics[width=0.49\textwidth]{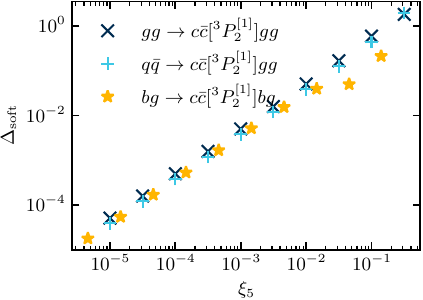}
\includegraphics[width=0.49\textwidth]{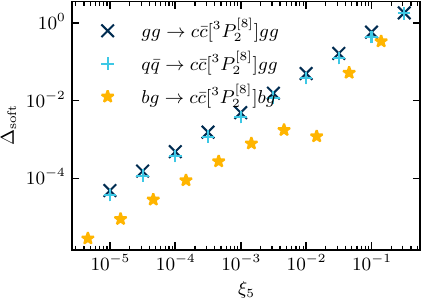}
\caption{Soft limit tests of real processes $gg\to c\bar{c}[n]gg$ (black cross), $q\bar{q}\to c\bar{c}[n]gg$ (blue plus) and $bg\to c\bar{c}[n]bg$ (orange star), where the Fock states $n$ include $8$ P-wave states.\label{fig:softchecks}}
\end{figure}

We present a few cross-checks of our new formalism derived in the last sections. Concerning the local FKS counterterms, we have verified the initial collinear and soft limits of all $2\to 1$ partonic Born processes $gg\to c\bar{c}[n]$ and $q\bar{q}\to c\bar{c}[n]$ with $n$ covering all S- and P-wave charmonium states, including both colour-singlet and colour-octet. One of the most non-trivial checks we performed involves the $2\to 2$ Born processes illustrated in fig.~\ref{fig:softchecks}. We consider the real processes $gg\to c\bar{c}[n]gg$, $q\bar{q}\to c\bar{c}[n]gg$ and $bg\to c\bar{c}[n]bg$, where both the charm quark $c$ and the bottom quark $b$ are massive, and the light quark $q$ is massless. The real matrix elements are numerically evaluated with \helaconia~\cite{Shao:2012iz,Shao:2015vga}, while the (reduced) Born and colour-linked Born matrix elements are computed analytically. The relative difference of the matrix elements is defined as
\begin{eqnarray}
\Delta_{\rm soft}&=&\abs{\dfrac{\ampsqnto(\dot{r})-\lim_{k_i\to 0}{\ampsqnto(\dot{r})}}{\ampsqnto(\dot{r})}}\propto\xii+\mathcal{O}(\xii^2),
\end{eqnarray}
where the expression of $\lim_{k_i\to 0}{\ampsqnto(\dot{r})}$ corresponds to the local soft counterterm of the real-emission counterpart $\ampsqnto(\dot{r})$ defined on the r.h.s of eq.\eqref{eq:softampsq0}. In fig.~\ref{fig:softchecks}, we demonstrate the soft gluon limit as $\xi_5\to 0$ for all $8$ P-wave Fock states. The relative differences $\Delta_{\rm soft}$ linearly vanish with $\xi_5$ (the rescaled final gluon energy defined in eq.\eqref{eq:xiidef}) asymptotically approaching zero. For S-wave states, similar soft limit tests were conducted, and the same behaviour was observed. These non-trivial checks confirm the correctness of our local soft counterterms defined in sect.~\ref{sec:softamp}. Additionally, we have performed initial and final collinear tests for some single quarkonium production processes.

In order to test the integrated counterterms, we have verified the universal IR poles of the one-loop virtual matrix elements outlined in app.~\ref{sec:IRpoles4virt}, stemming from the derived integrated counterterms due to the KLN theorem. These checks are carried out explicitly by computing the analytic UV-renormalised one-loop matrix elements for the processes $q\bar{q}\to c\bar{c}[n]g$ with all eight P-wave and four S-wave Fock states $c\bar{c}[n]$. The IR poles of the virtual matrix elements for these $12$ processes cancel perfectly with the formulas given in app.~\ref{sec:IRpoles4virt} at the analytic level. 

\begin{figure}[htbp!]
\centering
\includegraphics[width=\textwidth]{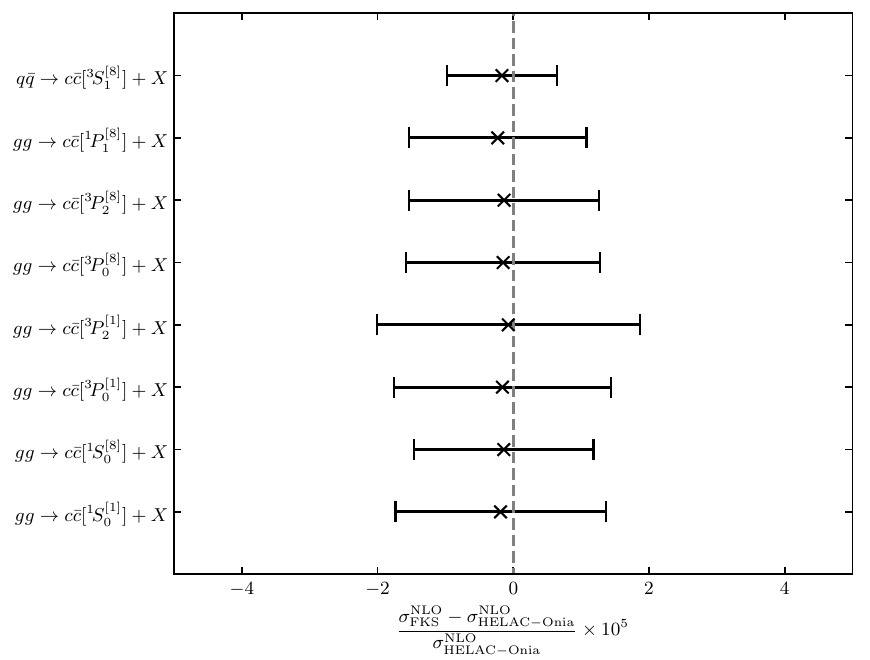}
\caption{Comparisons of inclusive NLO hadronic cross sections for $8$ selected $2\to 1$ processes computed using \helaconia\ and our FKS subtraction implementation. The crosses denote the differences between both results normalised to the central values of the cross sections calculated by \helaconia, while the error bars indicate numerical uncertainties arising from the Monte Carlo integration.\label{fig:totalxsec}}
\end{figure}

Finally, we consider the NLO QCD corrections to the $8$ physical hadronic cross sections of $2\to 1$ processes, as shown in fig.~\ref{fig:totalxsec}. The NLO partonic cross sections~\footnote{The analytic results for majority processes, excluding $gg\to c\bar{c}[{\bigl.^1\hspace{-1mm}P^{[8]}_1}]+X$, can also be found in ref.~\cite{Petrelli:1997ge}.} implemented in \helaconia\ are derived analytically, and additionally, we have computed the NLO cross sections within the FKS subtraction approach numerically. The relative differences between them are below $10^{-5}$, demonstrating perfect agreement within numerical errors.

\section{Summary\label{sec:summary}}

In this paper, we have generalised the FKS subtraction formalism to processes involving S- or P-wave quarkonium and elementary particles at NLO QCD in NRQCD factorisation. Our main new results are the local soft counterterms in sect.~\ref{sec:softampsquare} and the integrated soft counterterms in sect.~\ref{sec:integsoftCTs}. As a byproduct, we have also derived the universal IR poles of one-loop matrix elements in app.~\ref{sec:IRpoles4virt}. This serves as a cornerstone to achieve the NLO automation of cross section computations of quarkonium production processes~\footnote{Note that the factorisation-breaking effect due to colour transfer between a quarkonium and a heavy quark~\cite{Nayak:2007mb} only occurs in a particular phase-space corner, \ie\, the mass threshold region, where the relative velocity of the heavy quark and the quarkonium tends to zero. Therefore, it would be harmless for our general purpose unless we are probing that particular phase-space region, where, in any case, we need additional theoretical care.}. Our next steps would be to implement the new formulas in the \mgamc\ framework, and to generalise the formulation presented here to processes involving more-than-one quarkonia. The latter case, however, involves additional subtleties, such as the breakdown of NRQCD factorisation for processes with two P-wave bound states elucidated in ref.~\cite{He:2018hwb}.

\begin{acknowledgments}
We thank Fabio Maltoni and Michelangelo Mangano for reading the manuscript and for the comments. 
This work is supported by the grants from the ERC (grant 101041109 `BOSON'), the French ANR (grant ANR-20-CE31-0015 `PrecisOnium'), and the French LIA FCPPN. Views and opinions expressed are however those of the authors only and do not necessarily reflect those of the European Union or the European Research Council Executive Agency. Neither the European Union nor the granting authority can be held responsible for them.
\end{acknowledgments}

\appendix

\section{Eikonal tensor integrals\label{sec:eikonal}}

From the integrated soft counterterms, we need to solve the general eikonal tensor integral
\begin{eqnarray}
\bar{\mathcal{E}}^{\alpha_1\ldots \alpha_{n_1-1} \beta_1\ldots \beta_{n_2-1}}(\{n_1,n_2\},\{k_k,k_l\})&=&-\frac{\xi_{{\rm cut}}^{-2\epsilon}}{2\epsilon}\frac{2^{2\epsilon}}{(2\pi)^{1-2\epsilon}}\left(\frac{s}{\mu^2}\right)^{-\epsilon}\left(k_i^0\right)^2k_k\cdot k_l\nonumber\\
&&\times\underbrace{\int{d\Omega_i \frac{k_i^{\alpha_1}\ldots k_i^{\alpha_{n_1-1}}k_i^{\beta_1}\ldots k_i^{\beta_{n_2-1}}}{(k_k\cdot k_i)^{n_1} (k_l\cdot k_i)^{n_2}}}}_{\equiv I^{\alpha_1\ldots \alpha_{n_1-1} \beta_1\ldots \beta_{n_2-1}}(\{n_1,n_2\},\{k_k,k_l\})},\quad n_1,n_2 \geq 1,\nonumber\\
\end{eqnarray} 
where we have $m_k^2=k_k^2$ and $m_l^2=k_l^2$. We have followed the conventions in ref.~\cite{Frederix:2009yq}. The energy of the parton $i$ and the  measure (in $d-1=3-2\epsilon$ dimensions) over its angular variables $d\Omega_i$ ({\it cf.} eq.\eqref{eq:solidanglemeasure}) are defined in the center of mass frame of the colliding partons. We can rewrite the eikonal integral into the phase-space measure as
\begin{eqnarray}
\bar{\mathcal{E}}^{\alpha_1\ldots \alpha_{n_1-1} \beta_1\ldots \beta_{n_2-1}}(\{n_1,n_2\},\{k_k,k_l\})&=&8\pi^2\mu^{2\epsilon}k_k\cdot k_l\nonumber\\
&&\times\int{\frac{d^{3-2\epsilon}{\bold k}_i}{(2\pi)^{3-2\epsilon}2k_i^0} \frac{k_i^{\alpha_1}\ldots k_i^{\alpha_{n_1-1}}k_i^{\beta_1}\ldots k_i^{\beta_{n_2-1}}}{(k_k\cdot k_i)^{n_1} (k_l\cdot k_i)^{n_2}}}\nonumber\\
&&{\times\Theta(\xi_{{\rm cut}}-\xi_i)}\,.
\end{eqnarray} 
The introduction of $\Theta(\xi_{{\rm cut}}-\xi_i)$ spoils the Lorentz covariance.

In order to solve these tensor integrals, we are based on the observation that~\footnote{Note that in the case $l=k$ the derivative acts on both momenta.}
\begin{eqnarray}
&&\frac{\partial }{\partial k_{k, \alpha_{n_1}}}I^{\alpha_1\ldots \alpha_{n_1-1} \beta_1\ldots \beta_{n_2-1}}(\{n_1,n_2\},\{k_k,k_l\})\notag\\
&=&-n_1I^{\alpha_1\ldots \alpha_{n_1-1} \alpha_{n_1} \beta_1\ldots \beta_{n_2-1}}(\{n_1+1,n_2\},\{k_k,k_l\})\,,\nonumber\\
&&\frac{\partial }{\partial k_{l, \beta_{n_2}}}I^{\alpha_1\ldots \alpha_{n_1-1} \beta_1\ldots \beta_{n_2-1}}(\{n_1,n_2\},\{k_k,k_l\})\notag\\
&=&-n_2 I^{\alpha_1\ldots \alpha_{n_1-1} \beta_1\ldots \beta_{n_2-1}\beta_{n_2}}(\{n_1,n_2+1\},\{k_k,k_l\})\,.
\end{eqnarray}
Then, we can derive the tensor integrals from the scalar integral via
\begin{eqnarray}
I^{\alpha_1}(\{2,1\},\{k_k,k_l\})&=&-\frac{\partial}{\partial k_{k, \alpha_1}}I(\{1,1\},\{k_k,k_l\}),\nonumber\\
I^{\beta_1}(\{1,2\},\{k_k,k_l\})&=&-\frac{\partial}{\partial k_{l, \beta_1}}I(\{1,1\},\{k_k,k_l\}),\nonumber\\
I^{\alpha_1\beta_1}(\{2,2\},\{k_k,k_l\})&=&\frac{\partial^2}{\partial k_{k,\alpha_1}\partial k_{l, \beta_1}}I(\{1,1\},\{k_k,k_l\}).
\end{eqnarray}
The analytic expressions for $I(\{1,1\},\{k_k,k_l\})$ are known in the literature and are given in appendix A of ref.~\cite{Frederix:2009yq}. 

In the following, we will present the concrete analytic expressions of the eikonal tensor integrals that we need. In general, it would be convenient to further split the tensor integrals into the pole and finite parts:
\begin{eqnarray}
\bar{\mathcal{E}}^{\alpha_1\ldots \alpha_{n_1-1} \beta_1\ldots \beta_{n_2-1}}(\{n_1,n_2\},\{k_k,k_l\})&=&\hat{\mathcal{E}}^{\alpha_1\ldots \alpha_{n_1-1} \beta_1\ldots \beta_{n_2-1}}(\{n_1,n_2\},\{k_k,k_l\}) \nonumber\\
&&+\mathcal{E}^{\alpha_1\ldots \alpha_{n_1-1} \beta_1\ldots \beta_{n_2-1}}(\{n_1,n_2\},\{k_k,k_l\}),
\end{eqnarray}
where $\hat{\mathcal{E}}$ and $\mathcal{E}$ represent the IR poles and the finite component of $\bar{\mathcal{E}}$, respectively. We discuss their expressions in four categories.

\subsection{Two-massless case}

We first consider the case of two massless external legs ($m_k=m_l=0$). Since neither $\ident_k$ nor $\ident_l$ can be a quarkonium, we only need to consider the scalar integral ($n_1=n_2=1$), which has been known (\eg\ eqs.(A.5-A.6) in ref.~\cite{Frederix:2009yq}). In this case, if $l=k$, the tensor integrals are zero. If $l\neq k$, for the completeness, the expressions are
\begin{eqnarray}
\hat{\mathcal{E}}(\{1,1\},\{k_k,k_l\})&=&\frac{(4\pi)^\epsilon}{\Gamma(1-\epsilon)}
\left(\frac{\mu^2}{Q_{\rm ES}^2}\right)^\epsilon\left[
\frac{1}{\epsilon^2}-\frac{1}{\epsilon}
\left(\log\!\left(\frac{2 k_k \mydot k_l}{Q_{\rm ES}^2}\right)-
\log\!\left(\frac{4E_kE_l}{\xicut^2 s}\right)\right)\right]\,,~~~~~
\\
\mathcal{E}(\{1,1\},\{k_k,k_l\})&=& 
\dfrac{1}{2}\log^2\!\left(\frac{\xicut^2 s}{Q_{\rm ES}^2}\right)+
\log\!\left(\frac{\xicut^2 s}{Q_{\rm ES}^2}\right)\log\!\left(\frac{k_k \mydot k_l}{2E_kE_l}\right)
-{\rm Li}_2\!\left(\frac{k_k \mydot k_l}{2E_kE_l}\right)
\nonumber\\*&&+
\dfrac{1}{2}\log^2\!\left(\frac{k_k \mydot k_l}{2E_kE_l}\right)
-\log\!\left(1-\frac{k_k \mydot k_l}{2E_kE_l}\right)\log\!\left(\frac{k_k \mydot k_l}{2E_kE_l}\right)\,,
\end{eqnarray}
where ${\rm Li}_2()$ is the dilogarithm.

\subsection{One-massive-one-massless case}

The second case we consider involves one-massive and one-massless external leg ($m_k=0$ and $m_l\neq 0$). The scalar case has been given in eqs.(A.7-A.8) in ref.~\cite{Frederix:2009yq}:
\begin{eqnarray}
\hat{\mathcal{E}}(\{1,1\},\{k_k,k_l\})&=&\frac{(4\pi)^\epsilon}{\Gamma(1-\epsilon)}
\left(\frac{\mu^2}{Q_{\rm ES}^2}\right)^\epsilon\left[
\frac{1}{2\epsilon^2}-\frac{1}{\epsilon}
\left(\log\!\left(\frac{2k_k \mydot k_l}{Q_{\rm ES}^2}\right)-
\dfrac{1}{2}\log\!\left(\frac{4m_l^2E_k^2}{\xicut^2 s Q_{\rm ES}^2}\right)\right)\right]\,,\nonumber\\
\\
\mathcal{E}(\{1,1\},\{k_k,k_l\})&=& 
\log\!\left(\xicut\right)\left(\log\!\left(\frac{\xicut s}{Q_{\rm ES}^2}\right)+2\log\!\left(\frac{k_k \mydot k_l}{m_lE_k}\right)\right)
-\frac{\pi^2}{12}+\dfrac{1}{4}\log^2\!\left(\frac{s}{Q_{\rm ES}^2}\right)
\nonumber\\*&&-
\dfrac{1}{4}\log^2\!\left(\frac{1+\beta_l}{1-\beta_l}\right)
+\dfrac{1}{2}\log^2\!\left(\frac{k_k \mydot k_l}{(1-\beta_l)E_k E_l}\right)
+\log\!\left(\frac{s}{Q_{\rm ES}^2}\right)\log\!\left(\frac{k_k \mydot k_l}{m_lE_k}\right)
\nonumber\\*&&-
{\rm Li}_2\!\left(1-\frac{(1+\beta_l)E_kE_l}{k_k \mydot k_l}\right)
+{\rm Li}_2\!\left(1-\frac{k_k \mydot k_l}{(1-\beta_l)E_kE_l}\right)\,,
\end{eqnarray}
where 
\begin{equation}
\beta_l=\sqrt{1-\frac{m_l^2}{E_l^2}}.
\end{equation}
The IR pole part of the rank-1 tensor integral needed in the P-wave soft integrated counterterms is
\begin{eqnarray}
&&\hat{\mathcal{E}}^{\mu}(\{1,2\},\{k_k,k_l\}) \nonumber\\
&=&\dfrac{(4\pi)^\epsilon}{\Gamma(1-\epsilon)}\left(\dfrac{\mu^2}{Q_{\rm ES}^2}\right)^\epsilon\!\left\lbrace \dfrac{1}{2\epsilon^2} \dfrac{k_k^{\mu }}{k_k \mydot k_l}  + \dfrac{1}{\epsilon} \left[\dfrac{k_k^{\mu }}{k_k \mydot k_l}\! \left(1-\log\!\left(\dfrac{k_k \mydot k_l}{E_k m_l}\right)-\dfrac{1}{2}\log\!\left(\dfrac{\xicut^2 s}{Q_{\rm ES}^2}\right)\right)-\dfrac{k_l^{\mu}}{m_l^2} \right]\right\rbrace\,,\nonumber\\
\end{eqnarray}
while its finite part can be decomposed into 3 tensor structures
\begin{equation}
\mathcal{E}^{\mu}(\{1,2\},\{k_k,k_l\})= \sum_{i=1}^3{T_{i,12}^{(0,m_l),\mu}},
\end{equation}
where
\begin{eqnarray}
T_{1,12}^{(0,m_l),\mu}&=& 
\dfrac{k_k^{\mu}}{k_k \mydot k_l}\left\lbrace-\dfrac{\pi ^2}{12}-\log\! \left(\dfrac{\xicut^2 s}{Q_{\rm ES}^2}\right)+\dfrac{1}{4} \log^2\!\left(\dfrac{\xicut^2 s}{Q_{\rm ES}^2}\right)+\log\! \left(\dfrac{k_k \mydot k_l}{E_k m_l}\right)\log\! \left(\dfrac{\xicut^2 s}{Q_{\rm ES}^2}\right)\right.\nonumber\\
&&-\log\! \left(\dfrac{k_k \mydot k_l}{E_k E_l \left(1-\beta_l\right)}\right)+\dfrac{1}{2} \log\! ^2\left(\dfrac{k_k \mydot k_l}{E_k E_l \left(1-\beta_l\right)}\right)-\dfrac{1}{4} \log^2\!\left(\dfrac{1+\beta_l}{1-\beta_l}\right)\nonumber\\
&&+\mathrm{Li}_2\!\left(1-\dfrac{k_k \mydot k_l}{E_k E_l \left(1-\beta _l\right)}\right)-\mathrm{Li}_2\!\left(1-\dfrac{E_k E_l \left(1+\beta_l\right)}{k_k \mydot k_l}\right)\nonumber\\
&&+\dfrac{k_k \mydot k_l}{k_k \mydot k_l-E_k E_l(1-\beta_l)}\log\! \left(\dfrac{k_k \mydot k_l}{E_k E_l \left(1-\beta_l\right)}\right)+ \dfrac{E_k E_l(1+\beta_l)}{k_k \mydot k_l-E_k E_l(1+\beta_l)}\nonumber\\
&&\left.\times\log\! \left(\dfrac{k_k \mydot k_l}{E_k E_l \left(1+\beta_l\right)}\right)\right\rbrace,\nonumber
\end{eqnarray}
\begin{eqnarray}
T_{2,12}^{(0,m_l),\mu} &=& \dfrac{k_l^{\mu}}{m_l^2}\left\lbrace \log\!\left(\dfrac{\xicut^2 s}{Q_{\rm ES}^2}\right) + \dfrac{m_l^2}{\beta_l E_l^2}\left[-\dfrac{1}{1-\beta_l^2}\log\!\left(\dfrac{1+\beta_l}{1-\beta_l}\right) -\dfrac{E_k E_l}{k_k \mydot k_l - E_k E_l (1-\beta_l)} \right.\right.\nonumber\\
&&\left.\left.\times\log\!\left(\dfrac{k_k \mydot k_l}{E_k E_l(1-\beta_l)}\right)+\dfrac{E_kE_l}{k_k \mydot k_l - E_k E_l (1+\beta_l)}\log\!\left(\dfrac{k_k \mydot k_l}{E_k E_l(1+\beta_l)}\right)\right] \right\rbrace,\nonumber\\
T_{3,12}^{(0,m_l),\mu}&=&
\dfrac{\delta^{\mu0}}{k_k \mydot k_l-E_k E_l \left(1-\beta_l\right)}\left\lbrace E_k \left\lbrack\log\! \left(\dfrac{E_k E_l \left(1+\beta _l\right)}{k_k \mydot k_l}\right)-\log\! \left(\dfrac{k_k \mydot k_l}{E_k E_l \left(1-\beta _l\right)}\right)\right\rbrack\right.\nonumber\\
&&+\dfrac{k_k \mydot k_l \, m_l^2}{E_l^3 \beta _l \left(1-\beta _l^2\right)}\log\! \left(\dfrac{1+\beta _l}{1-\beta _l}\right) +\dfrac{E_k \left(E_l^2 \beta _l \left(1+\beta _l\right)+m_l^2\right)}{E_l^2 \left(k_k \mydot k_l-E_k E_l \left(1+\beta _l\right)\right)}\nonumber\\
&&\left.\times\left\lbrack E_k E_l\! \left(\log\! \left(\dfrac{E_k E_l \left(1+\beta _l\right)}{k_k \mydot k_l}\right)\!-\!\log\! \left(\dfrac{k_k \mydot k_l}{E_k E_l \left(1-\beta _l\right)}\right)\right)+\dfrac{k_k \mydot k_l}{1+\beta _l}\log\! \left(\dfrac{1+\beta _l}{1-\beta _l}\right)\right\rbrack\right\rbrace.\nonumber\\
\end{eqnarray}

\subsection{Massive self-eikonal case}

In the case of the massive self-eikonal integrals, we have $l=k$ and $m_k=m_l\neq 0$. The scalar integral corresponds to eqs.(A.9-A.10) in ref.~\cite{Frederix:2009yq}: 
\begin{eqnarray}
\hat{\mathcal{E}}(\{1,1\},\{k_k,k_k\})&=&\frac{(4\pi)^\epsilon}{\Gamma(1-\epsilon)}
\left(\frac{\mu^2}{Q_{\rm ES}^2}\right)^\epsilon\left(-\frac{1}{\epsilon}\right)\,,\\
\mathcal{E}(\{1,1\},\{k_k,k_k\})&=& 
\log\!\left(\frac{\xicut^2 s}{Q_{\rm ES}^2}\right)
-\frac{1}{\beta_k}\log\!\left(\frac{1+\beta_k}{1-\beta_k}\right)\,.
\end{eqnarray}
For the rank-1 tensor integral, the poles are
\begin{equation}
\hat{\mathcal{E}}^{\mu}(\{1,2\},\{k_k,k_k\})= \dfrac{(4\pi)^\epsilon}{\Gamma(1-\epsilon)}\left(\dfrac{\mu^2}{Q_{\rm ES}^2}\right)^\epsilon\dfrac{k_k^{\mu }}{m_k^2} \left(-\dfrac{1}{\epsilon} \right),
\end{equation}
and the finite term is
\begin{eqnarray}
\mathcal{E}^{\mu}(\{1,2\},\{k_k,k_k\})&=&\dfrac{k_k^{\mu }}{2\beta_k^3 m_k^2} \left\lbrack \log\!\left(\dfrac{1+\beta_k}{1-\beta_k}\right) -2 \beta_k -3\log\!\left(\dfrac{1+\beta_k}{1-\beta_k}\right)\beta_k^2 + 2 \log\!\left(\dfrac{\xicut^2 s}{Q_{\rm ES}^2}\right)\beta_k^3 \right\rbrack\nonumber\\
&& + \dfrac{\delta^{\mu0}}{2\beta_k^3 E_k}\left\lbrack-\log\!\left(\dfrac{1+\beta_k}{1-\beta_k}\right)+2 \beta_k+\log\!\left(\dfrac{1+\beta_k}{1-\beta_k}\right)\beta_k^2\right\rbrack.
\end{eqnarray}
The pole of the rank-2 integral is
\begin{equation}
\hat{\mathcal{E}}^{\mu\nu}(\{2,2\},\{k_k,k_k\})= \dfrac{(4\pi)^\epsilon}{\Gamma(1-\epsilon)}\left(\dfrac{\mu^2}{Q_{\rm ES}^2}\right)^\epsilon\dfrac{1}{3 m_k^2} \left\lbrack -\dfrac{1}{\epsilon} \left(-g^{\mu\nu}+\dfrac{4 k_k^{\mu}k_k^{\nu}}{m_k^2}\right)\right\rbrack.
\end{equation}
The finite piece is
\begin{equation}
\mathcal{E}^{\mu\nu}(\{2,2\},\{k_k,k_k\})= \sum_{i=1}^4{T^{(m_k,m_k),\mu\nu}_{i,22}},
\end{equation}
where
\begin{eqnarray}
T_{1,22}^{(m_k,m_k),\mu\nu}&=&\dfrac{\delta^{\mu0}\delta^{\nu0}}{6\beta_k^5 m_k^2}(1-\beta_k^2)\left\lbrack - 3\log\!\left(\dfrac{1+\beta_k}{1-\beta_k}\right) + 6\beta_k + 3 \log\!\left(\dfrac{1+\beta_k}{1-\beta_k}\right)\beta_k^2 - 4 \beta_k^3 \right\rbrack,\nonumber\\
T_{2,22}^{(m_k,m_k),\mu\nu} &=& \dfrac{g^{\mu\nu}}{6\beta_k^3 m_k^2}\left\lbrack -\log\!\left(\dfrac{1+\beta_k}{1-\beta_k}\right) + 2\beta_k + 3\log\!\left(\dfrac{1+\beta_k}{1-\beta_k}\right)\beta_k^2 - 2\log\!\left(\dfrac{\xicut^2 s}{Q^2_{\rm ES}}\right)\beta_k^3 \right\rbrack,\nonumber\\
T_{3,22}^{(m_k,m_k),\mu\nu}&=&\dfrac{\delta^{\mu0}k_k^{\nu}+k_k^{\mu}			\delta^{\nu0}}{6\beta_k^5 E_k m_k^2}\left\lbrack 3\log\!\left(\dfrac{1+\beta_k}{1-\beta_k}\right) - 6\beta_k - 6\log\!\left(\dfrac{1+\beta_k}{1-\beta_k}\right)\beta_k^2 \right.\nonumber\\
&&\left.+ 10\beta_k^3 + 3\log\!\left(\dfrac{1+\beta_k}{1-\beta_k}\right)\beta_k^4 \right\rbrack,\nonumber\\
T_{4,22}^{(m_k,m_k),\mu\nu}&=&\dfrac{k_k^{\mu}k_k^{\nu}}{6\beta_k^5 m_k^4}\left\lbrack -3\log\!\left(\dfrac{1+\beta_k}{1-\beta_k}\right) + 6\beta_k + 10\log\!\left(\dfrac{1+\beta_k}{1-\beta_k}\right)\beta_k^2 \right.\nonumber\\
&&\left.- 18\beta_k^3 - 15\log\!\left(\dfrac{1+\beta_k}{1-\beta_k}\right)\beta_k^4 + 8\log\!\left(\dfrac{\xicut^2 s}{Q_{\rm ES}^2}\right)\beta_k^5 \right\rbrack.
\end{eqnarray}

\subsection{Two-massive case}
Finally, let us consider the most complicated case with $l\neq k$, $m_k\neq0$ and $m_l\neq 0$. The scalar integral case can be referred to eqs.(A.11-A.12) in ref.~\cite{Frederix:2009yq}. Its expression is
\begin{eqnarray}
\hat{\mathcal{E}}(\{1,1\},\{k_k,k_l\})&=&\frac{(4\pi)^\epsilon}{\Gamma(1-\epsilon)}
\left(\frac{\mu^2}{Q_{\rm ES}^2}\right)^\epsilon\left(
-\frac{1}{2\epsilon}\frac{1}{v_{kl}}\log\!\left(\frac{1+v_{kl}}{1-v_{kl}}\right)\right)\,,
\\
\mathcal{E}(\{1,1\},\{k_k,k_l\})&=& 
\frac{1}{2v_{kl}}\log\!\left(\frac{1+v_{kl}}{1-v_{kl}}\right)
\log\!\left(\frac{\xicut^2 s}{Q_{\rm ES}^2}\right)
\nonumber\\*&&+
\frac{(1+v_{kl})(k_k \mydot k_l)^2}{2m_k^2}
\left({\rm J}^{(A)}\left(\alpha_{kl} E_k,\alpha_{kl} E_k\beta_k\right)
-{\rm J}^{(A)}\left(E_l,E_l\beta_l\right)\right)\,,\nonumber\\
\end{eqnarray}
where the introduced function is
\begin{eqnarray}
{\rm J}^{(A)}\left(x,y\right)&=&\frac{1}{2\lambda\nu}\left\lbrack\log^2\!\left(\frac{x-y}{x+y}\right)+4\mathrm{Li}_2\left(1-\frac{x+y}{\nu}\right)+4\mathrm{Li}_2\left(1-\frac{x-y}{\nu}\right)\right\rbrack
\end{eqnarray}
and
\begin{eqnarray}
v_{kl}&=&\sqrt{1-\left(\dfrac{m_km_l}{k_k\mydot k_l}\right)^2},\quad \alpha_{kl}=\dfrac{1+v_{kl}}{m_k^2}k_k\mydot k_l,\nonumber\\
\lambda&=&\alpha_{kl}E_k-E_l,\quad \nu=\dfrac{\alpha_{kl}^2m_k^2-m_l^2}{2\lambda}.
\end{eqnarray}

For the two-massive rank-1 eikonal tensor integral, its pole is
\begin{eqnarray}
\hat{\mathcal{E}}^{\mu}(\{1,2\},\{k_k,k_l\}) &=& \dfrac{(4\pi)^\epsilon}{\Gamma(1-\epsilon)}\left(\dfrac{\mu^2}{Q_{\rm ES}^2}\right)^\epsilon\dfrac{1}{2 \epsilon}\dfrac{m_k^2}{(k_k \mydot k_l)^2 v_{kl}^3} \left\lbrace -k_k^{\mu}\left\lbrack \dfrac{k_k \mydot k_l}{m_k^2} \log\!\left(\dfrac{1+v_{kl}}{1-v_{kl}}\right)v_{kl}^2 \right.\right.\nonumber\\
&&\left.\left. + \dfrac{m_l^2}{k_k \mydot k_l}\left(\log\!\left(\dfrac{1+v_{kl}}{1-v_{kl}}\right) - \dfrac{2v_{kl}}{1-v_{kl}^2}  \right)\right\rbrack + k_l^{\mu} \left(\log\!\left(\dfrac{1+v_{kl}}{1-v_{kl}}\right) - \dfrac{2v_{kl}}{1-v_{kl}^2} \right) \right\rbrace.\nonumber\\
\end{eqnarray}
The finite term can be decomposed into $8$ tensorial components
\begin{equation}
\mathcal{E}^{\mu}(\{1,2\},\{k_k,k_l\})= \sum_{i=1}^8{T_{i,12}^{(m_k,m_l),\mu}},
\end{equation}
where
\begin{eqnarray}
T_{1,12}^{(m_k,m_l),\mu} &=& k_k^{\mu} \left\lbrack \dfrac{1}{2(k_k \mydot k_l)v_{kl}}\log\!\left(\dfrac{\xicut^2 s}{Q_{\rm ES}^2}\right) \log\!\left(\dfrac{1+v_{kl}}{1-v_{kl}}\right) \right.\nonumber\\
&&\left. - \dfrac{(1+v_{kl})(k_k \mydot k_l)}{2 m_k^2} \left( \mathrm{J}^{(A)}(\alpha_{kl}E_k,\alpha_{kl} \beta_k E_k) - \mathrm{J}^{(A)}(E_l,\beta_l E_l) \right) \right\rbrack,\nonumber\\
T_{2,12}^{(m_k,m_l),\mu} &=& \dfrac{1}{2v_{kl}}\left(k_k^{\mu}\dfrac{m^2_l}{(k_k \mydot k_l)} - k_l^{\mu}\right) \left\lbrace \dfrac{m_k^2}{(k_k \mydot k_l)^2 v_{kl}^2} \log\!\left(\dfrac{\xicut^2 s}{Q_{\rm ES}^2}\right) \left\lbrack \log\!\left(\dfrac{1+v_{kl}}{1-v_{kl}}\right) - \dfrac{2v_{kl}}{1-v_{kl}^2} \right\rbrack \right.\nonumber\\
&&\left. - \left( \mathrm{J}^{(A)}(\alpha_{kl}E_k,\alpha_{kl} \beta_k E_k) - \mathrm{J}^{(A)}(E_l,\beta_l E_l) \right) \right\rbrace,\nonumber\\
T_{3,12}^{(m_k,m_l),\mu} &=& \dfrac{(1+v_{kl})E_k (k_k \mydot k_l)}{\lambda\nu \left\lbrack(\nu-\alpha_{kl}E_k)^2-(\alpha_{kl}\beta_k E_k)^2\right\rbrack v_{kl} m_k^2}\left(-k_k^{\mu}\left(\alpha_{kl}v_{kl}+\dfrac{m_l^2}{k_k \mydot k_l}\right) + k_l^{\mu}\right)\nonumber\\
&&\left\lbrace \left(\nu-\alpha_{kl}E_k(1-\beta_k^2)\right)\left\lbrack \log\!\left(\dfrac{\alpha_{kl}E_k(1-\beta_{k})}{\nu}\right) + \log\!\left(\dfrac{\alpha_{kl}E_k(1+\beta_{k})}{\nu}\right) \right\rbrack \right.\nonumber\\
&&\left.+ \nu \beta_k \log\!\left(\dfrac{1+\beta_k}{1-\beta_k}\right) \right\rbrace,\nonumber\\
T_{4,12}^{(m_k,m_l),\mu}&=&\dfrac{1+v_{kl}}{2 \lambda^2 \nu^2 v_{kl}m_k^2}\left\lbrack-k_k^{\mu}\left(\nu E_k - \alpha_{kl}m_k^2\right)\left(\alpha_{kl}v_{kl}(k_k \mydot k_l)+m_l^2\right)\right.\nonumber\\
&& \left.+k_l^{\mu}(k_k \mydot k_l)(\nu E_k-v_{kl}(k_k \mydot k_l)-\alpha_{kl} m_k^2) + \delta^{\mu0} \nu v_{kl} (k_k \mydot k_l)^2 \right\rbrack \nonumber\\
&& \times \left\lbrack \dfrac{2\alpha_{kl}E_k(1-\beta_k)}{\nu-\alpha_{kl}E_k(1-\beta_k)}\log\!\left(\dfrac{\alpha_{kl}E_k(1-\beta_k)}{\nu}\right) + \dfrac{2\alpha_{kl}E_k(1+\beta_k)}{\nu-\alpha_{kl}E_k(1+\beta_k)} \right.\nonumber\\
&&\left. \times \log\!\left(\dfrac{\alpha_{kl}E_k(1+\beta_k)}{\nu}\right)+ \lambda \nu \mathrm{J}^{(A)}(\alpha_{kl}E_k,\alpha_{kl}\beta_k E_k) \right\rbrack,\nonumber\\
T_{5,12}^{(m_k,m_l),\mu} &=&\dfrac{1+v_{kl}}{2\lambda v_{kl} m_k^2} \left\lbrack k_k^{\mu}E_k(\alpha_{kl}v_{kl}(k_k \mydot k_l)+m_l^2) - k_l^{\mu}E_k(k_k \mydot k_l) - \delta^{\mu0} v_{kl} (k_k \mydot k_l)^2 \right\rbrack \nonumber\\
&&\times\left\lbrack \mathrm{J}^{(A)}(\alpha_{kl}E_k,\alpha_{kl} \beta_k E_k) - \mathrm{J}^{(A)}(E_l,\beta_l E_l) \right\rbrack,\nonumber\\
T_{6,12}^{(m_k,m_l),\mu} &=& \dfrac{(1+v_{kl})(k_k \mydot k_l)^2}{\lambda \nu m_k^2\left\lbrack (\nu-E_l)^2-(\beta_l E_l)^2\right\rbrack} \delta^{\mu0} \left\lbrace \left( \nu - E_l(1-\beta_l^2) \right) \left\lbrack \log\!\left(\dfrac{E_l(1-\beta_l)}{\nu}\right)\right. \right.\nonumber\\
&&\left.\left.+ \log\!\left(\dfrac{E_l(1+\beta_l)}{\nu}\right) \right\rbrack + \nu \beta_l \log\!\left(\dfrac{1+\beta_l}{1-\beta_l}\right) \right\rbrace,\nonumber
\end{eqnarray}
\begin{eqnarray}
T_{7,12}^{(m_k,m_l),\mu}&=&\dfrac{(1+v_{kl})(k_k \mydot k_l)^2}{\lambda \nu \beta_l E_l^2 m_k^2} \bigg(k_l^{\mu} E_l - \delta^{\mu0} m_l^2\bigg) \left\lbrace \dfrac{1}{\nu - E_l(1-\beta_l)}\log\!\left(\dfrac{E_l(1-\beta_l)}{\nu}\right) \right.\nonumber\\
&&\left. - \dfrac{1}{\nu - E_l(1+\beta_l)}\log\!\left(\dfrac{E_l(1+\beta_l)}{\nu}\right) - \dfrac{1}{E_l(1-\beta_l^2)}\log\!\left(\dfrac{1+\beta_l}{1-\beta_l}\right) \right\rbrace,\nonumber\\
T_{8,12}^{(m_k,m_l),\mu}&=& \dfrac{1+v_{kl}}{2 \lambda^2 \nu^2 v_{kl}m_k^2}\left\lbrack k_k^{\mu}\left(\nu E_k - \alpha_{kl}m_k^2\right)\left(\alpha_{kl}v_{kl}(k_k \mydot k_l)+m_l^2\right) \right.\nonumber\\
&& \left.- k_l^{\mu}(k_k \mydot k_l)(\nu E_k-v_{kl}(k_k \mydot k_l)-\alpha_{kl} m_k^2) - \delta^{\mu0} \nu v_{kl} (k_k \mydot k_l)^2 \right\rbrack \nonumber\\
&&\times \left\lbrack \dfrac{2 E_l(1-\beta_l)}{\nu-E_l(1-\beta_l)}\log\!\left(\dfrac{E_l(1-\beta_l)}{\nu}\right) + \dfrac{2 E_l(1+\beta_l)}{\nu-E_l(1+\beta_l)}\log\!\left(\dfrac{E_l(1+\beta_l)}{\nu}\right) \right.\nonumber\\
&&\left. + \lambda \nu \mathrm{J}^{(A)}(E_l,\beta_l E_l) \right\rbrack.
\end{eqnarray}
We have verified that if we take $k=l$, the two-massive eikonal integrals are reduced to the massive self-eikonal case.

\section{Infrared poles of one-loop matrix elements\label{sec:IRpoles4virt}}

We present here the general IR poles of the UV renormalised one-loop matrix element at NLO QCD for arbitrary process that involves a quarkonium and elementary particles by assuming the validity of the KLN theorem. The one-loop virtual matrix element can be written as
\begin{eqnarray}
\ampsqo^{(n-1,1)}(\dot{r}^{\isubrmv})=\frac{\alpha_s}{2\pi}\frac{(4\pi)^\ep}{\Gamma(1-\ep)}
\left(\frac{\mu^2}{Q_{\rm ES}^2}\right)^\ep {\mathds V}(\dot{r}^{\isubrmv})\,.
\end{eqnarray}
We can derive the IR poles of ${\mathds V}(\dot{r}^{\isubrmv})$ generally from our integrated counterterms with its finite term denoted as ${\mathds V}^{(n-1,1)}_{\rm FIN}(\dot{r}^{\isubrmv})$. We remind readers that in the process $\dot{r}^{\isubrmv}$, the particles with their indices from $\nini$ to $\nlightB+2$ are coloured and massless, while those from $\nlightB+3$ to $\nlightB+\nheavy$ ($\nheavy\geq 2$) are massive coloured elementary particles. The index of the quarkonium is $\nlightB+\nheavy+1$.

\subsection{Colour-singlet S-wave state}

For a colour-singlet S-wave state with $\dot{r}^{\isubrmv}$ being eq.\eqref{eq:BornCSSwave}, the IR poles can be obtained by taking the bound state as a colour-singlet elementary particle, which amount to
\begin{eqnarray}
{\mathds V}(\dot{r}^{\isubrmv})&=&-\Bigg(
\frac{1}{\ep^2}\sum_{k=\nini}^{\nlightB+2}C(\ident_k)
+\frac{1}{\ep}\sum_{k=\nini}^{\nlightB+2}\gamma(\ident_k)
+\frac{1}{\ep}\sum_{k=\nlightB+3}^{\nlightB+\nheavy}C(\ident_k)
\Bigg)\ampsqnmoo(\dot{r}^{\isubrmv})
\nonumber\\
&&+\frac{1}{\ep}\sum_{k=\nini}^{\nlightB+2}
\sum_{l=k+1}^{\nlightB+\nheavy}\log\!\left(\frac{2k_k\mydot k_l}{Q_{\rm ES}^2}\right)
\ampsqnmoo_{kl}(\dot{r}^{\isubrmv})
\nonumber\\
&&+\frac{1}{2\ep}\sum_{k=\nlightB+3}^{\nlightB+\nheavy-1}
\sum_{l=k+1}^{\nlightB+\nheavy}
\frac{1}{v_{kl}}\log\!\left(\frac{1+v_{kl}}{1-v_{kl}}\right)
\ampsqnmoo_{kl}(\dot{r}^{\isubrmv})
\nonumber\\
&&-\frac{1}{2\ep}\sum_{k=\nlightB+3}^{\nlightB+\nheavy}
\log\!\left(\frac{m_k^2}{Q_{\rm ES}^2}\right)
\sum_{l=\nini}^{\nlightB+2}\ampsqnmoo_{kl}(\dot{r}^{\isubrmv})
+{\mathds V}^{(n-1,1)}_{\rm FIN}(\dot{r}^{\isubrmv})\,.
\end{eqnarray}
This equation is equivalent to eq.(B.2) in ref.~\cite{Frederix:2009yq}.\vfill

\subsection{Colour-octet S-wave state}

The colour-octet S-wave case with $\dot{r}^{\isubrmv}$ being eq.\eqref{eq:BornCOSwave} has the IR poles as follows:
\begin{eqnarray}
{\mathds V}(\dot{r}^{\isubrmv})&=&-\Bigg(
\frac{1}{\ep^2}\sum_{k=\nini}^{\nlightB+2}C(\ident_k)
+\frac{1}{\ep}\sum_{k=\nini}^{\nlightB+2}\gamma(\ident_k)
+\frac{1}{\ep}\sum_{k=\nlightB+3}^{\nlightB+\nheavy+1}C(\ident_k)
\Bigg)\ampsqnmoo(\dot{r}^{\isubrmv})
\nonumber\\
&&+\frac{1}{\ep}\sum_{k=\nini}^{\nlightB+2}
\sum_{l=k+1}^{\nlightB+\nheavy+1}\log\!\left(\frac{2k_k\mydot k_l}{Q_{\rm ES}^2}\right)
\ampsqnmoo_{kl}(\dot{r}^{\isubrmv})
\nonumber\\
&&+\frac{1}{2\ep}\sum_{k=\nlightB+3}^{\nlightB+\nheavy}
\sum_{l=k+1}^{\nlightB+\nheavy+1}
\frac{1}{v_{kl}}\log\!\left(\frac{1+v_{kl}}{1-v_{kl}}\right)
\ampsqnmoo_{kl}(\dot{r}^{\isubrmv})
\nonumber\\
&&-\frac{1}{2\ep}\sum_{k=\nlightB+3}^{\nlightB+\nheavy+1}
\log\!\left(\frac{m_k^2}{Q_{\rm ES}^2}\right)
\sum_{l=\nini}^{\nlightB+2}\ampsqnmoo_{kl}(\dot{r}^{\isubrmv})
+{\mathds V}^{(n-1,1)}_{\rm FIN}(\dot{r}^{\isubrmv})\,.
\end{eqnarray}

\subsection{Colour-singlet spin-singlet P-wave state}

The IR poles of the virtual matrix elements for a colour-singlet spin-singlet P-wave quarkonium production are
\begin{eqnarray}
{\mathds V}(\dot{r}^{\isubrmv})&=&-\Bigg(
\frac{1}{\ep^2}\sum_{k=\nini}^{\nlightB+2}C(\ident_k)
+\frac{1}{\ep}\sum_{k=\nini}^{\nlightB+2}\gamma(\ident_k)
+\frac{1}{\ep}\sum_{k=\nlightB+3}^{\nlightB+\nheavy}C(\ident_k)
\Bigg)\ampsqnmoo(\dot{r}^{\isubrmv})
\nonumber\\
&&+\frac{1}{\ep}\sum_{k=\nini}^{\nlightB+2}
\sum_{l=k+1}^{\nlightB+\nheavy}\log\!\left(\frac{2k_k\mydot k_l}{Q_{\rm ES}^2}\right)
\ampsqnmoo_{kl}(\dot{r}^{\isubrmv})
\nonumber\\
&&+\frac{1}{2\ep}\sum_{k=\nlightB+3}^{\nlightB+\nheavy-1}
\sum_{l=k+1}^{\nlightB+\nheavy}
\frac{1}{v_{kl}}\log\!\left(\frac{1+v_{kl}}{1-v_{kl}}\right)
\ampsqnmoo_{kl}(\dot{r}^{\isubrmv})
\nonumber\\
&&-\frac{1}{2\ep}\sum_{k=\nlightB+3}^{\nlightB+\nheavy}
\log\!\left(\frac{m_k^2}{Q_{\rm ES}^2}\right)
\sum_{l=\nini}^{\nlightB+2}\ampsqnmoo_{kl}(\dot{r}^{\isubrmv})
\nonumber\\
&&+\frac{1}{\ep}\sum_{k=\nini}^{\nlightB+2}\dfrac{k_{k,\mu}}{K\mydot k_k}\ampsqnmoo_{k[18]}(\dot{r}^{\isubrmv},\dot{r}_1^{\isubrmv})^\mu\nonumber\\
&&+\frac{1}{2\ep}\sum_{k=\nlightB+3}^{\nlightB+\nheavy}\dfrac{k_{k,\mu}}{K\mydot k_k} \frac{m_k^2K^2}{ v_{k}^3(K\mydot k_k)^2}\left[\frac{2v_k}{1-v_k^2}-\log\!\left(\frac{1+v_k}{1-v_k}\right)\right]\ampsqnmoo_{k[18]}(\dot{r}^{\isubrmv},\dot{r}_1^{\isubrmv})^\mu\nonumber\\
&&-\frac{1}{\ep}\left(\frac{1}{2\NC}\frac{8}{m_{\Quark}m_{\Antiquark}}-\frac{2}{K^2}C_{{\rm eff}}(\QQ_{[18]})\right)\ampsqnmoo(\dot{r}_1^{\isubrmv})^{\mu\nu}+{\mathds V}^{(n-1,1)}_{\rm FIN}(\dot{r}^{\isubrmv})\,,
\end{eqnarray}
where we have defined the abbreviation for $v_{kl}$ with $l=\nlightB+\nheavy+1$ (quarkonium)
\begin{eqnarray}
v_k&=&\sqrt{1-\frac{k_k^2K^2}{(k_k\cdot K)^2}}.
\end{eqnarray}
Note that, the two reduced Born processes $\dot{r}^{\isubrmv}$ and $\dot{r}_1^{\isubrmv}$ have been given in eq.\eqref{eq:BornCS1Pwave}.

\subsection{Colour-octet spin-singlet P-wave state}

For the colour-octet spin-singlet P-wave state, the IR poles of the virtual corrections are
\begin{eqnarray}
{\mathds V}(\dot{r}^{\isubrmv})&=&-\Bigg(
\frac{1}{\ep^2}\sum_{k=\nini}^{\nlightB+2}C(\ident_k)
+\frac{1}{\ep}\sum_{k=\nini}^{\nlightB+2}\gamma(\ident_k)
+\frac{1}{\ep}\sum_{k=\nlightB+3}^{\nlightB+\nheavy+1}C(\ident_k)
\Bigg)\ampsqnmoo(\dot{r}^{\isubrmv})
\nonumber\\*&&
+\frac{1}{\ep}\sum_{k=\nini}^{\nlightB+2}
\sum_{l=k+1}^{\nlightB+\nheavy+1}\log\!\left(\frac{2k_k\mydot k_l}{Q_{\rm ES}^2}\right)
\ampsqnmoo_{kl}(\dot{r}^{\isubrmv})
\nonumber\\*&&
+\frac{1}{2\ep}\sum_{k=\nlightB+3}^{\nlightB+\nheavy}
\sum_{l=k+1}^{\nlightB+\nheavy+1}
\frac{1}{v_{kl}}\log\!\left(\frac{1+v_{kl}}{1-v_{kl}}\right)
\ampsqnmoo_{kl}(\dot{r}^{\isubrmv})
\nonumber\\*&&
-\frac{1}{2\ep}\sum_{k=\nlightB+3}^{\nlightB+\nheavy+1}
\log\!\left(\frac{m_k^2}{Q_{\rm ES}^2}\right)
\sum_{l=\nini}^{\nlightB+2}\ampsqnmoo_{kl}(\dot{r}^{\isubrmv})
\nonumber\\*&&
+\frac{1}{\ep}\sum_{k=\nini}^{\nlightB+2}\dfrac{k_{k,\mu}}{K\mydot k_k}\left(\ampsqnmoo_{k[88]}(\dot{r}^{\isubrmv},\dot{r}_1^{\isubrmv})^\mu+\ampsqnmoo_{k[81]}(\dot{r}^{\isubrmv},\dot{r}_2^{\isubrmv})^\mu\right)\nonumber\\
&&+\frac{1}{2\ep}\sum_{k=\nlightB+3}^{\nlightB+\nheavy}\dfrac{k_{k,\mu}}{K\mydot k_k} \frac{m_k^2K^2}{ v_{k}^3(K\mydot k_k)^2}\left[\frac{2v_k}{1-v_k^2}-\log\!\left(\frac{1+v_k}{1-v_k}\right)\right]\nonumber\\
&&\times\left(\ampsqnmoo_{k[88]}(\dot{r}^{\isubrmv},\dot{r}_1^{\isubrmv})^\mu+\ampsqnmoo_{k[81]}(\dot{r}^{\isubrmv},\dot{r}_2^{\isubrmv})^\mu\right)\nonumber\\
&&-\frac{1}{\ep}\left(\BF\frac{8}{m_{\Quark}m_{\Antiquark}}-\frac{2}{K^2}C_{{\rm eff}}(\QQ_{[88]})\right)\ampsqnmoo(\dot{r}_1^{\isubrmv})\nonumber\\
&&-\frac{1}{\ep}\left(\CF\frac{8}{m_{\Quark}m_{\Antiquark}}-\frac{2}{K^2}C_{{\rm eff}}(\QQ_{[81]})\right)\ampsqnmoo(\dot{r}_2^{\isubrmv})+{\mathds V}^{(n-1,1)}_{\rm FIN}(\dot{r}^{\isubrmv})\,.
\end{eqnarray}
The three reduced Born processes $\dot{r}^{\isubrmv}$, $\dot{r}_1^{\isubrmv}$ and $\dot{r}_2^{\isubrmv}$ have been given in eq.\eqref{eq:BornCO1Pwave}.\vfill

\subsection{Colour-singlet spin-triplet P-wave state}
The IR poles of the one-loop virtual corrections for the processes involving a colour-singlet spin-triple P-wave quarkonium are
\begin{eqnarray}
{\mathds V}(\dot{r}^{\isubrmv})&=&-\Bigg(
\frac{1}{\ep^2}\sum_{k=\nini}^{\nlightB+2}C(\ident_k)
+\frac{1}{\ep}\sum_{k=\nini}^{\nlightB+2}\gamma(\ident_k)
+\frac{1}{\ep}\sum_{k=\nlightB+3}^{\nlightB+\nheavy}C(\ident_k)
\Bigg)\ampsqnmoo(\dot{r}^{\isubrmv})
\nonumber\\
&&+\frac{1}{\ep}\sum_{k=\nini}^{\nlightB+2}
\sum_{l=k+1}^{\nlightB+\nheavy}\log\!\left(\frac{2k_k\mydot k_l}{Q_{\rm ES}^2}\right)
\ampsqnmoo_{kl}(\dot{r}^{\isubrmv})
\nonumber\\
&&+\frac{1}{2\ep}\sum_{k=\nlightB+3}^{\nlightB+\nheavy-1}
\sum_{l=k+1}^{\nlightB+\nheavy}
\frac{1}{v_{kl}}\log\!\left(\frac{1+v_{kl}}{1-v_{kl}}\right)
\ampsqnmoo_{kl}(\dot{r}^{\isubrmv})
\nonumber\\
&&-\frac{1}{2\ep}\sum_{k=\nlightB+3}^{\nlightB+\nheavy}
\log\!\left(\frac{m_k^2}{Q_{\rm ES}^2}\right)
\sum_{l=\nini}^{\nlightB+2}\ampsqnmoo_{kl}(\dot{r}^{\isubrmv})
\nonumber\\
&&+\frac{1}{\ep}\sum_{k=\nini}^{\nlightB+2}\dfrac{k_{k,\mu}}{K\mydot k_k}\ampsqnmoo_{J,k[18]}(\dot{r}^{\isubrmv},\dot{r}_1^{\isubrmv})^\mu\nonumber\\
&&+\frac{1}{2\ep}\sum_{k=\nlightB+3}^{\nlightB+\nheavy}\dfrac{k_{k,\mu}}{K\mydot k_k} \frac{m_k^2K^2}{ v_{k}^3(K\mydot k_k)^2}\left[\frac{2v_k}{1-v_k^2}-\log\!\left(\frac{1+v_k}{1-v_k}\right)\right]\ampsqnmoo_{J,k[18]}(\dot{r}^{\isubrmv},\dot{r}_1^{\isubrmv})^\mu\nonumber\\
&&-\frac{1}{\ep}\frac{2}{3K^2}g_{\mu\nu}C_{{\rm eff}}(\QQ_{[18]})\ampsqnmoo_J(\dot{r}_1^{\isubrmv})^{\mu\nu}-\frac{1}{\ep}\frac{2J+1}{2\NC}\frac{8}{9m_{\Quark}m_{\Antiquark}}\ampsqnmoo(\dot{r}_1^{\isubrmv})\nonumber\\
&&+{\mathds V}^{(n-1,1)}_{\rm FIN}(\dot{r}^{\isubrmv})\,,
\end{eqnarray}
where the processes $\dot{r}^{\isubrmv}$ and $\dot{r}_1^{\isubrmv}$ are eq.\eqref{eq:BornCS3Pwave}.\vfill

\subsection{Colour-octet spin-triplet P-wave state}

Similarly, the IR poles of the virtual corrections for the colour-octet spin-triplet P-wave quarkonium are
\begin{eqnarray}
{\mathds V}(\dot{r}^{\isubrmv})&=&-\Bigg(
\frac{1}{\ep^2}\sum_{k=\nini}^{\nlightB+2}C(\ident_k)
+\frac{1}{\ep}\sum_{k=\nini}^{\nlightB+2}\gamma(\ident_k)
+\frac{1}{\ep}\sum_{k=\nlightB+3}^{\nlightB+\nheavy+1}C(\ident_k)
\Bigg)\ampsqnmoo(\dot{r}^{\isubrmv})
\nonumber\\*&&
+\frac{1}{\ep}\sum_{k=\nini}^{\nlightB+2}
\sum_{l=k+1}^{\nlightB+\nheavy+1}\log\!\left(\frac{2k_k\mydot k_l}{Q_{\rm ES}^2}\right)
\ampsqnmoo_{kl}(\dot{r}^{\isubrmv})
\nonumber\\*&&
+\frac{1}{2\ep}\sum_{k=\nlightB+3}^{\nlightB+\nheavy}
\sum_{l=k+1}^{\nlightB+\nheavy+1}
\frac{1}{v_{kl}}\log\!\left(\frac{1+v_{kl}}{1-v_{kl}}\right)
\ampsqnmoo_{kl}(\dot{r}^{\isubrmv})
\nonumber\\*&&
-\frac{1}{2\ep}\sum_{k=\nlightB+3}^{\nlightB+\nheavy+1}
\log\!\left(\frac{m_k^2}{Q_{\rm ES}^2}\right)
\sum_{l=\nini}^{\nlightB+2}\ampsqnmoo_{kl}(\dot{r}^{\isubrmv})
\nonumber\\*&&
+\frac{1}{\ep}\sum_{k=\nini}^{\nlightB+2}\dfrac{k_{k,\mu}}{K\mydot k_k}\left(\ampsqnmoo_{J,k[88]}(\dot{r}^{\isubrmv},\dot{r}_1^{\isubrmv})^\mu+\ampsqnmoo_{J,k[81]}(\dot{r}^{\isubrmv},\dot{r}_2^{\isubrmv})^\mu\right)\nonumber\\
&&+\frac{1}{2\ep}\sum_{k=\nlightB+3}^{\nlightB+\nheavy}\dfrac{k_{k,\mu}}{K\mydot k_k} \frac{m_k^2K^2}{ v_{k}^3(K\mydot k_k)^2}\left[\frac{2v_k}{1-v_k^2}-\log\!\left(\frac{1+v_k}{1-v_k}\right)\right]\nonumber\\
&&\times\left(\ampsqnmoo_{J,k[88]}(\dot{r}^{\isubrmv},\dot{r}_1^{\isubrmv})^\mu+\ampsqnmoo_{J,k[81]}(\dot{r}^{\isubrmv},\dot{r}_2^{\isubrmv})^\mu\right)\nonumber\\
&&-\frac{1}{\ep}\frac{2}{3K^2}g_{\mu\nu}\left(C_{{\rm eff}}(\QQ_{[88]})\ampsqnmoo_J(\dot{r}_1^{\isubrmv})^{\mu\nu}+C_{{\rm eff}}(\QQ_{[81]})\ampsqnmoo_J(\dot{r}_2^{\isubrmv})^{\mu\nu}\right)\nonumber\\
&&-\frac{1}{\ep}\frac{8(2J+1)}{9m_{\Quark}m_{\Antiquark}}\left(\BF\ampsqnmoo(\dot{r}_1^{\isubrmv})+\CF\ampsqnmoo(\dot{r}_2^{\isubrmv})\right)+{\mathds V}^{(n-1,1)}_{\rm FIN}(\dot{r}^{\isubrmv})\,.
\end{eqnarray}
The processes $\dot{r}^{\isubrmv}$, $\dot{r}_1^{\isubrmv}$ and $\dot{r}_2^{\isubrmv}$ are eq.\eqref{eq:BornCO3Pwave}.

%
%
%
%
%
%
\bibliographystyle{JHEP}        
\bibliography{paper}
%
%
%
%
%
\end{document}